\pdfoutput=1
\documentclass[12pt]{article}
\usepackage{graphicx}
\usepackage{latexsym}
\usepackage{amssymb, amsmath, amsfonts}
\usepackage{indentfirst}
\usepackage{float,times}
\usepackage{bbm}
\usepackage{subfigure}
\usepackage{slashed}
\usepackage{ifthen}
\usepackage{mathrsfs}
\usepackage{setspace}
\newskip\humongous \humongous=0pt plus 1000pt minus 1000pt

\newif\ifdtup

\def\oldreffmt#1{\rlap{[#1]} \hbox to 2\parindent{}}

\def\figfmt#1{\rlap{Figure {#1}} \hbox to 1in{}}

\def\beq{\begin{equation}}
\def\eeq{\end{equation}}
\def\bea{\begin{eqnarray}}
\def\eea{\end{eqnarray}}

\def\bq{\begin{quote}}
\def\eq{\end{quote}}

\def\GeV{\,{\rm GeV}}

\relax
\def\be{\begin{equation}} 
\def\ee{\end{equation}} 
\def\bea{\begin{eqnarray}}
\def\eea{\end{eqnarray}}
\def\ba{\begin{array}}
\def\ea{\end{array}}
\def\kslash{\raise.15ex\hbox{/}\kern-.57em k}

\def\GeV{{\rm GeV}}

\newcommand{\bear}{\begin{eqnarray}}
\newcommand{\eear}{\end{eqnarray}}

\newcommand{\ctfig}[3]{
  \begin{figure}[tp]
    \begin{center}
      \includegraphics[width=#1\linewidth]{#2}
      \caption{#3}
      \label{#2}
    \end{center}
\end{figure}
}
\newcommand{\nn}{~\nonumber\\}
\newcommand{\sfr}[2]{{\textstyle\frac{#1}{#2}}}

\def\beq{\begin{equation}}
\def\eeq{\end{equation}}
\def\bal{\begin{align}}
\def\eal{\end{align}}

\newcommand{\drawsquare}[2]{\hbox{%
\rule{#2pt}{#1pt}\hskip-#2pt% left vertical
\rule{#1pt}{#2pt}\hskip-#1pt% lower horizontal
\rule[#1pt]{#1pt}{#2pt}}\rule[#1pt]{#2pt}{#2pt}\hskip-#2pt%upper horizontal
\rule{#2pt}{#1pt}}% right vertical
\newcommand{\Yfund}{\raisebox{-.5pt}{\drawsquare{6.5}{0.4}}}% fund
\newcommand{\Ysymm}{\Yfund\hskip-0.4pt%
                    \Yfund}% symmetric second rank
\def\symm{\Ysymm}
\def\bsymm{\overline{\Ysymm}}
\def\drawbox#1#2{\hrule height#2pt
        \hbox{\vrule width#2pt height#1pt \kern#1pt
              \vrule width#2pt}
              \hrule height#2pt}
\def\Fund#1#2{\vcenter{\vbox{\drawbox{#1}{#2}}}}
\def\Asym#1#2{\vcenter{\vbox{\drawbox{#1}{#2}
              \kern-#2pt % line up boxes
              \drawbox{#1}{#2}}}}

\def\fund{\Fund{6.4}{0.3}}
\def\asymm{\Asym{6.4}{0.3}}
\def\bfund{\overline{\fund}}

\begin{document}
\begin{titlepage}
 \begin{center}
{\huge \bf
 Dynamical Stabilization  of the Fermi Scale}\\ [6mm]
{\Large 
 Phase Diagram of Strongly Coupled Theories \\[4mm] for \\[4mm] (Minimal) Walking Technicolor and Unparticles}
 \end{center}
 \par \vskip .2in \noindent
\begin{center}
{\bf  \Large Francesco Sannino}
\end{center}
\begin{center}
  \par \vskip .1in \noindent
{\large \it
University of Southern Denmark, Campusvej 55, DK-5230, Odense M.
}
   \par \vskip .1in \noindent
\end{center}     
\begin{center}{\large Abstract}\end{center}
\begin{quote}

We summarize basic features associated to dynamical breaking of the electroweak symmetry. The knowledge of the phase diagram of strongly coupled theories as function of the number of colors, flavors and matter representation plays a fundamental role when trying to construct viable extensions of the standard model (SM). Therefore we will report on the status of the phase diagram for $SU(N)$ gauge theories with fermionic matter transforming according to arbitrary representations of the underlying gauge group. We will discuss how the phase diagram can be used to construct unparticle models. 
We will then review Minimal Walking Technicolor (MWT) and other extensions, such as partially gauged and split technicolor. MWT is a sufficiently general, symmetry wise, model to be considered as a benchmark for any model aiming at breaking the electroweak symmetry dynamically.  The unification of the standard model gauge couplings will be revisited within technicolor extensions of the SM.  A number of appendices are added to review some basic methods and to provide useful details. In one of the appendices we will show how to gain information on the spectrum of strongly coupled theories relevant for new extensions of the SM by introducing and using alternative large N limits. 
 
\end{quote}
\par \vskip .1in
\vfill
\end{titlepage}
     
     \newpage

~
\vskip 7cm
\begin{flushright}
{\Large \it To Lene, Maria Sofia and Matthias}
\end{flushright}
\newpage

     \newpage

\def\baselinestretch{1.0}
\tiny
\normalsize

\tableofcontents

\newpage

\section{Introduction}
A number of possible generalizations of the standard model (SM) have been
conceived (see \cite{Giudice:2007qj} for a nice review). Such extensions are introduced on the base of one or more guiding principles or prejudices.

In the models we will consider here the electroweak symmetry breaks via a
fermion bilinear condensate. The Higgs sector of the standard model
becomes an effective description of a more fundamental fermionic
theory. This is similar to the Ginzburg-Landau theory of
superconductivity. If the force underlying the fermion condensate driving electroweak symmetry
breaking is due to a strongly interacting gauge theory these models are termed
technicolor. 

Technicolor, in brief, is an additional non-abelian and strongly interacting gauge theory augmented
with (techni)fermions transforming under a given representation of the gauge group.
The Higgs Lagrangian is replaced by a suitable
new fermion sector interacting strongly via a new gauge interaction (technicolor). Schematically:
\begin{eqnarray}
L_{Higgs} \rightarrow -\frac{1}{4}F_{\mu\nu}F^{\mu\nu} + i \bar{Q} \gamma_{\mu}D^{\mu} Q  + \cdots\ ,
\end{eqnarray}
where, to be as general as possible, we have left unspecified the underlying nonabelian gauge group and the associated technifermion ($Q$) representation. The dots represent new sectors which may even be needed to avoid, for example, anomalies introduced by the technifermions. 
The intrinsic scale of the new theory is expected to be less or of the order of one TeV.
The chiral-flavor symmetries of this theory, as for ordinary QCD,
break spontaneously when the technifermion condensate $\bar{Q} Q $ forms. It is
possible to choose the fermion charges in such a way that there is,
at least, a weak left-handed doublet of technifermions and the
associated right-handed one which is a weak singlet. The covariant derivative contains the new gauge field as well as the electroweak ones. The condensate
spontaneously breaks the electroweak symmetry down to the
electromagnetic and weak interactions.
The Higgs is now interpreted as the lightest scalar field with the same quantum numbers of the fermion-antifermion composite field. The Lagrangian part responsible for the mass-generation of the ordinary fermions will also be modified since the Higgs particle is no longer an elementary object. 

Models of electroweak symmetry breaking via new strongly interacting
theories of technicolor type \cite{Weinberg:1979bn,Susskind:1978ms}
are a mature subject (for recent reviews see
\cite{Hill:2002ap,Lane:2002wv}). One of the main difficulties in
constructing such extensions
of the standard model is the very
limited knowledge about generic strongly interacting theories. This
has led theorists to consider specific models of technicolor which
resemble ordinary quantum chromodynamics and for which the large
body of experimental data at low energies can be directly exported
to make predictions at high energies. Unfortunately the simplest version of this type of models are at odds with electroweak precision measurements. New strongly coupled theories with dynamics very different from the one featured by a scaled up version of QCD are needed \cite{Sannino:2004qp}. 

We will review models of dynamical electroweak symmetry breaking making use of new type of four dimensional gauge theories \cite{Sannino:2004qp,Dietrich:2005jn,Dietrich:2005wk} and their low energy effective description \cite{Foadi:2007ue} useful for collider phenomenology. The phase structure of a large number of  strongly interacting nonsupersymmetric theories, as function of number of underlying colors will be uncovered with traditional nonperturbative methods \cite{Dietrich:2006cm} as well as novel ones \cite{{Ryttov:2007cx}}. Finally we will explore the unification of the standard model couplings within technicolor extensions \cite{Farhi:1979zx, Gudnason:2006mk}. 

\newpage
\section{Dynamical Electroweak Symmetry Breaking}
Our goal is to review recent theoretical developments not present in earlier reviews on dynamical breaking of the electroweak theory. Hence we urge the reader to complement the present reading with earlier ones. The most recent reviews are the ones by Hill and Simmons \cite{Hill:2002ap} and Lane \cite{Lane:2002wv}. Other reviews are available \cite{Chanowitz:1988ae,Farhi:1980xs,Kaul:1981uk,Chivukula:2000mb}.

It is a fact that the standard model (SM) does not fail, when experimentally tested, to describe all of the known forces to a very high degree of experimental accuracy. This is true even if we include gravity. Why is it so successful?

The SM is a low energy effective theory valid up to a scale $\Lambda$ above which new interactions, symmetries, extra dimensional worlds or any possible extension can emerge. At sufficiently low energies with respect to the cutoff scale $\Lambda$ one expresses the existence of new physics via effective operators. The success of the SM is due to the fact that most of the corrections to its physical observable depend only logarithmically on the cutoff scale $\Lambda$. 

Superrenormalizable operators are very sensitive to the cut off scale. In the standard model there exists only one operator with naive mass dimension two which acquires corrections quadratic in $\Lambda$. This is the squared mass operator of the Higgs boson.  Since $\Lambda$ is expected to be the highest possible scale, in four dimensions the Planck scale, it is hard to explain {\it naturally}  why the mass of the Higgs is of the order of the electroweak scale. The Higgs is also the only particle predicted in the SM yet to be directly produced in experiments. Due to the occurrence of quadratic corrections in the cutoff this is the SM sector highly sensitve to the existence of new physics. 

In Nature we have already observed Higgs-type mechanisms. Ordinary superconductivity and chiral symmetry breaking in QCD are two time-honored examples. In both cases the mechanism has an underlying dynamical origin with the Higgs-like particle being a composite object of fermionic fields. 

\subsection{The Higgs Sector of the Standard Model}
 The Higgs sector of the standard model
possesses, when the gauge couplings are switched off, an
$SU_L(2)\times SU_R(2)$ symmetry. The full symmetry group can be made explicit when re-writing the Higgs
doublet field \begin{eqnarray} H=\frac{1}{\sqrt{2}}\left(%
\begin{array}{c}
  \pi_2 + i\, \pi_1 \\
  \sigma - i\, \pi_3 \\
\end{array}%
\right)\end{eqnarray} as the right column of the following two by two matrix:
\begin{eqnarray}
\left[i\,\tau_2H^{\ast}\
,\,H\right]=\frac{1}{\sqrt{2}}\left(\sigma + i\,
\vec{\tau}\cdot\vec{\pi} \right) \equiv M
 \ .
\end{eqnarray}
The $SU_L(2)\times SU_R(2)$ group acts linearly on $M$ according
to:
\begin{eqnarray}
M\rightarrow g_L M g_R^{\dagger} \qquad {\rm and} \qquad g_{L/R} \in SU_{L/R}(2)\ .
\end{eqnarray}
One can verify that:
\begin{eqnarray}
M\frac{\left(1-\tau^3\right)}{2} = \left[0\ , \, H\right] \ . \qquad
M\frac{\left(1+\tau^3\right)}{2} = \left[i\,\tau_2H^{\ast} \ , \, 0\right] \ .
\end{eqnarray}
The $SU_L(2)$ symmetry is gauged by introducing the weak gauge
bosons $W^a$ with $a=1,2,3$. The hypercharge generator is taken to
be the third generator of $SU_R(2)$. The ordinary covariant
derivative acting on the Higgs, in the present notation, is:
\begin{eqnarray}
D_{\mu}M=\partial_{\mu}M -i\,g\,W_{\mu}M + i\,g^{\prime}M\,B_{\mu} \ , \qquad {\rm
with}\qquad W_{\mu}=W_{\mu}^{a}\frac{\tau^{a}}{2} \ ,\quad
B_{\mu}=B_{\mu}\frac{\tau^{3}}{2} \ .
\end{eqnarray}
The Higgs Lagrangian is 
\begin{eqnarray}
{\cal L}&=&\frac{1}{2}{\rm Tr} \left[D_{\mu}M^{\dagger}
D^{\mu}M\right]-\frac{m^2}{2} {\rm
Tr}\left[M^{\dagger}M\right] - \frac{\lambda}{4}\,{\rm
Tr}\left[M^{\dagger}M\right]^2 \ .
\end{eqnarray}
At this point one {\it assumes} that the mass squared of
the Higgs field is negative and this leads to the electroweak
symmetry breaking. Except for the Higgs mass term the other SM operators have dimensionless couplings meaning that the natural scale for the SM is encoded in the Higgs mass\footnote{The mass of the proton is due mainly to strong interactions, however its value cannot be determined within QCD since the associated renormalization group invariant scale must be fixed to an hadronic observable.}  

At the tree level, when taking $m^2$ negative and the self-coupling $\lambda$ positive, one determines:
\begin{equation}
\langle \sigma \rangle^2 \equiv v_{weak}^2=\frac{|m^2|}{\lambda} \qquad {\rm and} \qquad \sigma = v_{weak} + h \ ,
\end{equation} 
where $h$ is the Higgs field. 
The global symmetry breaks to its diagonal subgroup:
\begin{eqnarray}
SU_L(2)\times SU_R(2) \rightarrow SU_V(2) \ .
\end{eqnarray}
To be more precise the $SU_R(2)$ symmetry is already broken explicitly by our choice of gauging only an $U_Y(1)$ subgroup of it and hence the actual symmetry breaking pattern is:
\begin{eqnarray}
SU_L(2)\times U_Y(1) \rightarrow U_Q(1) \ ,
\end{eqnarray}
with $U_Q(1)$ the electromagnetic abelian gauge symmetry. According to the Nambu-Goldstone's theorem three massless degrees of freedom appear, i.e. $\pi^{\pm}$ and $\pi^0$. In the unitary gauge these Goldstones become the longitudinal degree of freedom of the massive elecetroweak gauge-bosons. Substituting the vacuum value for $\sigma$ in the Higgs Lagrangian the gauge-bosons quadratic terms read:
\begin{equation}
\frac{v_{weak}^2}{8}\, \left[g^2\,\left(W_{\mu}^1
W^{\mu,1} +W_{\mu}^2 W^{\mu,2}\right)+ \left(g\,W_{\mu}^3 -
g^{\prime}\,B_{\mu}\right)^2\right]  \ . \end{equation}
 The $Z_{\mu}$ and the photon $A_{\mu}$ gauge bosons are:
\begin{eqnarray}
Z_{\mu}&=&\cos\theta_W\, W_{\mu}^3 - \sin\theta_{W}B_{\mu} \ ,\nonumber \\
A_{\mu}&=&\cos\theta_W\, B_{\mu} + \sin\theta_{W}W_{\mu}^3 \ ,
\end{eqnarray}
with $\tan\theta_{W}=g^{\prime}/g$ while the charged massive vector bosons are
$W^{\pm}_{\mu}=(W^1\pm i\,W^2_{\mu})/\sqrt{2}$. 
The bosons masses $M^2_W=g^2\,v_{weak}^2/4$ due to
the custodial symmetry satisfy the tree level relation $M^2_Z=M^2_W/\cos^2\theta_{W}$. 
Holding fixed the EW scale $v_{weak}$ the mass squared of the Higgs boson is $2\lambda v^2_{weak}$ and hence it increases with $\lambda$.  We recall that the Higgs Lagrangian has a familiar form since it is identical to the linear $\sigma$ Lagrangian which was introduced long ago to describe chiral symmetry breaking in QCD with two light flavors.

Besides breaking the electroweak symmetry dynamically the ordinary Higgs serves also the purpose to provide mass to all of the standard model particles via the Yukawa terms of the type:
\beq -Y_d^{ij}\bar{Q}_L^i H d_R^j  - Y_u^{ij}\bar{Q}_L^i (i\tau_2 H^{\ast}) u_R^j  + {\rm h.c.}\ , \eeq
where $Y_{q}$ is the Yukawa coupling constant, $Q_L$ is the
left-handed Dirac spinor of quarks, $H$ the Higgs doublet and
 $q$ the right-handed Weyl spinor for the quark and $i,j$ the flavor indices. The $SU_L(2)$ weak and spinor indices are suppressed. 

When considering quantum corrections the Higgs mass acquires large quantum corrections  proportional to the scale of the cut-off squared. 
\begin{equation}
m^2_{\rm ren} - m^2 \propto \Lambda^2 \ .
\end{equation}
 $\Lambda$ is the highest energy above which the SM is no longer a valid description of Nature and a large fine tuning of the parameters of the Lagrangian is needed to offset the effects of the cut-off. This large fine tuning is needed because there are no symmetries protecting the Higgs mass operator from large corrections which would hence destabilize the Fermi scale (i.e. the electroweak scale). This problem is often referred as the hierarchy problem of the SM.

\subsection{Superconductivity versus Electroweak Symmetry Breaking }
The breaking of the electroweak theory is a relativistic screening effect. It is useful to parallel it to ordinary superconductivity which is also a screening phenomenon albeit non-relativistic. The two phenomena happen at a temperature lower than a critical one. In the case of superconductivity one defines a density of superconductive electrons $n_s$ and to it one associates a macroscopic wave function $\psi$ such that  its modulus squared
\begin{eqnarray}
|\psi|^2 = n_C = \frac{n_s}{2} \ ,
\end{eqnarray}
is the density of Cooper's pairs. That we are describing a nonrelativistic system is manifest in the fact that the macroscopic wave function squared, in natural units, has mass dimension three while the modulus squared of the Higgs wave function evaluated at the minimum is equal to $<|H|^2> = v_{weak}^2/2$ and has mass dimension two, i.e. is a relativistic wave function. One can adjust the units by considering, instead of the wave functions, the Meissner-Mass of the photon in the superconductor which is
\begin{equation}
 M^2=q^2n_s/(4m_e) \ ,
 \end{equation}
  with $q=-2e$ and $2m_e$ the charge and the mass of a Cooper pair which is constituted by two electrons. In the electroweak theory the Meissner-Mass of the photon  is compared  with the relativistic mass of the $W$ gauge boson
  \begin{equation}
  M^2_W=g^2{v_{weak}^2}/{4}\ ,
  \end{equation}
  with $g$ the weak coupling constant and $v_{weak}$ the electroweak scale.  In a superconductor the relevant scale is given by the density of superconductive electrons typically of the order of $n_s\sim 4\times 10^{28}m^{-3}$ yielding a screening length of the order of $
\xi = 1/M\sim 10^{-6}{\rm cm}$. In the weak interaction case we measure directly the mass of the weak gauge boson which is of the order of $80$~GeV yielding a weak screening length $\xi_W=1/M_W\sim 10^{-15}{\rm cm}$.  

{}For a superconductive system it is clear from the outset that the wave function $\psi$ is not a fundamental degree of freedom, however for the Higgs we are not yet sure about its origin. The Ginzburg-Landau effective theory in terms of $\psi$ and the photon degree of freedom describes the spontaneous breaking of the $U_Q(1)$ electric symmetry and it is the equivalent of the Higgs Lagrangian.  

If the Higgs is due to a macroscopic relativistic screening phenomenon we expect it to be an effective description of a more fundamental system with possibly an underlying new strong gauge dynamics replacing the role of the phonons in the superconductive case. A dynamically generated Higgs system solves the problem of the quadratic divergences by replacing the cutoff $\Lambda$ with the weak energy scale itself, i.e. the scale of  compositness.  An underlying strongly coupled asymptotically free gauge theory, a la QCD,  is an example. 
 
 \subsection{From Color to Technicolor}
In fact even in complete absence of the Higgs sector in the SM the electroweak symmetry breaks \cite{Farhi:1980xs} due to the condensation of the following quark bilinear in QCD: 
\beq \langle\bar u_Lu_R + \bar d_Ld_R\rangle \neq 0 \ . \label{qcd-condensate}\eeq
  This mechanism, however, cannot account for the whole contribution to the weak gauge bosons masses. If QCD would be the only source contributing to the spontaneous breaking of the electroweak symmetry one would have
\beq M_W = \frac{gF_\pi}{2} \sim 29 {\rm MeV}\ , \eeq 
with $F_{\pi}\simeq 93$~MeV the pion decay constant. This contribution is very small with respect to the actual value of the $M_W$ mass that one typically neglects it. 

According to the original idea of technicolor \cite{Weinberg:1979bn,Susskind:1978ms} one augments the SM with another gauge interaction similar to QCD but with a new dynamical scale of the order of the electroweak one. It is sufficient that the new gauge theory is asymptotically free and has global symmetry able to contain the SM $SU_L(2)\times U_Y(1)$ symmetries. It is also required that the new global symmetries break dynamically in such a way that the embedded $SU_L(2)\times U_Y(1)$  breaks to the electromagnetic abelian charge $U_{Q}(1)$ . The dynamically generated scale will then be fit to the electroweak one. 

Note that, excepet in certain cases, dynamical behaviors are typically nonuniversal which means that different gauge groups and/or matter representations will, in general, posses very different dynamics. 

The simplest example of technicolor theory is the scaled up version of QCD, i.e. an $SU(N_{TC})$  nonabelian gauge theory with two Dirac Fermions transforming according to the fundamental representation or the gauge group. We need at least two Dirac flavors  to realize the $SU_L(2) \times SU_R(2)$ symmetry of the standard model discussed in the SM Higgs section. One simply chooses the scale of the theory to be such that the new pion decaying constant is: \beq F_\pi^{TC} = v_{\rm weak} \simeq 246~ {\rm GeV} \ . \eeq 
The flavor symmetries, for any $N_{TC}$ larger than 2 are $SU_L(2) \times SU_R(2)\times U_V(1)$ which spontaneously break to $SU_V(2)\times U_V(1)$. It is natural to embed the electroweak symmetries within the present technicolor model in a way that the hypercharge corresponds to the third generator of $SU_R(2)$. This simple dynamical model correctly accounts for the electroweak symmetry breaking. The new technibaryon number $U_V(1)$ can break due to not yet specified new interactions. 
In order to get some indication on the dynamics and spectrum of this theory one can use the 't Hooft large N limit  \cite{'t Hooft:1973jz, Witten:1979kh}.  {}For example the intrinsic scale of the theory is related to the QCD one via:
\beq \Lambda_{\rm TC} \sim
\sqrt{\frac{3}{N_{TC}}}\frac{F_\pi^{TC}}{F_\pi}\Lambda_{\rm QCD} \ . \eeq
At this point it is straightforward to use the QCD phenomenology for describing the experimental signatures and dynamics of a composite Higgs.  

\subsection{Constraints from Electroweak Precision Data}
\label{5}

The relevant corrections due to the presence of new physics trying to modify the electroweak breaking sector of the SM appear in the vacuum polarizations of the electroweak gauge bosons. These can be parameterized in terms of the three 
quantities $S$, $T$, and $U$ (the oblique parameters) 
\cite{Peskin:1990zt,Peskin:1991sw,Kennedy:1990ib,Altarelli:1990zd}, and confronted with the electroweak precision data. Recently, due to the increase precision of the measurements reported by LEP II, the list of interesting parameters to compute has been extended \cite{hep-ph/9306267,Barbieri:2004qk}.  We show below also the relation with the traditional one \cite{Peskin:1990zt}.  Defining with  $Q^2\equiv -q^2$ the Euclidean transferred momentum entering in a generic two point function vacuum polarization associated to the electroweak gauge bosons, and denoting derivatives with respect to $-Q^2$ with a prime we have \cite{Barbieri:2004qk}: 
\begin{eqnarray}
\hat{S} &\equiv & g^2 \ \Pi_{W^3B}^\prime (0) \ , \\
\hat{T} &\equiv & \frac{g^2}{M_W^2}\left[ \Pi_{W^3W^3}(0) -
\Pi_{W^+W^-}(0) \right] \ , \\
W &\equiv & \frac{g^2M_W^2}{2} \left[\Pi^{\prime\prime}_{W^3W^3}(0)\right] \ , \\
Y &\equiv & \frac{g'^2M_W^2}{2} \left[\Pi^{\prime\prime}_{BB}(0)\right] \ , \\
\hat{U} &\equiv & -g^2 \left[\Pi^\prime_{W^3W^3}(0)-
\Pi^\prime_{W^+W^-}(0)\right]\ , \\
V &\equiv & \frac{g^2 \, M^2_W}{2}\left[\Pi^{\prime\prime}_{W^3W^3}(0)-
\Pi^{\prime\prime}_{W^+W^-}(0)\right] \ , \\
X &\equiv & \frac{g g'\,M_W^2}{2} \ \Pi_{W^3B}^{\prime\prime}(0) \ .
\end{eqnarray}
Here $\Pi_V(Q^2)$ with $V=\{W^3B,\, W^3W^3,\, W^+W^-,\, BB\}$ represents the
self-energy of the vector bosons. Here the
electroweak couplings are the ones associated to the physical electroweak gauge bosons:
\begin{eqnarray}
\frac{1}{g^2} \equiv  \Pi^\prime_{W^+W^-}(0)
 \ , \qquad \frac{1}{g'^2}
\equiv  \Pi^\prime_{BB}(0) \ ,
\end{eqnarray}
while $G_F$ is
\begin{eqnarray}
\frac{1}{\sqrt{2}G_F}=-4\Pi_{W^+W^-}(0) \ ,
\end{eqnarray}
as in \cite{Chivukula:2004af}. $\hat{S}$ and $\hat{T}$ lend their name
from the well known Peskin-Takeuchi parameters $S$ and $T$ which are related to the new ones via
\cite{Barbieri:2004qk,Chivukula:2004af}:
\begin{eqnarray}
\frac{\alpha S}{4s_W^2} =  \hat{S} - Y - W  \ , \qquad 
\alpha T = \hat{T}- \frac{s_W^2}{1-s_W^2}Y \ .
\end{eqnarray}
Here $\alpha$ is the electromagnetic structure constant and $s_W=\sin \theta_W $
is the weak mixing angle. Therefore in the case where $W=Y=0$ we
have the simple relation
\begin{eqnarray}
\hat{S} &=& \frac{\alpha S}{4s_W^2} \ , \qquad 
\hat{T}= \alpha T \ .
\end{eqnarray}

\begin{figure}[t]
\begin{center}
\includegraphics[width=10truecm,height=10truecm]{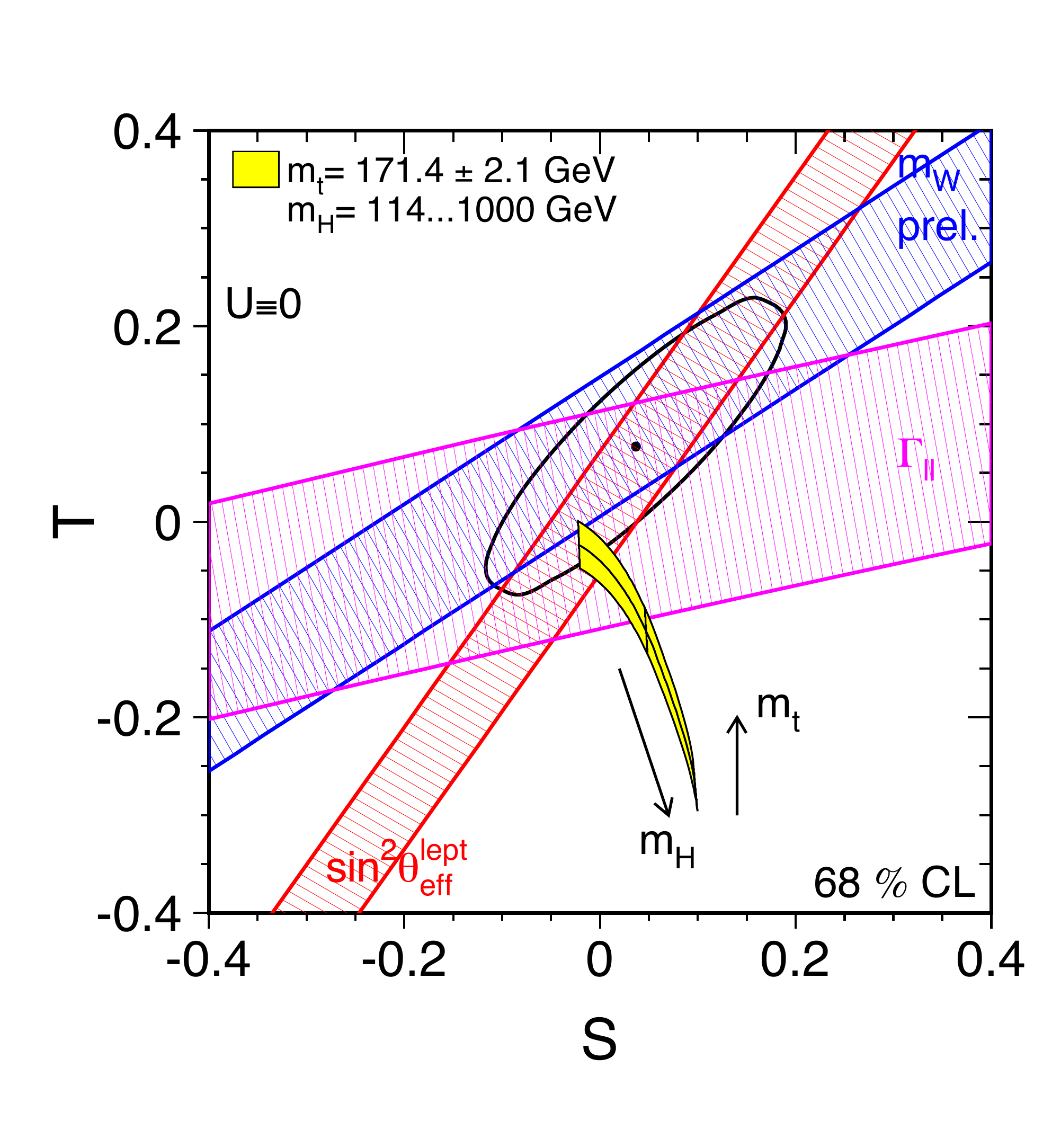}
\caption{$T$ versus $S$ one standard deviation ellipses as a result of a fit to the SM observables.} \label{s06_stu_contours}
\end{center}
\end{figure}
\noindent
The result of the current fit to precision data \cite{EWWG} is shown in
Fig.~\ref{s06_stu_contours}. If the value of the Higgs mass increases the central value of the $S$ parameters moves to the left towards negative values. 

\begin{figure}[t]
\begin{center}
\includegraphics[width=7truecm,height=4.5truecm]{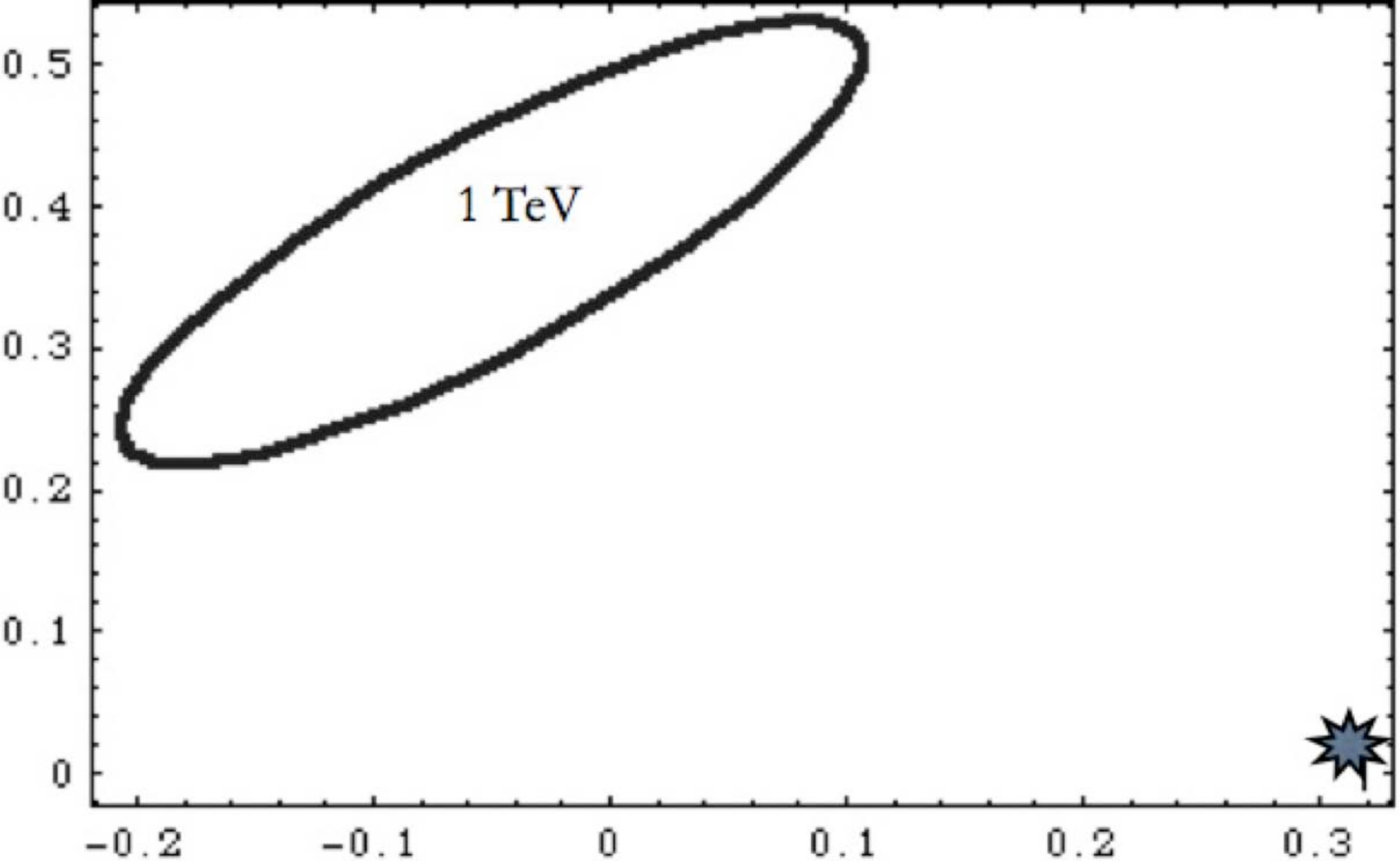}
\caption{$T$ versus $S$  for $SU(3)$ technicolor with one technifermion doublet (the full asterisk) versus precision data for a one TeV composite Higgs mass.} \label{TCSproblem}
\end{center}
\end{figure}

\noindent
In technicolor it is easy to have a vanishing $T$ parameter while typically $S$ is positive. Besides, the composite Higgs is typically heavy with respect to the Fermi scale, at least for technifermions in the fundamental representation of the gauge group and for a small number of techniflavors. The oldest technicolor models featuring QCD dynamics with three technicolors and a doublet of electroweak gauged techniflavors deviate a few sigma from the current precision tests as summarized in the figure \ref{TCSproblem}.
Clearly it is desirable to reduce the tension between the precision data and a possible dynamical mechanism underlying the electroweak symmetry breaking. It is possible to imagine different ways to achieve this goal and some of the earlier attempts have been summarized by Peskin and Wells in \cite{Peskin:2001rw}. 

The computation of the $S$ parameter in technicolor theories requires the knowledge of nonperturbative dynamics rendering difficult the precise knowledge of the contribution to  $S$. {}For example, it is not clear what is the exact value of the composite Higgs mass relative to the Fermi scale and, to be on the safe side, one typically takes it to be quite large, of the order at least of the TeV. However in certain models it may be substantially lighter due to the intrinsic dynamics. We will discuss the spectrum of different strongly coupled theories in the Appendix and its relation to the electroweak parameters later in this chapter. 

 It is, however, instructive to provide a simple estimate of the contribution to $S$ which allows to guide model builders. Consider a one-loop exchange of $N_D$ doublets of techniquarks transforming according to the representation $R_{TC}$ of the underlying technicolor gauge theory and with dynamically generated mass $\Sigma_{(0)}$ assumed to be larger than the weak intermediate gauge bosons masses. Indicating with $d(R_{\rm TC})$ the dimension of the techniquark representation, and to leading order in $M_{W}/\Sigma(0)$ one finds:
 \begin{eqnarray}
S_{\rm naive} = N_D \frac{d(R_{\rm TC})}{6\pi} \ .
\end{eqnarray} 
This naive value provides, in general, only a rough estimate of the exact value of $S$. 
However, it is clear from the formula above that, the more technicolor matter is gauged under the electroweak theory the larger is the $S$ parameter and that the final $S$ parameter is expected to be positive. 

Attention must be paid to the fact that the specific model-estimate of the whole $S$ parameter, to compare with the experimental value, receives contributions also from other sectors. Such a contribution can be taken sufficiently large and negative to compensate for the positive value from the composite Higgs dynamics. To be concrete: Consider an extension of the standard model in which the Higgs is composite but we also have new heavy (with a mass of the order of the electroweak) fourth family of Dirac leptons. In this case a sufficiently large splitting of the new lepton masses can strongly reduce and even offset the positive value of $S$. We will discuss this case in detail when presenting the Minimal Walking Technicolor model. The contribution of the new sector ($ S_{\rm NS}$) above, and also in  many other cases, is perturbatively under control and the total $S$ can be written as:
\begin{eqnarray}
S = S_{\rm TC} + S_{\rm NS} \ .
\end{eqnarray}
 The parameter $T$ will be, in general, modified and one has to make sure that the corrections do not spoil the agreement with this parameter.  From the discussion above it is clear that technicolor models can be constrained, via precision measurements, only model by model and the effects of possible new sectors must be properly included. 
 
We presented the constraints coming from $S$ using the underlying gauge theory information. However, in practice, these constraints apply directly to the physical spectrum. 
\subsection{Standard Model Fermion Masses}

Since in a purely technicolor model  the Higgs is a composite particle the Yukawa terms, when written in terms of the underlying technicolor fields, amount to four-fermion operators. The latter can be naturally interpreted as a low energy operator induced by a new strongly coupled gauge interaction emerging at energies higher than the electroweak theory. These type of theories have been termed extended technicolor interactions (ETC) \cite{Eichten:1979ah,Dimopoulos:1979es}. 

\begin{figure}[tp]
\begin{center}
\mbox{
\subfigure{\resizebox{!}{0.23\linewidth}{\includegraphics[clip=true]{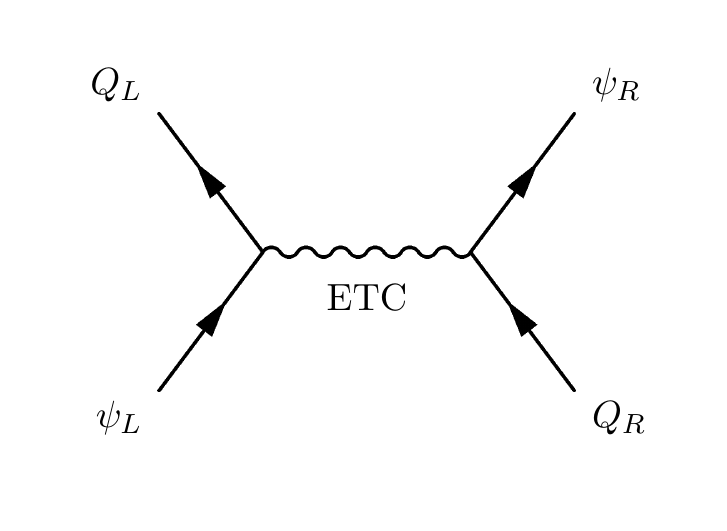}}}\qquad \qquad
\subfigure{\resizebox{!}{0.23\linewidth}{\includegraphics[clip=true]{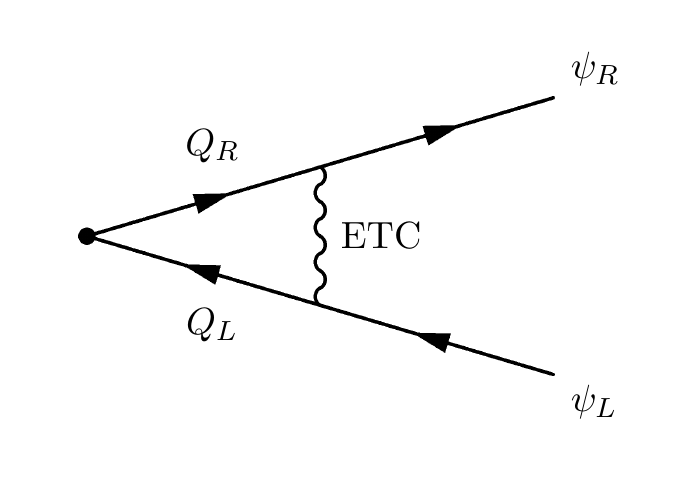}}}
}
\caption{Left Panel: ETC  gauge boson interaction involving
techniquarks and SM fermions. Right Panel: Diagram contribution to the mass to the SM fermions.}
\label{etcint}
\end{center}
\end{figure}

In the literature various extensions have been considered and we will mention them later in the text.  Here we will describe the simplest ETC model in which the ETC interactions connect the chiral symmetries of the techniquarks to those of the SM fermions (see Left panel of Fig.~\ref{etcint}).

When TC chiral symmetry breaking occurs it leads to the diagram drawn in
Fig.~\ref{etcint}b. Let's start with the case in which the ETC dynamics is represented by a $SU(N_{ETC})$ gauge group with: 
\beq N_{ETC} = N_{TC} + N_g \ , \eeq
and $N_g$ is the number of SM
generations. In order to give masses to all of the SM fermions, in this scheme, one needs a condensate for each SM fermion. This can be achieved by using as technifermion matter a complete generation of quarks and leptons (including a neutrino right) but now gauged with respect to the technicolor interactions.  

The ETC gauge group is assumed to spontaneously break $N_g$ times down to
$SU(N_{TC})$ permitting
three different mass scales, one  for each SM family. This type of technicolor with associated ETC is
termed the \emph{one family model} \cite{Farhi:1979zx}.
The heavy masses are provided by the
breaking at low energy and the light masses are provided by breaking
at higher energy scales. 
This model does not, per se, explain how the
gauge group is broken several times, neither is the breaking of weak isospin
symmetry accounted for. {}For example we cannot explain why the neutrino have masses much smaller than the associated electrons. See, however, \cite{Appelquist:2004ai} for progress on these issues. Schematically one has $SU(N_{TC} + 3)$ which breaks to  $SU(N_{TC} + 2)$ at the scale 
$\Lambda_1$ providing the first generation of fermions with a typical mass $m_1 \sim {4\pi
  (F_\pi^{TC})^3}/{\Lambda_1^2}$ at this point the gauge group breaks to $SU(N_{TC} + 1)$ with dynamical scale $\Lambda_2 $ leading to a second generation mass of the order of $m_2 \sim{4\pi
  (F_\pi^{TC})^3}/{\Lambda_2^2}$ finally the last breaking
$SU(N_{TC} )$ at scale 
$\Lambda_3$ leading to the last generation mass $m_3 \sim {4\pi
  (F_\pi^{TC})^3}/{\Lambda_3^2}$. 
 
Without specifying an ETC one can write down the most general type of four-fermion operators involving technicolor particles $Q$ and ordinary fermionic fields $\psi$.  {}Following the notation of Hill and Simmons \cite{Hill:2002ap} we write:
\beq \alpha_{ab}\frac{\bar Q\gamma_\mu T^aQ\bar\psi \gamma^\mu
  T^b\psi}{\Lambda_{ETC}^2} +
\beta_{ab}\frac{\bar Q\gamma_\mu T^aQ\bar Q\gamma^\mu
  T^bQ}{\Lambda_{ETC}^2} + 
\gamma_{ab}\frac{\bar\psi\gamma_\mu T^a\psi\bar\psi\gamma^\mu
  T^b\psi}{\Lambda_{ETC}^2} \ , \eeq
where the $T$s are unspecified ETC generators. After performing a Fierz rearrangement one has:
\beq \alpha_{ab}\frac{\bar QT^aQ\bar\psi T^b\psi}{\Lambda_{ETC}^2} +
\beta_{ab}\frac{\bar QT^aQ\bar QT^bQ}{\Lambda_{ETC}^2} +
\gamma_{ab}\frac{\bar\psi T^a\psi\bar\psi T^b\psi}{\Lambda_{ETC}^2}
+ \ldots \ , \label{etc} \eeq
The coefficients parametrize the ignorance on the specific ETC physics. To be more specific, the $\alpha$-terms, after the technicolor particles have condensed, lead to mass terms for the SM fermions
\beq m_q \approx \frac{g_{ETC}^2}{M_{ETC}^2}\langle \bar
QQ\rangle_{ETC} \ , \eeq
where $m_q$ is the mass of e.g.~a SM quark, $g_{ETC}$ is the ETC gauge 
coupling constant evaluated at the ETC scale, $M_{ETC}$ is the mass of
an ETC gauge boson and $\langle \bar QQ\rangle_{ETC}$ is the
technicolor condensate where the operator is evaluated at the ETC
scale. Note that we have not explicitly considered the different scales for the different generations of ordinary fermions but this should be taken into account for any realistic model. 

\begin{figure}[tbh]
\begin{center}
\includegraphics[width=7truecm,height=4truecm]{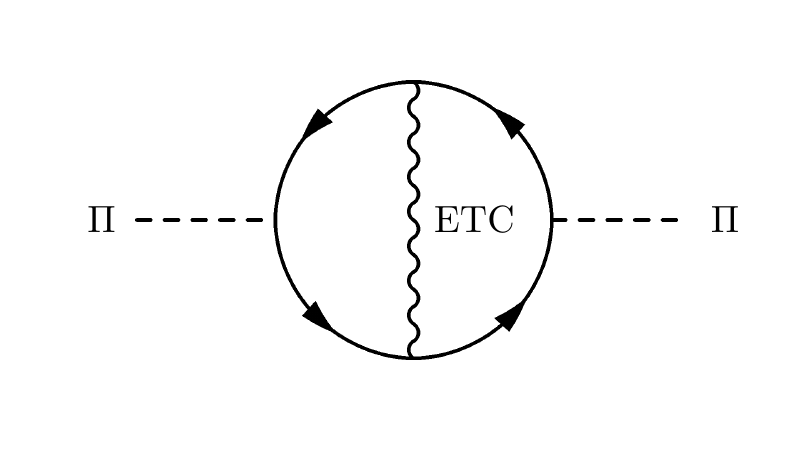}
\caption{Leading contribution to the mass of the TC pseudo
  Goldstone bosons via an exchange of an ETC gauge boson.} \label{masspgb}
\end{center}
\end{figure}

The $\beta$-terms of Eq.~(\ref{etc}) provide masses for
pseudo Goldstone bosons and also provide masses for techniaxions
\cite{Hill:2002ap}, see figure \ref{masspgb}. 
The last class of terms, namely the $\gamma$-terms of
Eq.~(\ref{etc}) induce flavor changing neutral currents. {}For example it may generate the following terms:
\beq \frac{1}{\Lambda_{ETC}^2}(\bar s\gamma^5d)(\bar s\gamma^5d) +
\frac{1}{\Lambda_{ETC}^2}(\bar \mu\gamma^5e)(\bar e\gamma^5e) + 
\ldots \ , \label{FCNC} \eeq
where $s,d,\mu,e$ denote the strange and down quark, the muon
and the electron, respectively. The first term is a $\Delta S=2$
flavor-changing neutral current interaction affecting the
$K_L-K_S$ mass difference which is measured accurately. The experimental bounds on these type of operators together with the very {\it naive} assumption that ETC will generate these operators with $\gamma$ of order one leads to a constraint on the ETC scale to be of the order of or larger than $10^3$
TeV \cite{Eichten:1979ah}. This should be the lightest ETC scale which in turn puts an upper limit on how large the ordinary fermionic masses can be. The naive estimate is that  one can account up to around 100 MeV mass for a QCD-like technicolor theory, implying that the Top quark mass value cannot be achieved.

The second term of Eq.~(\ref{FCNC}) induces flavor
changing processes in the leptonic sector such as $\mu\to e\bar ee,
e\gamma$ which are not observed.
It is clear that, both for the precision measurements  and the fermion masses, that a better theory of the flavor is needed. 

\subsection{ Walking}
To better understand in which direction one should go to modify the QCD dynamics we analyze the TC condensate. 
The value of the technicolor condensate used when giving mass to the ordinary fermions should be evaluated not at the technicolor scale but at the extended technicolor one. Via the renormalization group one can relate the condensate at the two scales via:
\beq \langle\bar QQ\rangle_{ETC} =
\exp\left(\int_{\Lambda_{TC}}^{\Lambda_{ETC}}
d(\ln\mu)\gamma(\alpha(\mu))\right)\langle\bar QQ\rangle_{TC} \ ,
\label{rad-cor-tc-cond}
\eeq 
where $\gamma$ is the anomalous dimension of the techniquark mass-operator. The boundaries of the integral 
 are at the ETC scale and the TC one.
{}For TC theories with a running of the coupling constant similar to the one in QCD, i.e.
\beq \alpha(\mu) \propto \frac{1}{\ln\mu} \ , \quad {\rm for}\ \mu >
\Lambda_{TC} \ , \eeq
this implies that the anomalous dimension of the techniquark masses $\gamma \propto
\alpha(\mu)$. When computing the integral one gets
\beq \langle\bar QQ\rangle_{ETC} \sim
\ln\left(\frac{\Lambda_{ETC}}{\Lambda_{TC}}\right)^{\gamma}
\langle\bar QQ\rangle_{TC} \ , \label{QCD-like-enh} \eeq
which is a logarithmic enhancement of the operator. We can hence neglect this correction and use directly the value of the condensate at the TC scale when estimating the generated fermionic mass:
\beq m_q \approx \frac{g_{ETC}^2}{M_{ETC}^2}\Lambda_{TC}^3 \ , \qquad 
 \langle \bar QQ\rangle_{TC} \sim \Lambda_{TC}^3 \ . \eeq

The tension between having to reduce the FCNCs and at the same time provide a sufficiently large mass for the heavy fermions in the SM as well as the pseudo-Goldstones can be reduced if the dynamics of the underlying TC theory is different from the one of QCD. The computation of the TC condensate at different scales shows that  if the dynamics is such that the TC coupling does not {\it run} to the UV fixed point but rather slowly reduces to zero one achieves a net enhancement of the condensate itself with respect to the value estimated earlier.  This can be achieved if the theory has a near conformal fixed point. This kind of dynamics has been denoted of {\it walking} type. In this case 
\beq \langle\bar QQ\rangle_{ETC} \sim
\left(\frac{\Lambda_{ETC}}{\Lambda_{TC}}\right)^{\gamma(\alpha^*)}
\langle\bar QQ\rangle_{TC} \ , \label{walking-enh} \eeq
which is a much larger contribution than in QCD dynamics  \cite{Yamawaki:1985zg,Holdom:1984sk,Holdom:1981rm,Appelquist:1986an}. Here $\gamma$ is evaluated at the would be fixed point value $\alpha^*$.  Walking can help resolving the problem of FCNCs in
technicolor models since with a large enhancement of the $\langle\bar
QQ\rangle$ condensate the four-fermi operators involving SM fermions and
technifermions and the ones involving technifermions are enhanced by a factor of
$\Lambda_{ETC}/\Lambda_{TC}$ to the $\gamma$  power while the one involving only standard model fermions is not enhanced.

\begin{figure}
\centering
\begin{tabular}{cc}
\resizebox{6.0cm}{!}{\includegraphics{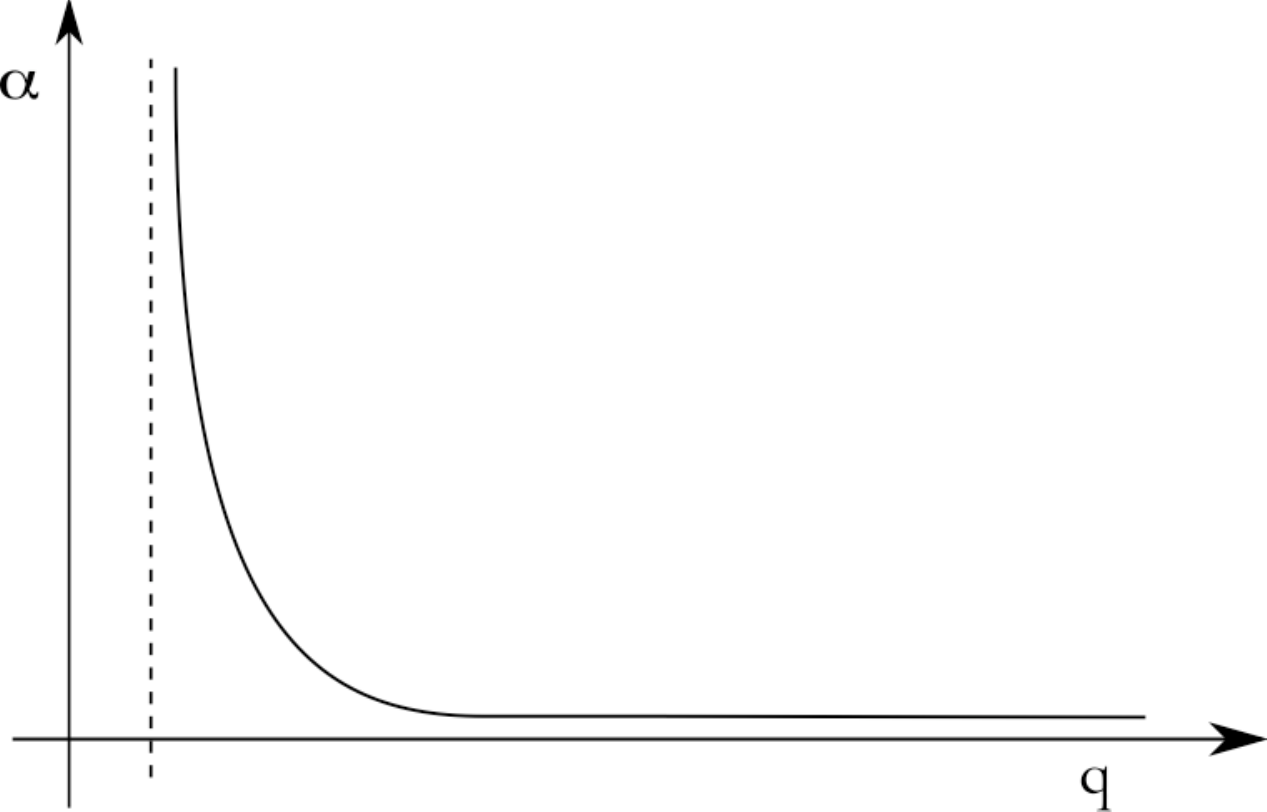}} ~~~&~~ \resizebox{6.0cm}{!}{\includegraphics{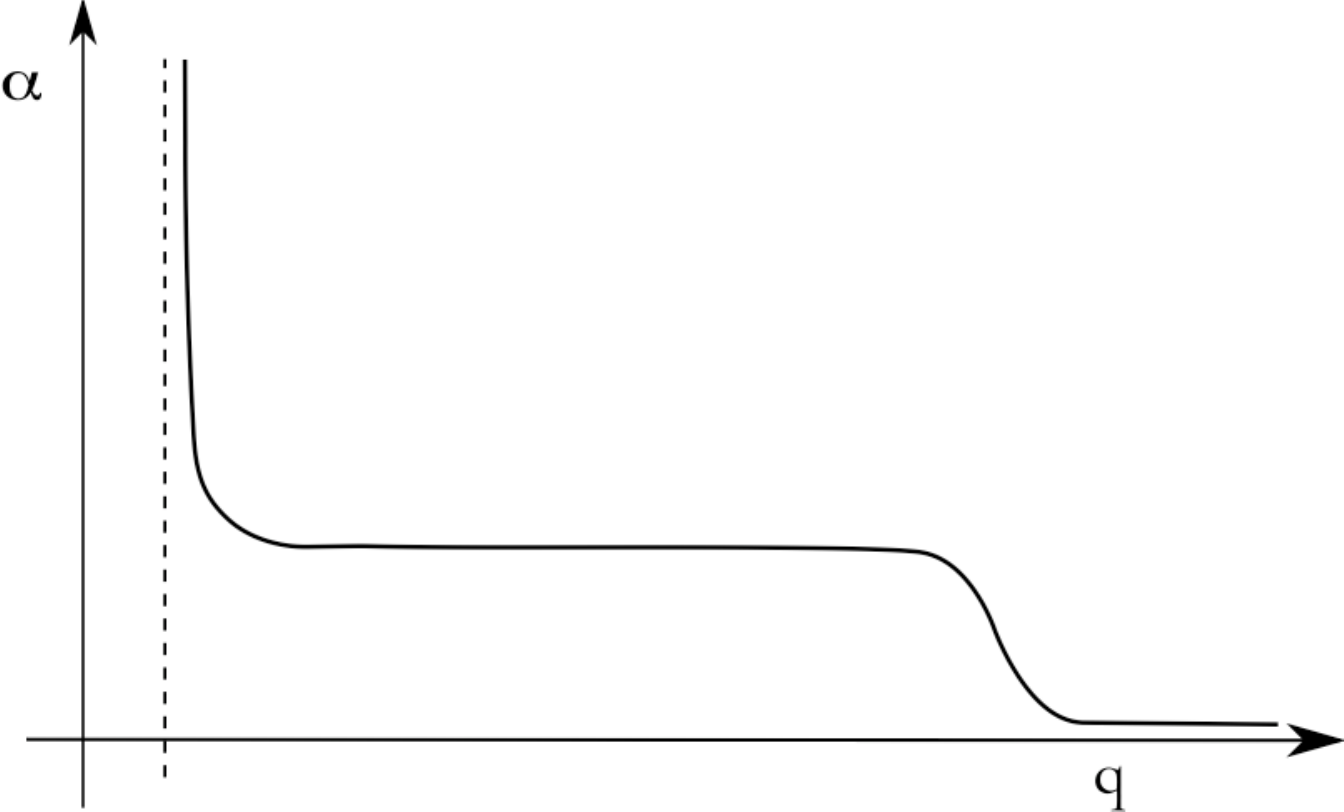}} \\&\\
&~~~~\resizebox{6.0cm}{!}{\includegraphics{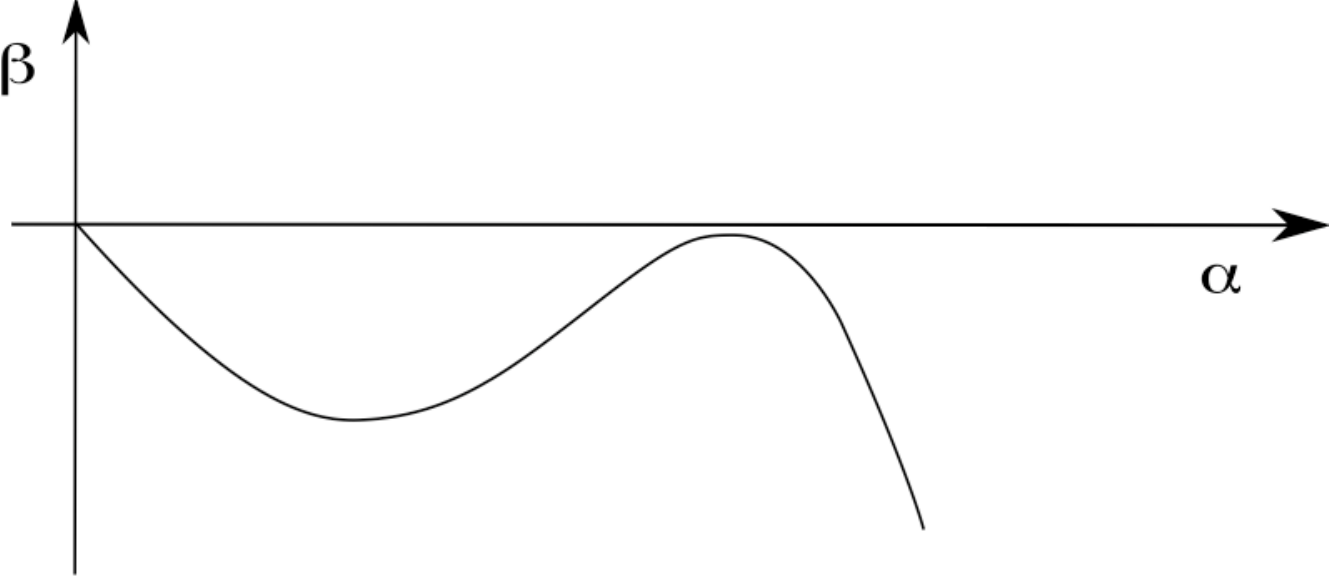}} 
\end{tabular}
\caption{Top Left Panel: QCD-like behavior of the coupling constant as function of the momentum (Running). Top Right Panel: Walking-like behavior of the coupling constant as function of the momentum (Walking). Bottom Right Panel: Cartoon of the beta function associated to a generic walking theory.}
\label{walkbeta}
\end{figure}

In the figure \ref{walkbeta} the comparison between a running and walking behavior of the coupling is qualitatively represented. 

\subsection{Weinberg Sum Rules and Electroweak Parameters}\label{sec:electroweak}

Any strongly coupled dynamics, even of walking type, will generate a spectrum of resonances whose natural splitting in mass is of the order of the intrinsic scale of the theory which in this case is the Fermi scale. In order to extract predictions  for the composite vector spectrum and couplings in presence of a strongly interacting sector and an asymptotically free gauge theory, we make use of the time-honored Weinberg sum rules (WSR) \cite{Weinberg:1967kj} but we will also use the results found in \cite{Appelquist:1998xf}  allowing us to treat walking and running theories in a unified way.

\subsubsection{Weinberg sum rules}
The Weinberg sum rules (WSRs)  are linked to the  two point vector-vector minus axial-axial vacuum polarization which is known to be sensitive to chiral symmetry breaking.  We define
\begin{equation}
i\Pi_{\mu \nu}^{a,b}(q)\equiv \int\!d^4x\, e^{-i qx}
\left[<J_{\mu,V}^a(x)J_{\nu,V}^b(0)> -
 <J_{\mu,A}^a(x)J_{\nu,A}^b(0)>\right] \ ,
\label{VA}
\end{equation}
within the underlying strongly coupled gauge theory, where
\begin{equation}
\Pi_{\mu \nu}^{a,b}(q)=\left(q_{\mu}q_{\nu} - g_{\mu\nu}q^2 \right) \,
\delta^{a b} \Pi(q^2) \ .
\end{equation}
Here $a,b=1,...,N_f^2-1$, label the flavor
currents and the SU(N$_f$) generators are normalized according to
$\rm{Tr} \left[T^a T^b\right]= (1/2) \delta^{ab} $.  The
function $\Pi(q^2)$ obeys the unsubtracted dispersion relation
\begin{equation}
\frac{1}{\pi} \int_0^{\infty}\!ds\, \frac{{\rm Im}\Pi(s)}{s + Q^2}
=\Pi(Q^2) \ ,
\label{integral}
\end{equation}
where $Q^2=-q^2 >0$, and the constraint
$\displaystyle{-Q^2 \Pi(Q^2)>0}$ holds for $0 < Q^2 < \infty$~\cite{Witten:1983ut}. The discussion above is for the standard chiral symmetry breaking pattern SU(N$_f$)$\times$ SU(N$_f$)$ \rightarrow $SU(N$_f$) but it is generalizable to any breaking pattern.

Since we are taking the underlying theory to be asymptotically
free, the behavior of $\Pi(Q^2)$ at asymptotically high momenta is
the same as in ordinary QCD, i.e. it scales like $Q^{-6}$~\cite{Bernard:1975cd}. Expanding the left hand side of the dispersion relation
thus leads to the two conventional spectral function sum rules
\begin{equation}
\frac{1}{\pi} \int_0^{\infty}\!ds\,{\rm Im}\Pi(s) =0
\label{spectral1}
\quad {\rm and} \quad
\frac{1}{\pi} \int_0^{\infty}\!ds\,s \,{\rm Im}\Pi(s) =0 \ .
\end{equation}
Walking dynamics affects only the second sum rule \cite{Appelquist:1998xf} which is more sensitive to large but not asymptotically large momenta due to fact that the associated integrand contains an extra power of $s$.

We now saturate the absorptive part of the vacuum
polarization. We follow reference \cite{Appelquist:1998xf} and hence divide the
energy range of integration in three parts. The light resonance part. In this regime, the    
integral is saturated by the
Nambu-Goldstone pseudoscalar along with massive vector and
axial-vector states. If we assume, for example, that there is only a
single, zero-width vector multiplet and a single, zero-width axial
vector multiplet, then
\begin{equation}
{\rm Im}\Pi(s)=\pi F^2_V \delta \left(s -M^2_V \right) - \pi F^2_A
\delta \left(s - M^2_A \right) - \pi F^2_{\pi} \delta \left(s \right)
\ .
\label{saturation}
\end{equation}
The zero-width approximation is valid to leading order in the large
$N$ expansion for fermions in the fundamental representation of the gauge group and it is even narrower for fermions in higher dimensional representations. Since we are working near a conformal fixed point the large $N$ argument for the width is not directly applicable. We will nevertheless use this simple model for the spectrum
to  illustrate the effects of a near critical IR fixed point. 

The first WSR implies:
\begin{equation}
F^2_V - F^2_A = F^2_{\pi}\ ,
\label{1rule}
\end{equation}
where $F^2_V$ and $F^2_A$ are the vector and axial mesons decay
constants.  This sum rule holds for walking and running dynamics. {A
more general representation of the resonance spectrum would, in principle, replace
the left hand side of this relation with a sum over vector and axial
states. However the heavier resonances should not be included since in the approach of \cite{Appelquist:1998xf} the walking dynamics in the intermediate energy range is already approximated by the exchange of underlying fermions. The walking is encapsulated in the dynamical mass dependence on the momentum dictated by the gauge theory. The introduction of heavier resonances is, in practice, double counting. Note that the approach is in excellent agreement with the Weinberg approximation for QCD, since in this case, the  approximation automatically returns the known results.  }

The second sum rule receives important contributions from throughout
the near conformal region and can be
expressed in the form of:
\begin{equation}
F^2_V M^2_V - F^2_A M^2_A = a\,\frac{8\pi^2}{d(R)}\,F_{\pi}^4,
\label{2rule-2}
\end{equation}
where $a$ is expected to be positive and $O(1)$ and $d(R)$ is the dimension of the representation of the underlying fermions.  
We have generalized the result of reference \cite{Appelquist:1998xf}  to the case in which the fermions belong to a generic representation of the gauge group. In the case of running dynamics the right-hand side of the previous equation vanishes.  

{We stress that $a$ is a non-universal quantity depending on the details of the underlying gauge theory. A reasonable measure of how large $a$ can be is given by a function of the amount of walking which is the ratio of the scale above which the underlying coupling constant start running divided by the scale below which chiral symmetry breaks.} 
The fact that $a$ is positive and of order one in walking dynamics is supported, indirectly, also via the results of Kurachi and Shrock \cite{Kurachi:2006ej}. At the onset of conformal dynamics the axial and the vector will be degenerate, i.e. $M_A=M_V=M$, using the first sum rule one finds via the second sum rule $a = d({\rm R})M^2/(8\pi^2 F^2_{\pi})$ leading to a numerical value of about 4- 5 from the approximate results in \cite{Kurachi:2006ej}.  We will however use only the constraints coming from the generalized WSRs  expecting them to be less model dependent.
\subsubsection{Relating WSRs to the Effective Theory \& $S$ parameter }
The $S$ parameter is related to the absorptive part  of the
vector-vector minus axial-axial vacuum polarization as follows:
\begin{equation}
S=4\int_0^\infty \frac{ds}{s} {\rm Im}\bar{\Pi}(s)= 4\pi
\left[\frac{F^2_V}{M^2_V} - \frac{F^2_A}{M^2_A} \right] \ ,
\label{s-def}
\end{equation}
where ${\rm Im}\bar{\Pi}$ is obtained from ${\rm Im}\Pi$ by
subtracting the Goldstone boson contribution.

Other attempts to estimate the $S$ parameter for walking technicolor
theories have been made in the past \cite{Sundrum:1991rf} showing reduction of the $S$ parameter. $S$ has also been evaluated using computations inspired by the original AdS/CFT correspondence \cite{Maldacena:1997re} in \cite{Hong:2006si,Hirn:2006nt,Piai:2006vz,Agashe:2007mc,Carone:2007md}. 

Kurachi, Shrock and Yamawaki \cite{Kurachi:2007at} have further confirmed the results presented in \cite{Appelquist:1998xf} with their computations tailored for describing four dimensional gauge theories near the conformal window.
The present approach \cite{Appelquist:1998xf} is more physical since it is based on the
nature of the spectrum of states associated directly to the underlying gauge theory.  

Note that we will be assuming a rather conservative approach in which the $S$ parameter, although reduced with respect to the case of a running theory, is positive and not small.  After all, other sectors of the theory such as new leptons further reduce or completely offset a positive value of $S$ due solely to the technicolor theory. 

\subsection{Technibaryon Dark Matter and Electroweak Phase Transition}
Technicolor  interactions bind
techniquarks in technimesons and
technibaryon bound states. The spin of the technibaryons depends on the representation according to which the technifermion transform, the number of flavors and colors. The lightest technimeson is
 short-lived thus evading BBN
constraints, but the lightest technibaryon has typically \footnote{There may be situation in which the technibaryon is a goldstone boson of an enhanced flavor symmetry.} a mass of the order 
\begin{equation}
 m_{TB}  \sim  1-2\,{\rm TeV} \ .
\end{equation} 
It may be unstable, by analogy with ordinary proton. 
As in QCD, only non-renormalizable interaction can cause
technibaryon to decay.  Technicolor models must in
any case be extended to incorporate quark and lepton masses, and the technicolor gauge theory may be unified in some Grand Unified Theory (GUT) as explored in the last chapter. Therefore one expects the existence of higher-order effective
non-renormalizable interactions which cause technibaryon decay, of the
form
\begin{equation}
 {L}_{{TB}} \sim \frac{Q^{N_{TC}} \psi^n}{\Lambda_{{TB}}^{3/2 
  (N_{TC} + n) - 4}}\ ,
\end{equation}
where $\psi$ is a quark or lepton field and $\Lambda_{TB}$ is some mass
scale $\gg\,\Lambda_{TC}$ at which the effective interaction is
generated.  We are using the fundamental representation for the technifermions. This interaction would imply a technibaryon lifetime \cite{Sarkar:1995dd}
\begin{equation}
\label{tauTB}
 \tau_{TB} \sim \frac{1}{\Lambda_{TC}} 
  \left(\frac{\Lambda_{TB}}{\Lambda_{TC}}\right)^{3(N_{TC}+n) - 8}\ , 
\end{equation} 
i.e. $\sim10^{-27}(\Lambda_{TB}/\Lambda_{TC})^4\sec$, for $N_{TC}=4$ with the minimal choice $n=0$.

Estimating the self-annihilation cross-section of technibaryons to be \cite{Chivukula:1989qb}
\begin{equation}
\label{TBann}
 {\langle \sigma v \rangle}_{TB} \simeq 
  {\langle\sigma v \rangle}_{{p}\bar{p}} 
  \left(\frac{m_{p}}{m_{TB}}\right)^2 \simeq 3 \times 10^{-5} {\rm GeV}^2 ,
\end{equation}
the minimum expected relic abundance is about
$m_{TB}\,n_{TB}/n_\gamma\,=\,3\times10^{-13}\GeV$. Unstable technibaryons with such a small abundance
are not constrained by nucleosynthesis \cite{Dodelson:1988yv}. However
technibaryons may have a much higher relic density if they possess an
asymmetry  \cite{Nussinov:1985xr}. 
If the latter is due to a net $B-L$ generated at some high
energy scale, then this would be subsequently distributed among {\em
all} electroweak doublets by fermion-number violating processes in the
standard model at temperatures above the electroweak scale \cite{Shaposhnikov:1991cu,Kuzmin:1991ft,Shaposhnikov:1991wi}, thus naturally generating a technibaryon
asymmetry as well. If such $B+L$ violating processes cease being
important below a temperature
$T_{*}\approx\ 200-300 {\rm GeV}$, then the
technibaryon-to-baryon ratio, which is suppressed by a factor
$[m_{TB}(T_{*})/T_{*}]^{3/2}e^{-m_{TB}(T_{*})/T_{*}}$, is just
right to give \cite{Barr:1990ca,Gudnason:2006yj}
\begin{equation}
\frac{\Omega_{TB}}{\Omega_B} = \frac{m_{TB}}{m_p} \frac{TB}{B}  \sim  {\cal O} (1) \ .
\end{equation}
Hence, such a mechanism can lead to a natural explanation of the observed fraction of dark to bright matter in the universe. To avoid experimental constraints the technibaryon can be constructed to be a complete singlet under the electroweak interactions \cite{Barr:1990ca,Dietrich:2006cm} while still having the underlying guage theory to be walking \cite{Dietrich:2006cm}. In this case it would be hard to detect it \cite{Bagnasco:1993st,Gudnason:2006yj} in current earth based experiment  such as CDMS \cite{ Akerib:2004fq,Akerib:2005kh}.  Other possibilities have been envisioned in \cite{Kouvaris:2007iq,Khlopov:2007ic} and possible astrophysical effects studied in \cite{ Kouvaris:2007ay}.  One can use also particles for dark matter not coming from the technicolor sector but rather from possible associate new sectors \cite{Kainulainen:2006wq}. Another interesting possibility is to investigate candidates emerging from sectors of the technicolor model not gauged under the electroweak interactions \cite{Dietrich:2006cm} (see the partially gauged technicolor models section for a description of the models).

 The order of the electroweak phase transition depends on the underlying type of strong dynamics and  plays an important role for baryogenesis \cite{Cline:2002aa,Cline:2006ts}. The technicolor chiral phase transition at finite temperature is mapped into the electroweak one. Attention must be paid to the way the electroweak is embedded into the global symmetries of the underlying technicolor theory. Wilson renomralization group methods suggest that the larger the global symmetries are the more strongly first order one expects the phase transition to be. An interesting analysis dedicated to these issues has been performed in \cite{Kikukawa:2007zk}. 

 A possibility  not yet studied in the literature is associated to the fact that one can have two independent phase transitions at nonzero temperature in technicolor theories whenever the theory possesses a non trivial center symmetry.  The two phase transitions are the chiral one, directly related to the electroweak phase transition, and a confining one at lower temperatures.  During the history of the universe we would predict a phase transition around the electroweak scale and another one at lower temperatures with a jump in the entropy proportional to the number of degrees of freedom liberated (or gapped) when increasing (decreasing) the temperature (see \cite{Mocsy:2003qw} for a simple understanding of this phenomenon and a list of relevant references). This may have very interesting cosmological consequences, which we hope will be studied in the future.

\newpage
\section{Phase Diagram of Non Supersymmetric Strongly Coupled Theories}

Constructing dynamical models of electroweak symmetry breaking is hard due to the limited number of analytic methods available when exploring strong dynamics in the regime when perturbation theory fails. First principle lattice simulations are now capable to accurately investigate the spectrum and the dynamics of various four dimensional gauge theories which are of great interest in our pursue of a dynamical origin of the stabilization of the Fermi scale \cite{DelDebbio:2008wb,Catterall:2007yx,Appelquist:2007hu}.  It is, however, instructive and useful to study the dynamics and/or spectrum of a generic nonsupersymmetric gauge theory applying, for example, the proposal of an all order beta function for nonsupersymmetric gauge theories with fermionic matter \cite{Ryttov:2007cx} or the truncated Schwinger-Dyson equation (SD) \cite{Appelquist:1988yc,Cohen:1988sq,Miransky:1996pd} (referred also as the ladder approximation in the literature) or even conjectures such as the Appelquist-Cohen-Schmaltz (ACS) one  \cite{Appelquist:1999hr} which makes use of the counting of the thermal degrees of freedom at high and low temperature. The ACS conjecture is, however, not sufficiently constraining when studying vector-like theories with matter in higher dimensional representations as we  have shown in \cite{Sannino:2005sk} and hence will not be used here. The ACS conjecture has been tested also for chiral gauge theories \cite{Appelquist:1999vs}. There it was also found that to make some definite predictions a stronger requirement is needed \cite{Appelquist:2000qg}.  

We will start from the proposal of the beta function for nonsupersymmetric gauge theories with fermionic matter in any arbitrary representation of the underlying $SU(N)$ gauge group and for any number of colors \cite{Ryttov:2007cx}. It is  written in a form useful for constraining the phase diagram of strongly coupled theories. The form is inspired by the Novikov-Shifman-Vainshtein-Zakharov  (NSVZ) beta function for supersymmetric theories \cite{Novikov:1983uc,Shifman:1986zi} and the renormalization scheme coincides with the NSVZ one. We  will then review the phase diagram deduced with the SD approximation and discuss the lattice results while at the same time combine and confront with theoretical expectations. The predictions which make use of  the all order beta function are supported by lattice results and one can further test them \footnote{Note that the fact that the beta function beyond two loops is scheme dependent does not mean that its form is not relevant. In fact, being able to determine the anomalous dimensions at an infrared fixed point is important. Of course, any scheme should lead to the same result. In certain schemes, however, it is harder to determine certain quantities. As a relevant example consider the 't Hooft scheme, the one in which the beta function is two loops exact, here one cannot compute the anomalous dimensions to all orders at the infrared fixed point.}. Finally we  will show how the results can be used also to construct novel extensions of the SM. 

\subsection{Phases of  Gauge Theories}
We wish to study the phase diagram of any asymptotically free non-supersymmetric theories with fermionic matter transforming according to a generic representation of an SU($N$) gauge group as function of the number of colors and
flavors. 

We start by characterizing the possible phases via the potential $V(r)$ between
two electric test charges separated by a large distance r. The list of possible potentials is given below:
\begin{eqnarray}
 {\rm \mathbf{Coulomb:}} &~~~~~& V (r) \propto \frac{1}{r} \\
 &~~~~~& \nonumber \\
  {\rm \mathbf{Free~electric:}} &~~~~~& V (r) \propto \frac{1}{r\log(r)} \\
 &~~~~~& \nonumber \\
  {\rm \mathbf{ Free~magnetic:}} &~~~~~&  V (r) \propto \frac{\log(r)}{r}  \\
  &~~~~~& \nonumber \\
 {\rm \mathbf{ Higgs:}}&~~~~~&  V (r) \propto {\rm  constant}  \\  
 &~~~~~& \nonumber \\
   {\rm \mathbf{ Confining:}} &~~~~~&   V (r) \propto \sigma r \ .
  \end{eqnarray}
 % 
%  \begin{eqnarray}
% {\rm \mathbf{Coulomb:}} &~~~~~& V (r) \propto \frac{1}{r} 
% \end{eqnarray}\begin{eqnarray}
%  {\rm \mathbf{Free~electric:}} &~~~~~& V (r) \propto \frac{1}{r\log(r)} 
%  \end{eqnarray}
%  \begin{eqnarray}
%  {\rm \mathbf{ Free~magnetic:}} &~~~~~&  V (r) \propto \frac{\log(r)}{r} \end{eqnarray}
%  \begin{eqnarray} {\rm \mathbf{ Higgs:}}&~~~~~&  V (r) \propto {\rm  constant}  \end{eqnarray}  \begin{eqnarray} {\rm \mathbf{ Confining:}} &~~~~~&   V (r) \propto \sigma r \ .
%  \end{eqnarray}

  A nice review of these phases can be found in \cite{Intriligator:1995au} which here we re-review for completeness. 
In the Coulomb phase, the electric charge $e^2(r)$  is a constant while in
the free electric phase massless electrically charged fields renormalize the charge to
zero at long distances as, i.e. $e^2(r) \sim 1/\log(r )$. QED is an abelian example of a free electric phase. The free
magnetic phase occurs when massless magnetic monopoles renormalize
the electric coupling constant at large distance with $e^2(r) \sim \log(r)$. 
  
  In the Higgs phase, the condensate of an electrically
charged field gives a mass gap to the gauge fields by the Anderson-Higgs-Kibble mechanism
and screens electric charges, leading to a potential which, up to an additive constant, has an exponential Yukawa decay to zero at long distances.
In the confining phase, there is a mass gap with electric flux confined into a thin
tube, leading to the linear potential with string tension $\sigma$.

We will be mainly interested in finding theories possessing a
non-Abelian Coulomb phase or being close in the parameter space to these theories. In this phase we have massless interacting quarks and gluons exhibiting
the Coulomb potential. This phase occurs when there is a non-trivial, infrared
fixed point of the renormalization group. These are thus non-trivial, interacting, four
dimensional conformal field theories.

To guess the behavior of the magnetic charge, at large distance separation, between two test magnetic charges  one uses the Dirac condition:
\begin{equation}
e(r) g(r) \sim 1 \ .
\end{equation}
Then it becomes clear that $g(r)$ is constant in the Coulomb phase, increases with $\log(r)$ in the free electric phase and decreases as $1/\log(r)$ in the free magnetic phase. In these three phases the potential goes like $g^2(r)/r$. A linearly rising potential in the Higgs phase for magnetic test charges corresponds to the Meissner effect in the electric charges.

Confinement does not survive the presence of massless matter in the fundamental representation, such as light quarks in QCD. This is so since it is more convenient for the underlying theory to pop from the vacuum virtual quark-antiquark pairs when pulling two electric test charges apart. The potential for the confining phase will then change and there is no distinction between Higgs and confining phase.  

Under electric-magnetic duality one exchanges electrically charged
fields with magnetic ones then the behavior in the free electric phase is
mapped in that of the free magnetic phase. The Higgs and confining phases are  also expected to be exchanged under duality. 
Confinement can then be understood as the dual Meissner effect associated with
a condensate of monopoles.

\subsection{The all orders beta function conjecture}
All of the analytic methods used, so far, to determine the phase diagram of nonsupersymmetric gauge theories are based on guesses or rough approximations. Here we  will present first our bound on the phase diagram using the proposed beta function for {\it any}  $SU(N)$ gauge theory with $N_f $ Dirac fermions in a given, but arbitrary, representation $r$
of the gauge group. We  will then compare the bound with the conformal window obtained in the ladder approximation. 

The beta function compares well with numerical results as well as with scheme-independent physical results obtained via other model computations. 
We start by reviewing the two loops beta function which reads:
\begin{eqnarray}\beta (g) = -\frac{\beta_0}{(4\pi)^2} g^3 - \frac{\beta_1}{(4\pi)^4} g^5 \ ,
\label{perturbative}
\end{eqnarray}
where $g$ is the gauge coupling and the beta function coefficients are given by
\begin{eqnarray}
\beta_0 &=&\frac{11}{3}C_2(G)- \frac{4}{3}T(r)N_f \\
\beta_1 &=&\frac{34}{3} C_2^2(G)
- \frac{20}{3}C_2(G)T(r) N_f  - 4C_2(r) T(r) N_f  \ .\end{eqnarray}
To this order the two coefficients are universal,
i.e. do not depend on which renormalization group scheme one has used to determine them.
The perturbative expression for the anomalous dimension reads:
\begin{equation}
\gamma(g^2) = \frac{3}{2} C_2(r) \frac{g^2}{4\pi^2} + O(g^4) \ .
\end{equation}
With $\gamma =-{d\ln m}/{d\ln \mu}$ and $m$ the renormalized fermion mass.

The generators $T_r^a,\, a=1\ldots N^2-1$ of the gauge group in the
representation $r$ are normalized according to
$\text{Tr}\left[T_r^aT_r^b \right] = T(r) \delta^{ab}$ while the
quadratic Casimir $C_2(r)$ is given by $T_r^aT_r^a = C_2(r)I$. The
trace normalization factor $T(r)$ and the quadratic Casimir are
connected via $C_2(r) d(r) = T(r) d(G)$ where $d(r)$ is the
dimension of the representation $r$. The adjoint
representation is denoted by $G$. In order to evaluate the relevant group-theoretical factors we use
the Dynkin indices (the Dynkin labels of the highest weight of an
irreducible representation). The latter uniquely characterize the representations and
determine the relevant coefficients. We summarize the relevant formulae in the appendix  \cite{Dietrich:2006cm} while for the reader's convenience in Table \ref{factors} we list the
explicit group factors for the representations used here. A complete
list of all of the group factors for any representation and the way
to compute them is available in Table II of \cite{Dietrich:2006cm}
and the associated appendix \footnote{The normalization for the
generators here is different than the one adopted in
\cite{Dietrich:2006cm}.}.

\begin{table}
\begin{center}
    %\begin{minipage}{3.8in}
    \begin{tabular}{c||ccc }
    r & $ \quad T(r) $ & $\quad C_2(r) $ & $\quad
d(r) $  \\
    \hline \hline
    $ \fund $ & $\quad \frac{1}{2}$ & $\quad\frac{N^2-1}{2N}$ &\quad
     $N$  \\
        $\text{$G$}$ &\quad $N$ &\quad $N$ &\quad
$N^2-1$  \\
        $\symm$ & $\quad\frac{N+2}{2}$ &
$\quad\frac{(N-1)(N+2)}{N}$
    &\quad$\frac{N(N+1)}{2}$    \\
        $\asymm$ & $\quad\frac{N-2}{2}$ &
    $\quad\frac{(N+1)(N-2)}{N}$ & $\quad\frac{N(N-1)}{2}$
    \end{tabular}
    %\end{minipage}
    \end{center}
\caption{Relevant group factors for the representations used
throughout this paper. However, a complete list of all the group
factors for any representation and the way to compute them is
available in Table II and the appendix of
\cite{Dietrich:2006cm}.}\label{factors}
    \end{table}

The first observation is that the perturbative anomalous
dimension depends on $C_2(r)$ which
appears explicitly in the last term of the second coefficient of the beta function.
We proposed the following form \cite{Ryttov:2007cx} of the beta function:
\begin{eqnarray}
\beta(g) &=&- \frac{g^3}{(4\pi)^2} \frac{\beta_0 - \frac{2}{3}\, T(r)\,N_f \,
\gamma(g^2)}{1- \frac{g^2}{8\pi^2} C_2(G)\left( 1+ \frac{2\beta_0'}{\beta_0} \right)} \ ,
\end{eqnarray}
with
\begin{eqnarray}
\beta_0' &=& C_2(G) - T(r)N_f  \ .
\end{eqnarray}
It is a simple matter to show that the above beta function reduces
to Eq. (\ref{perturbative}) when expanding to $O(g^5)$. 

\subsubsection{Non-Abelian QED phase (Free Electric)}
This is the region of $N_f$ for which  $\beta_0$ is negative and asymptotic freedom is lost. The theory behaves like QED and hence it becomes strongly coupled at high energy. 

 We indicate with $N_f^\mathrm{I}$ the number of flavors above which the theory is no longer asymptotically free. This corresponds to \mbox{$\beta_0[N_f^\mathrm{I}]{=}0$}. For values of $N_f$ larger than $N_f^\mathrm{I}$ the theory is in a non-Abelian QED theory. We obtain
\begin{eqnarray}
N_f^{\rm{I}} = \frac{11}{4} \frac{C_2(G)}{T(r)} \ .
\end{eqnarray}

\subsubsection{IR Fixed Point or Coulomb Phase}
As we decrease the number of flavors from just below the point where asymptotic freedom is lost one expects a perturbative (in the coupling) zero in the beta function to occur \cite{Banks:1981nn}. From the expression proposed above one finds that at the zero of the beta function, barring zeros in the denominator,  one must have
\begin{eqnarray}
\gamma = \frac{11C_2(G)-4T(r)N_f}{2T(r)N_f} \ .\end{eqnarray}
The anomalous dimension at the IR fixed point is small for a value of $N_f$ such that:
\begin{equation}
N_f = N_f^I (1 - \epsilon) \ , \quad {\rm with} \quad {\epsilon > 0} \ ,
\end{equation}
and $\epsilon \ll 1$. Indeed, in this approximation we find:
\begin{equation}
\gamma = \frac{2\epsilon}{1-\epsilon} \ll 1 \ .
\end{equation}
It is also clear that the value of $\gamma$ increases as we keep decreasing the number of flavors.
Before proceeding let us also analyze in more detail the denominator of our beta function.
At the infrared fixed point we have:
\begin{equation}
1- \frac{g^2_{\ast}}{8\pi^2} \, C_2(G)\,\frac{1}{2}\,\left(5 - \frac{21}{11\epsilon}  \right) \ ,
\end{equation}
For very small $\epsilon$ the denominator is positive while staying
finite as $\epsilon$ approaches zero. The finiteness of the denominator
is due to the fact that from the perturbative expression of the
anomalous dimension (valid for small epsilon) the fixed point value
of $g_{\ast}$ is:
\begin{eqnarray}
\frac{g^2_{\ast}}{8\pi^2} = \epsilon \,\,\frac{2}{3C_2(r)} + O(\epsilon^2) \ .
\end{eqnarray}

Since a perturbative fixed point does exist we extend the analysis to a lower number of flavors. The dimension of the chiral condensate is $D(\bar{\psi} \psi)=3-\gamma$ which at the IR fixed point value reads
\begin{equation}
D (\bar{\psi} \psi)= \frac{10T(r)N_f - 11C_2(G)}{2T(r)N_f} \ .
\end{equation}
 To avoid negative norm states in a conformal field theory one must have $D\geq 1$ for non-trivial
 spinless operators \cite{Mack:1975je,Flato:1983te,Dobrev:1985qv}.

\begin{figure}[h]
\begin{center}\resizebox{12cm}{!}{\includegraphics{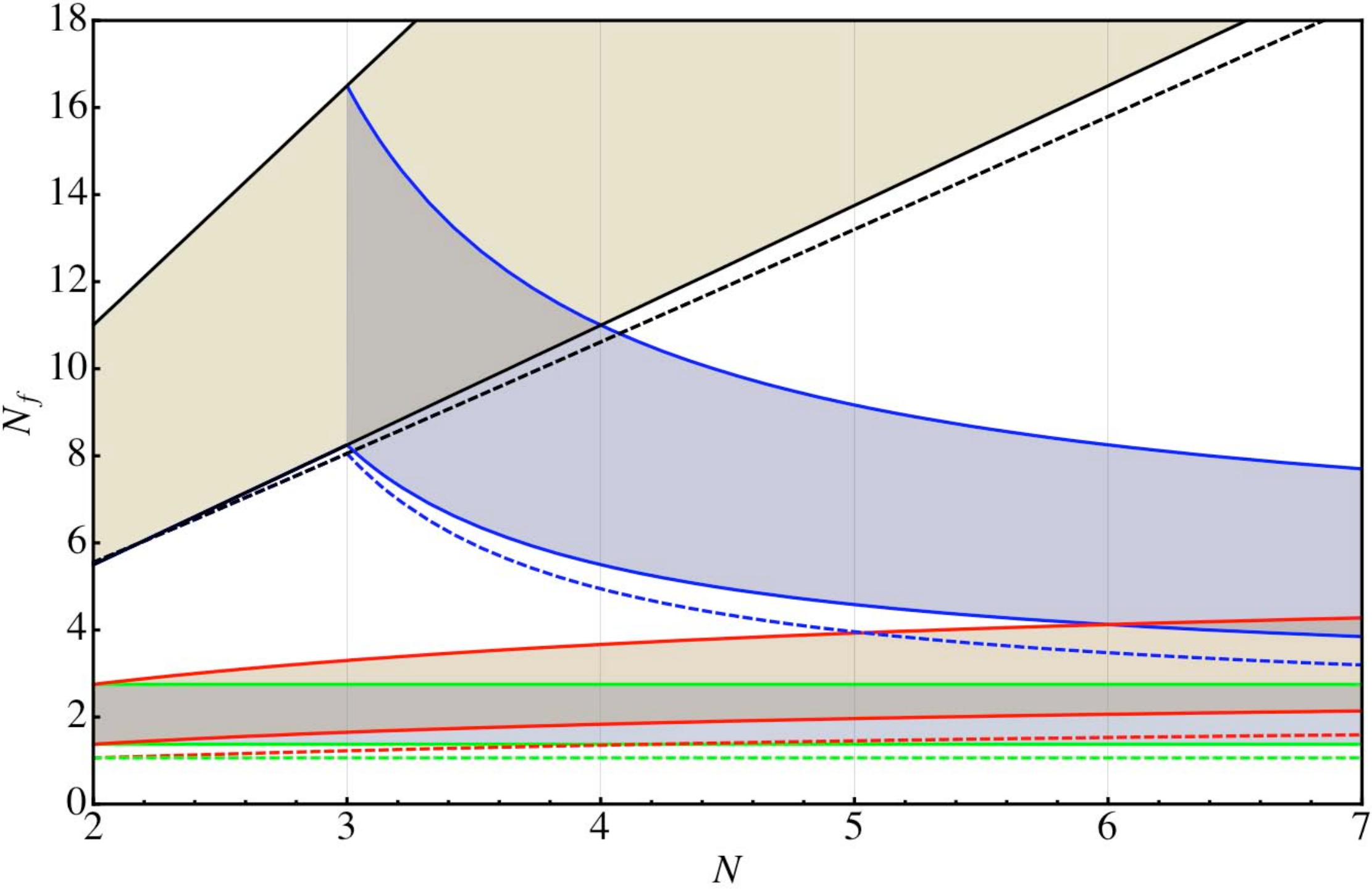}}
\caption{Phase diagram for nonsupersymmetric theories with fermions
in the: i) fundamental representation (black), ii) two-index
antisymmetric representation (blue), iii) two-index symmetric
representation (red), iv) adjoint representation (green) as a
function of the number of flavors and the number of colors. The
shaded areas depict the corresponding conformal windows. Above the
upper solid curve  the theories are no longer asymptotically free.
Between the upper and the lower solid curves the theories are
expected to develop an infrared fixed point according to the NSVZ
inspired beta function. The dashed curve represents the change of
sign in the second coefficient of the beta function.} \label{PHNew}\end{center}
\end{figure}

Hence the critical number of flavors below which the unitarity bound is violated is
\begin{eqnarray}
N_f^{\rm{II}} = \frac{11}{8} \frac{C_2(G)}{T(r)} \ ,
\end{eqnarray}
which corresponds to having set $\gamma=2$. One should note that the analysis above is
similar to the one done for supersymmetric gauge theories \cite{Seiberg:1994pq}.
However, the actual size of the conformal window may be smaller than the one
presented here which hence can be considered as a bound on the size of the window. The reason being that chiral symmetry breaking could be triggered for a value of $\gamma$ lower than two, as for example suggested by the ladder approximation. In Figure \ref{PHNew} we plot the phase diagram. 

\subsubsection{Comparison with the Ladder approximation}

 We now confront our bound for the conformal windows with the one obtained using the ladder approximation in \cite{Dietrich:2006cm}. To determine the number of flavors above which the theory becomes conformal, we
employ the criterion proposed in \cite{Appelquist:1988yc,Cohen:1988sq}.  

The idea behind this method is simple\footnote{The reader is urged to read the original papers for a more detailed explanation.}. One simply compares the two couplings in the infrared associated to i) an infrared zero in the $\beta$ function, call it $\alpha^{\ast}$  with ii) the critical coupling, denoted with $\alpha_c$, above which a dynamical mass for the fermions generates nonperturbatively and chiral symmetry breaking occurs. If $\alpha^{\ast}$ is less than $\alpha_c$ chiral symmetry does not occur and the theory remains conformal in the infrared, viceversa if $\alpha^{\ast}$ is larger than $\alpha_c$ then the fermions acquire a dynamical mass and the theory cannot be conformal in the infrared. The condition $\alpha^{\ast} = \alpha_c$ provides the desired ${N_f^\mathrm{II}}_{\rm Ladder} $ as function of $N$.  Because of the presence of fundamental scalars this procedure is not applicable to supersymmetric QCD.  In practice for point i) one uses the two-loop beta function for determining the zero and for point ii) the truncated SD equation.  

In the approximations of \cite{Appelquist:1988yc,Cohen:1988sq}, the critical value $\alpha_c$ of
the coupling constant for which chiral symmetry breaking occurs is given by:
\bear
\alpha_c=\frac{\pi }{3C_2(\mathrm{r})}.
\eear
This corresponds to when the anomalous dimension of the quark mass operator
becomes approximately unity\footnote{Alternative methods for calculating the
conformal window have been suggested, e.g. in
\cite{Banks:1981nn,Miransky:1996pd,Appelquist:1996dq,Braun:2006jd}.}.  In Appendix \ref{ra} we briefly summarized some of the relevant relations and how to derive them. 

Compared to that, the two-loop fixed point value of the coupling constant
reads \cite{Sannino:2004qp}
\bear
\frac{\alpha^*}{4\pi}=-\frac{\beta_0}{\beta_1}.
\eear
For a fixed number of colors the critical number of flavors for which the
order of $\alpha^{\ast}$ and $\alpha_c$ changes is defined by imposing 
$\alpha^*{=}\alpha_c$, and it is given by
\begin{eqnarray}
{N_f^\mathrm{II}}_{\rm Ladder} &=& \frac{17C_2(G)+66C_2(r)}{10C_2(G)+30C_2(r)}
\frac{C_2(G)}{T(r)} \ . \label{nonsusy}
\end{eqnarray}

{}The region of values of $N_f$ less than $N_f^\mathrm{I}$ but larger than $N_f^\mathrm{II}$ is  the { conformal window}. This is so since in this region of the number of flavors no dynamical invariant scale is generated. Comparing with the previous result obtained using the all order beta function we see that it is the coefficient of $C_2(G)/T(r)$ which is different.

\begin{figure}[h]
\begin{center}\resizebox{12cm}{!}{\includegraphics{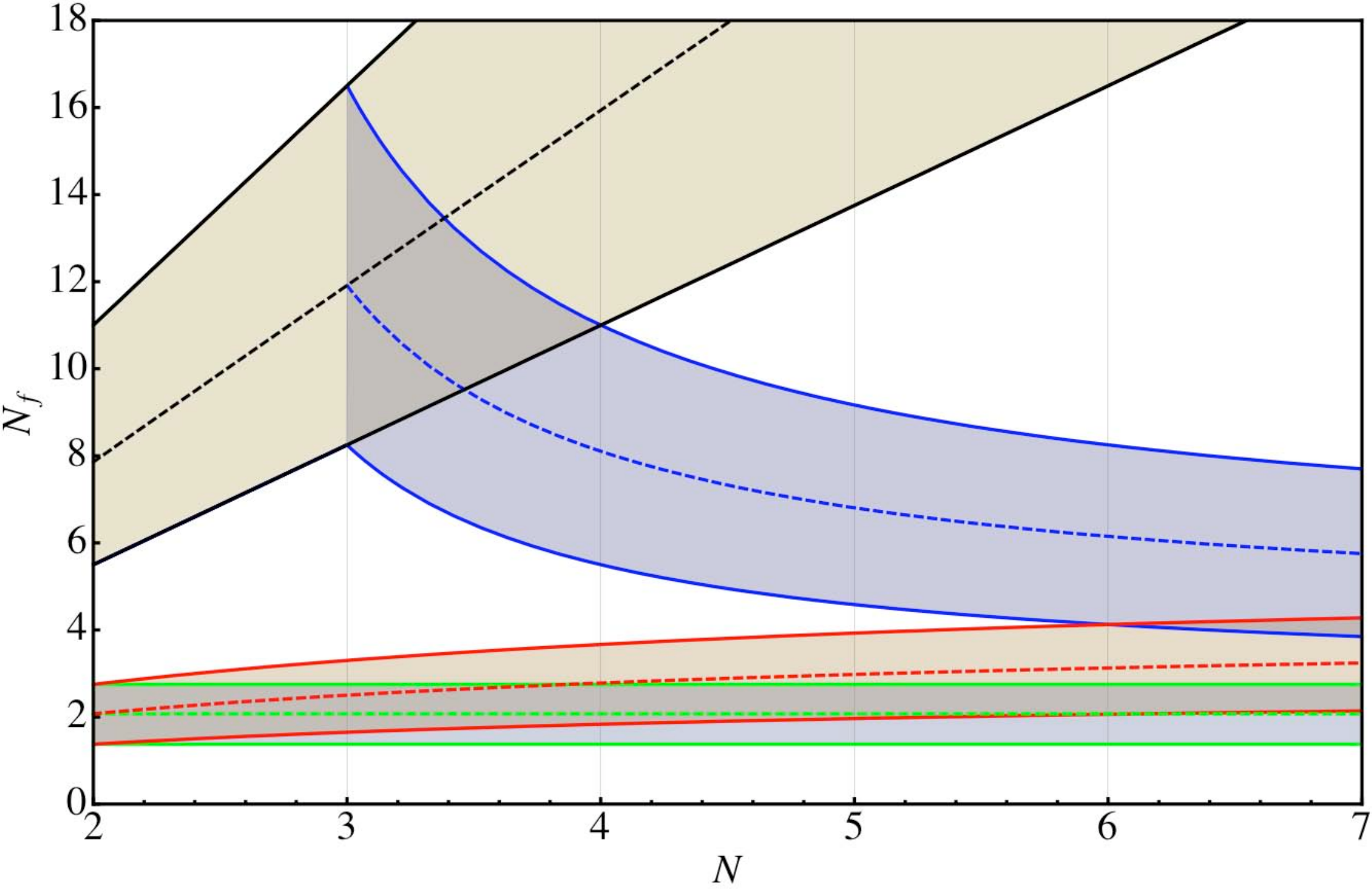}}
\caption{Phase diagram for nonsupersymmetric theories with fermions
in the: i) fundamental representation (black), ii) two-index
antisymmetric representation (blue), iii) two-index symmetric
representation (red), iv) adjoint representation (green) as a
function of the number of flavors and the number of colors. The
shaded areas depict the corresponding conformal windows. Above the
upper solid curve  the theories are no longer asymptotically free.
In between the upper and the lower solid curves the theories are
expected to develop an infrared fixed point according to the NSVZ
inspired beta function. The area between the upper solid curve and
the dashed curve corresponds to the conformal window obtained in the
ladder approximation.} \label{PHComparison}\end{center}
\end{figure}

To better appreciate the differences between these two results we plot the two
conformal windows predicted within these two methods in Figure
\ref{PHComparison} for four types of fermion representation.

The ladder result provides a size of the window, for every  fermion representation, smaller than the bound found earlier. This is a consequence of the value of the anomalous dimension at the lower bound of the window. The unitarity constraint corresponds to $\gamma =2$ while the ladder result is closer to $\gamma \sim 1$. Indeed if we pick $\gamma =1$ our conformal window approaches the ladder result. Incidentally, a value of $\gamma$ larger than one, still allowed by unitarity, is a welcomed feature when using this window to construct walking technicolor theories. It may allow for the physical value of the mass of the top while avoiding a large violation of flavor changing neutral currents \cite{Luty:2004ye} which were investigated in  \cite{Evans:2005pu} in the case of the ladder approximation for minimal walking models.

\subsubsection{Beta function for Multiple Representations}
The generalization for a generic gauge theory with massless fermions in $k$ different representations is:
\begin{eqnarray}
\beta(g) &=&- \frac{g^3}{(4\pi)^2} \frac{\beta_0 - \frac{2}{3}\, \sum_{i=1}^k T(r_i)\,N_{f}(r_i) \,\gamma_i}{1- \frac{g^2}{8\pi^2} C_2(G)\left( 1+ \frac{2\beta_0'}{\beta_0} \right)} \ ,
\end{eqnarray}
with
\begin{eqnarray}
\beta_0' &=& C_2(G) - \sum_{i=1}^k T(r_i)N_f(r_i)  \ ,
\end{eqnarray}
and
\begin{eqnarray}
\beta_0 &=&\frac{11}{3}C_2(G)- \frac{4}{3}\sum_{i=1}^k \,T(r_i)N_f(r_i)  \ .
\end{eqnarray}

\subsection{Testing the beta function}

We now take different limits in theory space and, in doing so, we will gain confidence on the validity of the all order beta function. 
We first recall how to relate the gauge singlet bilinear fermion condensate at different energy scales in the case of the canonically normalized fermion kinetic term 
$\bar{\psi}\gamma^{\mu}D_{\mu} \psi$:
\begin{equation}
\langle \bar{\psi}\psi\rangle _{Q}  =  \exp\left[\int_{\mu}^{Q} dg\, \frac{\gamma(g)}{\beta(g)}\right]  \,\langle \bar{\psi} \psi\rangle_{\mu} \ .
\end{equation}
Here $\bar{\psi} \psi$ is a gauge singlet operator and we have suppressed the color and flavor indices.
At the lowest order in perturbation theory one obtains the simple formula:
\begin{equation}
\langle \bar{\psi}\psi\rangle _{Q}  =  \left[ \frac{g(\mu)^2}{g(Q)^2}\right]^{\frac{3C_2(r)}{\beta_0}} \,\langle \bar{\psi} \psi\rangle_{\mu} \ ,
\end{equation}
with $r$ the representation of the Dirac fermion $\psi$. By construction and at the lowest order in perturbation theory the operator
\begin{equation}
\left[g(Q)^2\right]^{\frac{3C_2(r)}{\beta_0}}\langle\bar{\psi} \psi \rangle _{Q} \ ,
\end{equation} is renormalization group invariant.

\subsubsection{Recovering the Super Yang-Mills beta function at any N}
Consider the theory with one single Weyl fermion transforming according to the adjoint representation of the gauge group. The beta function reads:
\begin{equation}
\beta(g) = -\frac{g^3}{(4\pi)^2}3N\frac{1-\frac{\gamma_{\rm Adj}}{9}}{1-\frac{g^2}{8\pi^2}\frac{4N}{3}} \ ,
\label{SYM-Inspired}
\end{equation}
with $\gamma_{\rm Adj}$ the anomalous dimension of the fermion condensate. This theory corresponds to super Yang-Mills for which we know the result \cite{Jones:1983ip,Novikov:1983uc}:
\begin{equation}
\beta_{SYM}(g) = -\frac{g^3}{(4\pi)^2}\frac{3N}{1-\frac{g^2}{8\pi^2}N} \ .
\end{equation}
In the NSVZ expression above there is no explicit appearance of the anomalous dimension while this is manifest in Eq.~(\ref{SYM-Inspired}). The absence of the anomalous dimension in the NSVZ form of the beta function is due to the choice of normalization of the gluino condensate which renders the associated operator renormalization group invariant. Assuming that the two beta functions have been computed in the same renormalization scheme we can equate them. This provides the expression for the anomalous dimension of the fermion bilinear in the adjoint representation of the gauge group normalized in the standard way:
\begin{equation}
\gamma_{\rm {Adj}} = \frac{g^2}{8\pi^2} \frac{3N}{1-\frac{g^2}{8\pi^2} N} \ .
\end{equation}
Note that in our scheme we have that
\begin{equation}
g^2(Q) \langle\lambda \lambda\rangle_{Q} \ ,
\end{equation}
is a renormalization group invariant quantity to all orders. This is exactly the definition of the gaugino condensate used by NSVZ. One should also note that we recover the perturbative expression of $\gamma_{Adj}$ when expanding to $O(g^2)$.

There is another limit we can take. Consider
large N for one Dirac flavor in the 2 index antisymmetric/symmetric
representation of the gauge group. Then the beta function also goes into the
SYM beta function as shown in \cite{Armoni:2003gp}. Our proposal is, however,  for any number of flavors, colors and matter representation. 

\subsubsection{Pure Yang-Mills and Comparison with Lattice Data}
Pure Yang-Mills is an excellent study case, since it has been widely investigated in the literature, and much is known, especially via lattice simulations. Setting the number of flavors to zero we have:
\begin{equation}
\beta_{YM}(g)= -\frac{g^3}{(4\pi)^2}\frac{{\beta}_0}{1-\frac{g^2}{(4\pi)^2}\frac{{\beta}_1}{{\beta}_0}}\ ,
\end{equation}
with
\begin{equation}
\beta_0=\frac{11N}{3} \ , \qquad \beta_1=\frac{34N^2}{3} \ ,
\end{equation}
respectively for the one and two loop coefficients of the beta function. These are the only universal coefficients of a generic beta function in any scheme.
We now integrate the above beta function and compare our running coupling constant with the two loop result and find:
\begin{eqnarray}
\mu = \Lambda_{1}\exp\left[\frac{8\pi^2}{g^2\beta_0}\right]\left( g^2\beta_0\right)^\frac{\beta_1}{2\beta_0^2}\ , \qquad {\rm All~orders~beta~function}\end{eqnarray}
to be compared with the two loop beta function result:
\begin{eqnarray}
\mu = \Lambda_{2}\exp\left[\frac{8\pi^2}{g^2\beta_0}\right]\left( g^2\beta_0\right)^\frac{\beta_1}{2\beta_0^2}\left( 1 +\frac{ g^2}{16\pi^2}\frac{\beta_1}{\beta_0}\right)^\frac{-\beta_1}{2\beta_0^2} \ .\qquad {\rm 2~loops}\end{eqnarray}
Note that we have normalized the invariant scales $\Lambda_i$ in such a way that they do not depend on the number of colors. It is also clear that the two results do not depend on the number of colors when considering $g^2N$ as the coupling, i.e. the 't Hooft coupling.

\begin{figure}[h]
\begin{center}\resizebox{12cm}{!}{\includegraphics{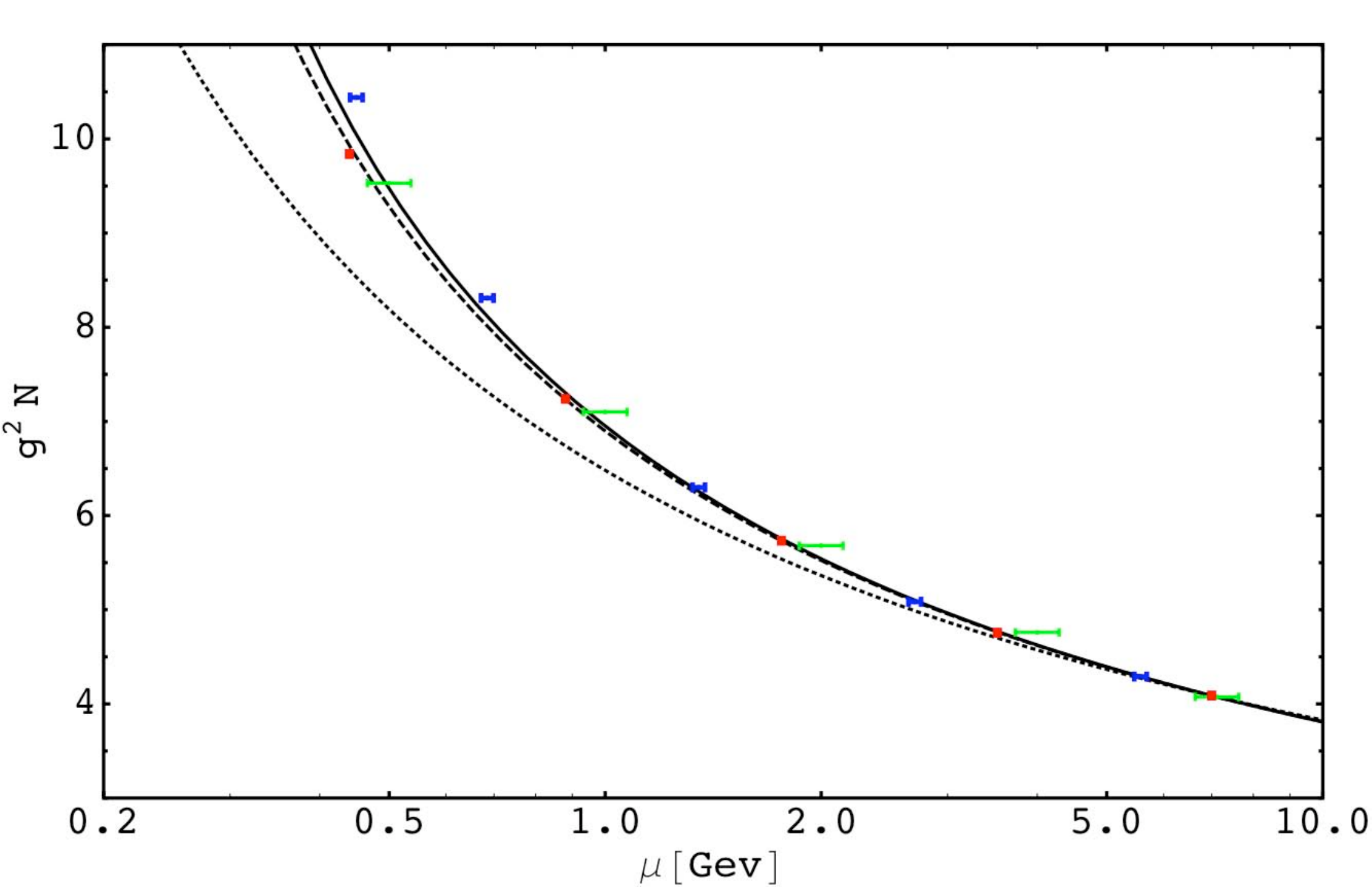}}
\caption{The evolution of the gauge coupling squared times the number of colors (i.e. the 't Hooft coupling) as a function of the energy scale for two, three and four colors. The solid curve is obtained using the susy inspired beta function, the dashed is obtained via the two loop beta function while the dotted curve is the one loop result. The green dots (biggest errorbars) correspond to lattice data for $SU(2)$ \cite{Luscher:1992zx}, the blue dots to $SU(3)$ \cite{Luscher:1993gh} and the red dots (smallest errorbars) to $SU(4)$ \cite{Lucini:2007sa}.}\end{center}
\label{SU(n)}
\end{figure}

It is instructive to compare the deviation from the two loop result of our beta function with the deviation of the lattice data also with respect to the two loop one.
In Figure \ref{SU(n)} we show the evolution of the 't Hooft coupling as a function of the energy scale and plot it together with the two, three and four colors  lattice data. The computation of the running of the coupling constant on the lattice is carried out using the Schr\"odinger functional.  The procedure to determine the beta function on the lattice also defines its renormalization scheme \cite{Luscher:1992an}. 

The solid curve is obtained via the all orders beta function, the dashed is obtained via the two loop beta function while the dotted curve is the one loop result. The green dots (biggest errorbars) correspond to lattice data for $SU(2)$ taken from \cite{Luscher:1992zx}, the blue dots to $SU(3)$ \cite{Luscher:1993gh} and the red dots (smallest errorbars) to $SU(4)$ \cite{Lucini:2007sa}. Despite the fact that the two renormalization schemes are different the size of the corrections with respect to the two loop coming from the lattice data and  the present beta function are similar. We find this result encouraging.

Another  form for the  large N pure Yang-Mills beta function is given in \cite{Bochicchio:2007za}.
\subsection{Conformal Window: Analytical versus Lattice Results}

We  now compare and combine analytical predictions for the conformal window with lattice results \cite{Catterall:2007yx,Shamir:2008pb,Appelquist:2007hu} for a surprisingly large number of gauge theories of great interest for dynamical breaking of the electroweak theory. In the meanwhile the first exhaustive perturbative analysis relevant to start a systematic study of gauge theories with fermions in any given representation of the $SU(N)$ on the lattice has just ppeared \cite{DelDebbio:2008wb}.

\subsubsection{Two Dirac fermions in the two-index symmetric of SU(2) and SU(3)}
Two and three colors with two Dirac flavors transforming according to the two index symmetric (2S) representation of the gauge group have been investigated on the lattice respectively in \cite{Catterall:2007yx} and \cite{Shamir:2008pb}.
 For $SU(2)$ the spectrum of the theory  \cite{Catterall:2007yx} has been studied and confronted with the theory with two colors and two Dirac flavors in the fundamental representation. The lattice studies indicate that either the theory is very near an infrared stable fixed point or the fixed point is already reached. These are only preliminary results and more refined  investigations are needed. Nevertheless let's compare them directly with analytical results. According to ladder results we should be below the conformal window but very near conformal \cite{Sannino:2004qp}.  According to the all orders beta function the anomalous dimension of the mass operator, if the IR fixed point is reached, assumes the value:
 \begin{equation}
 \gamma = \frac{3}{4}   \qquad SU(2)~~{\rm model~with~2~ (2S)~Flavors } \qquad {\rm All~orders~beta~function}   
 \end{equation} 
The all order beta function shows that one has not yet reached $\gamma$ equal one and suggests that the $SU(2)$ model is indeed conformal in the infrared if one uses $\gamma =1 $ as an indication of when the conformal window ceases to exist. However, as explained above, the constraint coming from unitarity of the conformal theories allows $\gamma$ to take even larger values, i.e. up to 2, before loosing conformality. 

The situation is very intriguing for the $SU(3)$ theory. Recent lattice results  \cite{Shamir:2008pb} suggest that this theory may already have achieved an IR fixed point. Here, as well, more studies are needed. The ladder approximation predicts, however, this theory to be near conformal (i.e. walking) but further away from conformality then the $SU(2)$ theory.  If the theory were indeed conformal in the infrared, via the all orders beta function, we predict the anomalous dimension of the fermion condensate to assume the following value:  
 \begin{equation}
 \gamma = 1.3   \qquad SU(3)~~{\rm model~with~2~ (2S)~Flavors} \qquad {\rm All~orders~beta~function} 
 \end{equation} 
The anomalous dimension of the mass operator turns out to be larger than one! This would be quite an important result since large anomalous dimensions are needed when constructing extended technicolor models able to account for the heavy quark masses. In fact the common lore is that the anomalous dimension of the quark operator does not exceed one.  If the $SU(3)$ generates an infrared fixed point then the $SU(2)$ would also generate it since fermions screen even more there.

 \subsubsection{Eight and twelve Dirac fermions in the fundamental of SU(3)}
The all orders beta function {\it predicts} that the conformal window cannot be achieved for a number of flavors less then 8.25 in the fundamental representation of $SU(3)$. This is confirmed by the latest lattice results \cite{Appelquist:2007hu}. In this work it is also suggested that the theory with twelve Dirac fermions has reached an infrared fixed point. The prediction of the anomalous dimension of the quark mass operator is:
\begin{equation}
 \gamma = \frac{3}{4}   \qquad SU(3)~~{\rm model~with~12~Fundamental~Flavors} \qquad {\rm All~orders~beta~function}   
 \end{equation} 
Amusingly the theories with 12 fundamental flavors in $SU(3)$ and 2 adjoint Dirac flavors in $SU(2)$ (adjoint fermions here correspond to the 2S in this case) have the same anomalous dimension if both develop the infrared fixed point. In the ladder approximation one would also expect a conformal fixed point to be reached in this case. What would be extremely interesting to know is if a fixed point is generated for a number of flavors less then eleven but higher than eight since according to the all orders beta function this would correspond to an anomalous dimension larger than one but still smaller than two. 
\begin{figure}[h]
\vskip -2cm
\begin{center}\resizebox{15cm}{!}{\includegraphics{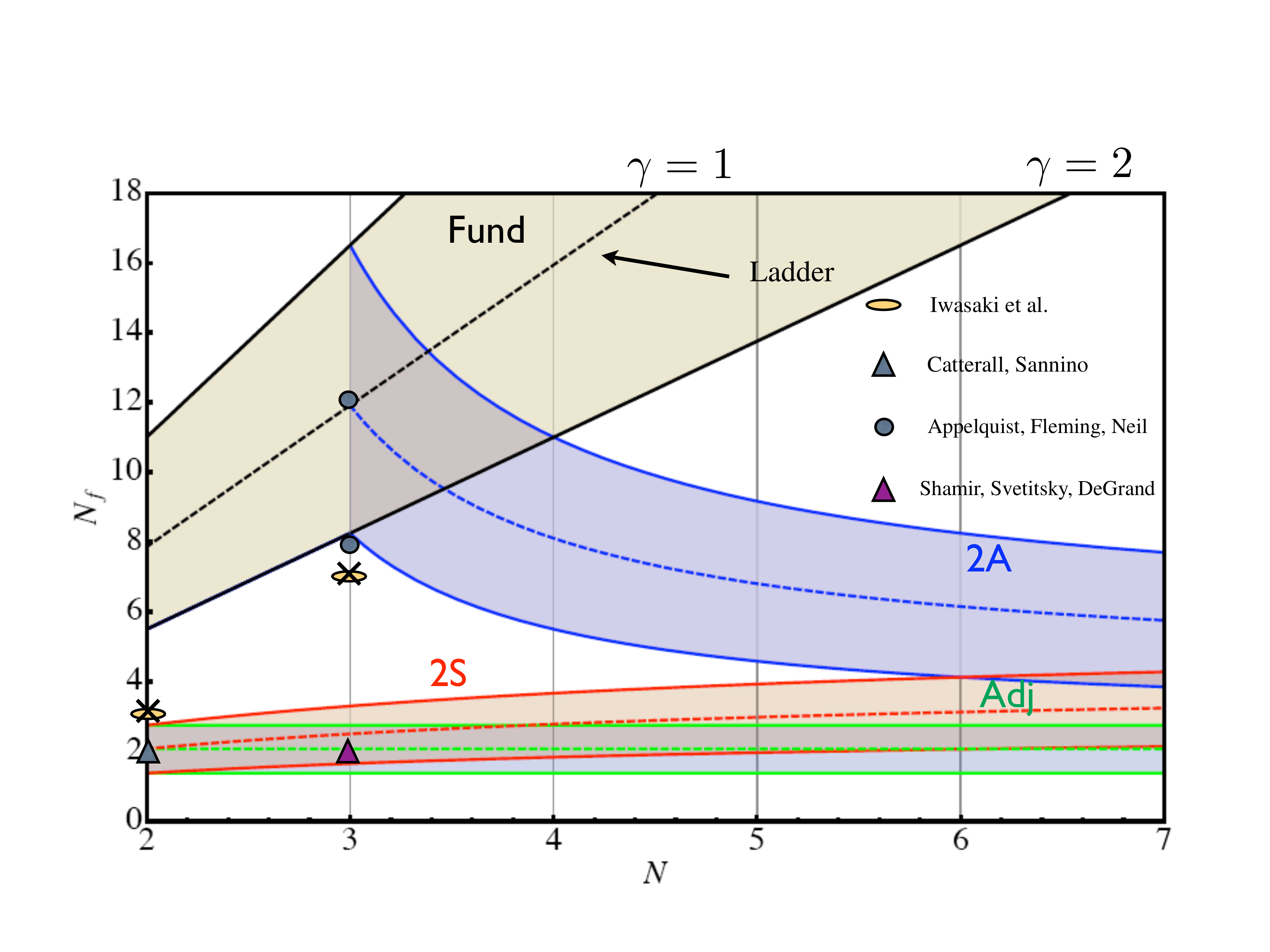}}
\vskip -1cm
\caption{Phase diagram for nonsupersymmetric theories with fermions in various representations with superimposed for which theories the lattice simulations were performed. The dashed lines correspond to the ladder approximation boundary of the conformal window which correspond to gamma about one. While the bound from the all order beta function is obtained for $\gamma=2$. Note that if we were to use $\gamma=1$ constraint with the all order beta function the conformal window would be a little larger than the ladder one. Oval and round circles denote early lattice studies \cite{Iwasaki:2003de} on the lattice with fermions in the fundamental representation. Triangles denote the lattice results for fermions in the two index representation. The cross on the ovals denote that the conclusion of the theories already being conformal  is in disagreement with the theoretical predictions as well as newer lattice results.} \label{LatticeandPD}\end{center}
\end{figure}

\subsection{Constructing Unparticle}

Georgi has put forward the possibility that one may test possible hidden scale invariant sectors weakly interacting with the standard model particles \cite{Georgi:2007ek,Georgi:2007si}. He has also suggested that one may be able to construct the unparticle sector via an asymptotically free gauge theory developing an infrared fixed point  a la Banks-Zaks (BZ) \cite{Banks:1981nn}. The BZ fixed point is, by construction, obtained in perturbation theory and hence all of the anomalous dimensions of the composite gauge singlet operators are small which means that for a gauge theory with underlying fermions all of the dimensions of the composite operators are large and close to the ones of the free theory. 

Here we  will make use of the all order beta function for any nonsupersymmetric SU(N) gauge theory  \cite{Ryttov:2007cx} to provide explicit UV unparticle examples constructed in terms of asymptotic free gauge theories running into an infrared fixed point at low energies. Since the anomalous dimension of the mass within our beta function is known at the infrared fixed point we  will be able to link the UV degrees of freedom with phenomenological constraints. This will  allow to severely constraint the landscape of possible underlying gauge theories which can be used to construct the unparticle world. 

The simplest and natural (i.e. do not lead to new type of fine tuning) four dimensional type of unparticle models to construct are the ones in which at a very large energy scale the standard model is augmented by at least one novel asymptotically free gauge theory which couples to the standard model fields via some heavy mediators of mass of the order of $M_{\cal U}$.  At energies below the mediator mass but still much larger then the electroweak scale  the following operators will then emerge:
\begin{eqnarray}
\frac{1}{M_{\cal U}^{d_{UV}+d_{SM}-4}}{\cal O}_{SM}\,\,{\cal O}_{UV} + {\rm contact~terms}\ ,
\end{eqnarray}
with $d_{UV}$ and $d_{SM}$ the dimensions of the two respective operators and the contact terms taking properly into account the decoupling of the heavy states \cite{Grinstein:2008qk}. Lorentz invariance is imposed and hence any Lorentz preserving structure of the operators is admitted.  Due to the asymptotically free nature of the UV realization of the unparticle sector $d_{UV}$ is the underlying perturbative dimension of the gauge singlet composite operators. Following Georgi we require near or above the electroweak scale the hidden sector to develop an infrared fixed point which is typically nonperturbative. The $UV$ operator is now replaced by its $IR$ operator with scaling dimensions $d_{IR}=d_{UV} - \gamma_{\cal U}$ and a new scale $\Lambda_{\cal U}$, smaller than $M_{\cal U}$, naturally emerges keeping track of the change in dimensions. The low energy effective operator reads: 
\begin{eqnarray}
\frac{\Lambda_{\cal U}^{\gamma_{\cal U}}}{M_{\cal U}^{d_{UV}+d_{SM}-4}}{\cal O}_{SM}\,{\cal O}_{IR} \equiv\frac{{\cal O}_{SM}\,{\cal O}_{IR}}{{M^{\prime}_{\cal U}}^{d_{UV}+d_{SM}-4 - \gamma_{\cal U}}}\ .
\end{eqnarray}
The goal here is to identify and constrain the landascape of nonsupersymmetric gauge theories which can be used to construct the UV completion of the ${\cal{O}}_{IR}$ using as tool the all order beta function presented in \cite{Ryttov:2007cx}. This tool allows to determine at the IR fixed point $\gamma_{\cal U}$, at least for  a primary scalar operator, directly in terms of the defining UV gauge degrees of freedom. By comparing the explicit expression for $\gamma_{\cal{U}}$ with the phenomenological constraints one can immediately constrain the underlying UV landscape. We observe that by using explicit realization of the unparticle world in terms of asymptotically free gauge theories one bypasses a number of potentially dangerous problems that an uncontrolled unparticle physical realization can provide such as inconsistent UV constructions.  A similar analysis using the supersymmetric conformal windows has been performed in \cite{Fox:2007sy,Nakayama:2007qu}.

\subsubsection{Scalar Unparticle}
As our starting point, to construct an underlying gauge theory able to provide the desired IR fixed point in the infrared, we consider a generic $SU(N)$ non supersymmetric gauge theory with $N_f $ Dirac fermions in a given representation $r$
of the gauge group. We are now ready to provide explicit {\it constructions} of the nonsupersymmetric unparticle sectors built out of ordinary gauge theories. 

The simplest operator to use is the scalar one constructed in the UV as follows:
\begin{equation}
{\cal O}_{UV} = \bar{\psi} \psi \ , \quad {\rm with} \quad d_{UV} = 3 \ .
\end{equation} 
At the infrared fixed point we have: 
\begin{equation}{\cal O}_{IR}  \quad {\rm with} \quad d_{IR} = 3 - \gamma_{\cal U} \ , 
\end{equation}
with $\gamma_{\cal U} = \gamma $. A compendium of operators emerging when coupling unparticle operators with a given scale dimension to the SM ones can be found in \cite{Chen:2007qr} while a comparision of constraints from astrophysics and collider physics has appeared in \cite{Freitas:2007ip,Deshpande:2007mf}. Allowing, for simplicity, only parity even couplings with our scalar operator one has\begin{eqnarray}
\frac{{\cal O}_{IR}}{M_Z^{3-\gamma}}\left[ c_{S1}\bar{f}\gamma_{\mu}D^{\mu}f -\frac{c_{\gamma\gamma}}{4}
F_{\mu\nu}F^{\mu\nu}\right] \ .
\end{eqnarray}
We  used the notation given in \cite{Freitas:2007ip} in which the underlying scales of dynamics have been arranged such that $M^{\prime}_{\cal{U}} =M_Z$ which is the $Z$ gauge boson mass with $c_{S1}$ and $c_{\gamma \gamma}$ dimensionless couplings. One can now directly export the phenomenological constraints onto the theory space featuring a concrete construction of the unparticle worlds. For example, from table 2 in \cite{Freitas:2007ip} one learns that the strongest constraints on $c_{S1}$ come from 5th force experiments while from  table 3 \cite{Freitas:2007ip} $c_{\gamma\gamma}$ is constrained studying the stars' energy loss. It turns out that for a $c_{S1}$ larger than $1.6\times 10^{-8}$ and $c_{\gamma\gamma}$ larger than $5.3\times 10^{-9}$ one has to take $\gamma < 4/3$ which is translated in theory space in the following range for $N_f$  and $N$:
\begin{eqnarray}
 \frac{33}{20 T(r)} \leq \frac{N_f}{N}< \frac{11}{6 T(r)} \ ,
\label{boundaries}
\end{eqnarray}
where the upper limit corresponding to the $\gamma=1$, i.e. $d_{IR}=2$,  is intended to allow for sizable low energy interactions. The maximum value of $N_f/N =  \frac{11}{4 T(r)}$ derives, instead, from the loss of  asymptotic freedom. The lower bound derives from the phenomenological constraints on $c_{S1}$ and $c_{\gamma \gamma}$ above. We have assumed the conformal window to extends till $\gamma =2$, i.e. till the lower bound of the conformal window. As an explicit example of an underlying theory for unparticle is the $SU(2)$ gauge theory with $N_f=7$ Dirac fermions in the fundamental representation yielding $N_f/N = 3.5$ which is a value within the boundaries in (\ref{boundaries}). 

If the conformal window does not extend till $\gamma=2$ but it is limited to the range  $0<\gamma \leq 1$ - where the upper bound for $\gamma$ derives from the ladder approximation -- then the experimental bounds are very weak. The relevance of the knowledge of the beta function resides here in the explicit relation between gamma at the IR fixed point and the parameters of the theory, i.e. representation, number of flavors and colors.

\subsubsection{Spinor Unparticle}
Spinorial unparticle operators emerge naturally in theories with adjoint matter. In this case, in fact, one can construct the following unparticle spinor operator:
\begin{eqnarray}
{{\cal O}_{\alpha,f}}_{UV} = G_{\mu\nu}^A\sigma^{\mu\nu}_{\alpha \beta} \lambda^{A\beta}_{f}  \ , \quad {\rm with} \quad d_{UV}^{\rm spinor}=\frac{5}{2} \ ,
\end{eqnarray}
with $A=1,\ldots N^2-1$ the gauge indices in the adjoint representation, $\alpha, \beta$ the spinor indices, $f$ the flavor index ranging from one to $N_{w}=2N_f$ which is the number of adjoint Weyl fermions $\lambda^{A\alpha}_f$. For fermions in the adjoint the scalar unparticle constraints found earlier lead to the following range for  $N_{w}$ which, for any number of colors, reads:
\begin{eqnarray}
3.3\leq N_{w} < 3.\bar{6} \ , \quad {\rm with} \quad 1<\gamma \leq 4/3 \ ,
\end{eqnarray}
which means that given that $N_{w}$ must be an integer there is {\it no} $SU(N)$ gauge theory with fermions transforming solely according to the adjoint  (with $1<\gamma \leq 4/3$ ) representation one can use to construct the unparticle world. If we extend $\gamma$ to $5/3$ corresponding to $d_{IR}=4/3$ we have $3 \leq N_{w} < 3.\bar{6} $ which means that one can construct an unparticle world with at least three Weyl fermions. Note that further extending the window to $\gamma =2$ does not increase the number of possible integer values of $N_{w}$.  For $N_w=3$ one can then construct the gauge singlet spinor operator ${{\cal O}_{\alpha,f}}_{UV}$ with the infrared nonperturbative dimension being constrained by unitarity arguments \cite{Mack:1975je,Flato:1983te,Dobrev:1985qv}. 
\begin{equation}
{{\cal O}_{\alpha,f}}_{IR} \ , \quad {\rm with}  \quad  \frac{3}{2}\leq d_{IR}^{\rm spinor}\leq\frac{5}{2} \ .
\end{equation}
In order to couple this operator to the SM we need to break the $SU(3)$ flavor symmetry acting on the three Weyl fermions.

In the case $0<\gamma <1$ then $ 3.\bar{6}<N_w<5.5$  and for 4 and 5 number of Weyl fermions an unparticle world having also spinorial unparticle can be constructed.

{}For the unparticle scenario presented here the vectorial currents constructed using the fermion bilinears must have conformal dimensions larger than two and hence they will not be considered here. Besides, if they are conserved currents, due to non-renormalization theorems will have dimension three even at the nonperturbative infrared fixed point leading to irrelevant operators.

\subsubsection{Sizing the unparticle world: A New Universal Ratio}
In \cite{Ryttov:2007sr} we suggested a measure of how large, in theory space, the fraction of the
unparticle world is. We showed that a reasonable
measure is then, for a given representation, the ratio of the area of the
conformal window to that of the total window for asymptotically free gauge
theories
\begin{equation}
R_{FP} = \frac{\int_{N_{min}}^{\infty} N_f^{\mathrm I} \,dN
-\int_{N_{min}}^{\infty}  N_f^{\mathrm {II}} \,
dN}{\int_{N_{min}}^{\infty} N_f^{\mathrm I}\, dN }  \ ,
\end{equation}
where $N_{min}$ is the smallest number of colors permitted for
the chosen representation.

Remarkably, the above ratio turned out to be universal, i.e. independent of the matter representation, for any ${\cal N}=1$ supersymmetric theory\footnote{To be precise we can discuss the universal ratio only for the bound on the supersymmetric conformal windows.}. The value being 1/2. Using our beta function for nonsupersymmetric theories we again find the same universal result:
\begin{eqnarray}
R_{FP} = \frac{\frac{11}{4}-\frac{11}{8}}{\frac{11}{4}} = \frac{1}{2} \ . \qquad \qquad {\rm All~orders}
\end{eqnarray}
This must be considered, in this case, as an upper bound. Choosing $\gamma =1$ for determining the conformal window we find again an universal ration but this time $R_{FP} = 1/3$. 

A generic gauge theory will, in general, have matter transforming
according to distinct representations of the gauge group. We follow
the analysis first performed in  \cite{Ryttov:2007sr} of the conformal region for a generic $SU(N)$
gauge theory with ${N_f}(r_i)$ vector-like matter fields
transforming according to the representation $r_i$ with
$i=1,\ldots,k$ . We shall consider the non-supersymmetric case here and will use our beta function to determine the fraction of conformal regions.

The generalization to $k$ different representations for
the expression determining the region in flavor space above which
asymptotic freedom is lost is simply
\begin{eqnarray}
\sum_{i=1}^{k}\frac{4}{11}T(r_i)N_f(r_i) = C_2(G) \ .
\end{eqnarray}

Following \cite{Ryttov:2007sr} we estimate the region above which the theories
develop an infrared fixed point via the following expression
\begin{eqnarray}
\sum_{i=1}^{k} \frac{8}{11} T(r_i) N_f(r_i) = C_2(G) \ ,
\end{eqnarray}

The volume, in flavor and color space,
occupied by a generic $SU(N)$ gauge theory is defined to be:
\begin{eqnarray}
V_{\eta}[N_{min}, N_{max}] &=& \int_{N_{min}}^{N_{max}} dN
\prod_{i=1}^{k} \int_{0}^{\frac{C_2(G) - \sum_{j=2}^{i} \eta
T(r_j)N_f(r_j)}{\eta T(r_{i+1})}} N_f(r_{i+1}) \ ,
\end{eqnarray}
with $\eta$ reducing to the number $4/11$ when the region to be evaluated is associated to the asymptotically free one and to $8/11$ when the region is the one below which one does not expect the occurrence of an infrared fixed point. The notation is such that $T(r_{k+1})\equiv
T(r_1)$, $N_f(r_{k+1}) \equiv N_f(r_1)$ and the sum
$\sum_{j=2}^{i}\eta \,T(r_j)N_f(r_j)$ in the upper limit of the flavor
integration vanishes for $i=1$. We defined the volume within a
fixed range of number of colors $N_{min}$ and $N_{{max}}$.

Hence the fraction of the conformal region to the region occupied by
the asymptotically free theories is, for a given number of
representations $k$:
\begin{eqnarray}
R_{FP} = \frac{V_{\frac{4}{11}}[N_{min}, N_{max}] -
V_{\frac{8}{11}}[N_{ {min}},N_{
{max}}]}{V_{\frac{4}{11}}[N_{ {min}},N_{ {max}}]} = 1-\left(\frac{1}{2}\right)^k\ .
\end{eqnarray}
Quite surprisingly the result obtained using the present beta function does not depend on which representation one uses but depends solely on the number $k$ of representations present. We recover $1/2$ for $k=1$. We estimated this ratio in the case of the ladder approximation first in \cite{Ryttov:2007sr}. We noticed then a small dependence which, however, can be related to the large uncertainty stemming from the ladder approximation. The bound of the conformal region is larger than the one computed using the ladder approximation.

We have presented a comprehensive analysis of the phase diagram of
non-supersymmetric vector-like and strongly coupled SU($N$) gauge theories 
with matter in various representations of the gauge group. 

Other approaches, such as the instanton-liquid model \cite{Schafer:1995pz} or the one developed in \cite{Grunberg:2000ap,Gardi:1998ch,Grunberg:1996hu} have also been used to investigate the QCD chiral 
phase transition as function of the number of flavors. According to the instanton-liquid model one expects for QCD the transition to occur for a very small number of flavors. This result is at odds with the bound for the conformal window found with our novel beta function \cite{Ryttov:2007cx} as well lattice data \cite{Appelquist:2007hu}.

\newpage
\section{Minimal Walking Theories}

The simplest technicolor model has $N_{T f}$ Dirac fermions in
the fundamental representation of $SU(N)$. These models, when
extended to accommodate the fermion masses through the extended technicolor interactions,
suffer from large flavor changing neutral currents. This problem is
alleviated if the number of flavors is sufficiently large
such that the theory is almost conformal. This is estimated to happen
for $N_{T f} \sim 4 N$ \cite{Yamawaki:1985zg} as also summarized in the section dedicated to the Phase Diagram of strongly interacting theories. This, in turn, implies a large
contribution to the oblique parameter $S$ (within naive estimates) \cite{Hong:2004td}. Although near the conformal window \cite{
Appelquist:1998xf,Sundrum:1991rf} the $S$ parameter is reduced due to
non-perturbative corrections, it is still too large if the
model has a large particle content. In addition, such models may
have a large number of pseudo Nambu-Goldstone bosons. By
choosing a higher dimensional technicolor representation for the new technifermions one
can overcome these problems \cite{Sannino:2004qp,Hong:2004td}. 

To have a very low $S$ parameter one would ideally have a technicolor theory which with only one doublet breaks dynamically the electroweak theory but at the same time being walking to reduce the $S$ parameter. The walking nature then also enhances the scale responsible for the fermion mass generation. 
 
According to the phase diagram exhibited earlier the promising candidate theories with the properties required are either theories with fermions in the adjoint representation or two index symmetric one. In Table \ref{symmetric} we present the
generic S-type theory.
\begin{table}[h]
\begin{center}
%\begin{minipage}{3in}
\begin{tabular}{c||ccccc }
 & $SU(N)$ & $SU_L(N_{T f})$& $SU_R(N_{T f})$&$U_V(1)$ & $U_A(1)$  \\
  \hline \hline \\
${Q_L}$& $\symm$ & $\fund$ & $1$ & $1$ & $1$  \\
 &&&\\
 $\bar{Q}_R$ & $\bsymm $ & $1$& $\bfund$& $-1$ & $1$ \\
 &&&\\
$G_{\mu}$ & {\rm Adj} & $0$&$0$ &$0$ & $0$    \\
\end{tabular}
%\end{minipage}
\end{center}
\caption{Schematic representation of a generic nonsupersymmetric vector like $SU(N)$ gauge theory with matter content 
in the two-index representation.  Here $Q_{L(R)}$ are Weyl fermions.}
\label{symmetric}
\end{table}

The relevant feature, found first in \cite{Sannino:2004qp} using the ladder approximation, is that the
S-type theories can be near conformal already at $N_{T f}=2$ when $N=2$ or $3$. This
should be contrasted with theories in which the fermions are in the fundamental
representation for which the minimum number of flavors required to
reach the conformal window is eight for $N=2$. This last statement is supported by the all order beta function results \cite{Ryttov:2007cx} as well as lattice simulations \cite{Catterall:2007yx,Shamir:2008pb,Appelquist:2007hu}. The critical value of flavors increases with the number of colors
for the gauge theory with S-type matter: the limiting value is
$4.15$ at large $N$. 

The situation is different for the theory with A-type matter. Here the critical number of flavors increases when decreasing the
number of colors. The maximum value of about $N_{Tf}=12$  is obtained - in the ladder approximation - for
$N=3$, i.e. standard QCD.  In reference \cite{Hong:2004td} it has been argued that 
the nearly conformal A-type theories have, already 
at the perturbative level, a very large $S$ parameter with respect to the experimental data. These theories can be re-considered if one gauges under the electroweak symmetry only a part of the flavor symmetries as we shall see in the section dedicated to {\it partially gauged} technicolor.

\subsection{Minimal Walking Technicolor (MWT)}

The dynamical sector we consider, which underlies the Higgs mechanism, is an SU(2) technicolor gauge theory with two adjoint
technifermions \cite{Sannino:2004qp}. The theory is asymptotically free if the number of flavors $N_f$ is less than $2.75$ according to the ladder approximation. Lattice results support the conformal or near conformal behavior of this theory. In any event the symmetries and properties of this model make it ideal for a comprehensive study for LHC physics. 

The two adjoint fermions are conveniently written as \beq Q_L^a=\left(\begin{array}{c} U^{a} \\D^{a} \end{array}\right)_L , \qquad U_R^a \
, \quad D_R^a \ ,  \qquad a=1,2,3 \ ,\eeq with $a$ being the adjoint color index of SU(2). The left handed fields are arranged in three
doublets of the SU(2)$_L$ weak interactions in the standard fashion. The condensate is $\langle \bar{U}U + \bar{D}D \rangle$ which
correctly breaks the electroweak symmetry as already argued for ordinary QCD in (\ref{qcd-condensate}).

The model as described so far suffers from the Witten topological anomaly \cite{Witten:fp}. However, this can easily be solved by
adding a new weakly charged fermionic doublet which is a technicolor singlet \cite{Dietrich:2005jn}. Schematically: 
\beq L_L =
\left(
\begin{array}{c} N \\ E \end{array} \right)_L , \qquad N_R \ ,~E_R \
. \eeq In general, the gauge anomalies cancel using the following
generic hypercharge assignment
\begin{align}
Y(Q_L)=&\frac{y}{2} \ ,&\qquad Y(U_R,D_R)&=\left(\frac{y+1}{2},\frac{y-1}{2}\right) \ , \label{assign1} \\
Y(L_L)=& -3\frac{y}{2} \ ,&\qquad
Y(N_R,E_R)&=\left(\frac{-3y+1}{2},\frac{-3y-1}{2}\right) \ \label{assign2} ,
\end{align}
where the parameter $y$ can take any real value \cite{Dietrich:2005jn}. In our notation
the electric charge is $Q=T_3 + Y$, where $T_3$ is the weak
isospin generator. One recovers the SM hypercharge
assignment for $y=1/3$.

To discuss the symmetry properties of the
theory it is
convenient to use the Weyl basis for the fermions and arrange them in the following vector transforming according to the
fundamental representation of SU(4)
\beq Q= \begin{pmatrix}
U_L \\
D_L \\
-i\sigma^2 U_R^* \\
-i\sigma^2 D_R^*
\end{pmatrix},
\label{SU(4)multiplet} \eeq where $U_L$ and $D_L$ are the left
handed techniup and technidown, respectively and $U_R$ and $D_R$ are
the corresponding right handed particles. Assuming the standard
breaking to the maximal diagonal subgroup, the SU(4) symmetry
spontaneously breaks to $SO(4)$. Such a breaking is driven by the
following condensate \beq \langle Q_i^\alpha Q_j^\beta
\epsilon_{\alpha \beta} E^{ij} \rangle =-2\langle \overline{U}_R U_L
+ \overline{D}_R D_L\rangle \ , \label{conde}
 \eeq
where the indices $i,j=1,\ldots,4$ denote the components
of the tetraplet of $Q$, and the Greek indices indicate the ordinary
spin. The matrix $E$ is a $4\times 4$ matrix defined in terms
of the 2-dimensional unit matrix as
 \beq E=\left(
\begin{array}{cc}
0 & \mathbbm{1} \\
\mathbbm{1} & 0
\end{array}
\right) \ . \eeq

We follow the notation of Wess and Bagger \cite{WessBagger}
$\epsilon_{\alpha \beta}=-i\sigma_{\alpha\beta}^2$ and $\langle
 U_L^{\alpha} {{U_R}^{\ast}}^{\beta} \epsilon_{\alpha\beta} \rangle=
 -\langle  \overline{U}_R U_L
 \rangle$. A similar expression holds for the $D$ techniquark.
The above condensate is invariant under an $SO(4)$ symmetry. This leaves us with nine broken  generators with associated Goldstone bosons.

Replacing the Higgs sector of the SM with the MWT the Lagrangian now
reads:
\begin{eqnarray}
\mathcal{L}_H &\rightarrow &  -\frac{1}{4}{\cal F}_{\mu\nu}^a {\cal F}^{a\mu\nu} + i\bar{Q}_L
\gamma^{\mu}D_{\mu}Q_L + i\bar{U}_R \gamma^{\mu}D_{\mu}U_R +
i\bar{D}_R \gamma^{\mu}D_{\mu}D_R \nonumber \\
&& +i \bar{L}_L \gamma^{\mu} D_{\mu} {L}_L +
i\bar{N}_R \gamma^{\mu}D_{\mu}N_R + i\bar{E}_R
\gamma^{\mu}D_{\mu}E_R
\end{eqnarray}
with the technicolor field strength ${\cal F}_{\mu\nu}^a =
\partial_{\mu}{\cal A}_{\nu}^a - \partial_{\nu}{\cal A}_{\mu}^a + g_{TC} \epsilon^{abc} {\cal A}_{\mu}^b
{\cal A}_{\nu}^c,\ a,b,c=1,\ldots,3$.
For the left handed techniquarks the covariant derivative is:

\begin{eqnarray}
D_{\mu} Q^a_L &=& \left(\delta^{ac}\partial_{\mu} + g_{TC}{\cal
A}_{\mu}^b \epsilon^{abc} - i\frac{g}{2} \vec{W}_{\mu}\cdot
\vec{\tau}\delta^{ac} -i g'\frac{y}{2} B_{\mu} \delta^{ac}\right)
Q_L^c \ .
\end{eqnarray}
${\cal A}_{\mu}$ are the techni gauge bosons, $W_{\mu}$ are the
gauge bosons associated to SU(2)$_L$ and $B_{\mu}$ is the gauge
boson associated to the hypercharge. $\tau^a$ are the Pauli matrices
and $\epsilon^{abc}$ is the fully antisymmetric symbol. In the case
of right handed techniquarks the third term containing the weak
interactions disappears and the hypercharge $y/2$ has to be replaced
according to whether it is an up or down techniquark. For the
left-handed leptons the second term containing the technicolor
interactions disappears and $y/2$ changes to $-3y/2$. Only the last
term is present for the right handed leptons with an appropriate
hypercharge assignment.

\subsubsection{Low Energy Theory for MWT}

We construct the effective theory for MWT including composite scalars and vector bosons, their self interactions, and their interactions with the electroweak gauge fields and the standard model fermions

\subsubsection{Scalar Sector} \label{sec:scalar}
The relevant effective theory for the Higgs sector at the electroweak scale consists, in our model, of a composite Higgs and its pseudoscalar partner, as well as nine pseudoscalar Goldstone bosons and their scalar partners. These
can be assembled in the matrix
\begin{eqnarray}
M = \left[\frac{\sigma+i{\Theta}}{2} + \sqrt{2}(i\Pi^a+\widetilde{\Pi}^a)\,X^a\right]E \ ,
\label{M}
\end{eqnarray}
which transforms under the full SU(4) group according to
\begin{eqnarray}
M\rightarrow uMu^T \ , \qquad {\rm with} \qquad u\in {\rm SU(4)} \ .
\end{eqnarray}
The $X^a$'s, $a=1,\ldots,9$ are the generators of the SU(4) group which do not leave  the vacuum expectation value (VEV) of $M$ invariant
\begin{eqnarray}
\langle M \rangle = \frac{v}{2}E
 \ .
\end{eqnarray}
Note that the notation used is such that $\sigma$ is a \emph{scalar}
while the $\Pi^a$'s are \emph{pseudoscalars}. It is convenient to
separate the fifteen generators of SU(4) into the six that leave the
vacuum invariant, $S^a$, and the remaining nine that do not, $X^a$.
Then the $S^a$ generators of the SO(4) subgroup satisfy the relation
\begin{eqnarray}
S^a\,E + E\,{S^a}^{T} = 0 \ ,\qquad {\rm with}\qquad  a=1,\ldots  ,  6 \ ,
\end{eqnarray}
so that $uEu^T=E$, for $u\in$ SO(4). The explicit realization of the generators is shown in appendix \ref{appgen}.
With the tilde fields included, the matrix $M$ is invariant in form under U(4)$\equiv$SU(4)$\times$U(1)$_{\rm
A}$, rather than just SU(4). However the U(1)$_{\rm A}$ axial symmetry is anomalous, and is therefore broken at the quantum level.

The connection between the composite scalars and the underlying techniquarks can be derived from the transformation properties under SU(4), by observing that the elements of the matrix $M$ transform like techniquark bilinears:
\begin{eqnarray}
M_{ij} \sim Q_i^\alpha Q_j^\beta \varepsilon_{\alpha\beta} \quad\quad\quad {\rm with}\ i,j=1\dots 4.
\label{M-composite}
\end{eqnarray}
Using this expression, and the basis matrices given in appendix \ref{appgen}, the scalar fields can be related to the wavefunctions of the techniquark bound states. This gives the following charge eigenstates:
\begin{eqnarray}
\begin{array}{rclcrcl}
v+H & \equiv & \sigma \sim  \overline{U}U+\overline{D}D  &,~~~~ &
\Theta  &\sim& i \left(\overline{U} \gamma^5 U+\overline{D} \gamma^5 D\right) \ ,  \\
A^0 & \equiv & \widetilde{\Pi}^3  \sim  \overline{U}U-\overline{D}D &,~~~~ &
\Pi^0 & \equiv & \Pi^3 \sim i \left(\overline{U} \gamma^5 U-\overline{D} \gamma^5 D\right) \ , \\
A^+ & \equiv & {\displaystyle \frac{\widetilde{\Pi}^1 - i \widetilde{\Pi}^2}{\sqrt{2}}} \sim \overline{D}U &,~~~~&
\Pi^+ & \equiv & {\displaystyle \frac{\Pi^1 - i \Pi^2}{\sqrt{2}}} \sim i \overline{D} \gamma^5 U \ , \\
A^- & \equiv & {\displaystyle \frac{\widetilde{\Pi}^1 + i \widetilde{\Pi}^2}{\sqrt{2}}} \sim \overline{U}D &,~~~~&
\Pi^- & \equiv & {\displaystyle \frac{\Pi^1 + i \Pi^2}{\sqrt{2}}} \sim i \overline{U} \gamma^5 D \ ,
\end{array}
\label{TM-eigenstates}
\end{eqnarray}
for the technimesons, and
\begin{eqnarray}
\begin{array}{rcl}
\Pi_{UU} & \equiv & {\displaystyle \frac{\Pi^4 + i \Pi^5 + \Pi^6 + i \Pi^7}{2}} \sim U^T C U \ , \\
\Pi_{DD} & \equiv & {\displaystyle \frac{\Pi^4 + i \Pi^5 - \Pi^6 - i \Pi^7}{2}} \sim D^T C D \ , \\
\Pi_{UD} & \equiv & {\displaystyle \frac{\Pi^8 + i \Pi^9}{\sqrt{2}}} \sim U^T C D \ , \\
\widetilde{\Pi}_{UU} & \equiv &
{\displaystyle \frac{\widetilde{\Pi}^4 + i \widetilde{\Pi}^5 + \widetilde{\Pi}^6 + i \widetilde{\Pi}^7}{2}} \sim i U^T C \gamma^5 U \ , \\
\widetilde{\Pi}_{DD} & \equiv &
{\displaystyle \frac{\widetilde{\Pi}^4 + i \widetilde{\Pi}^5 - \widetilde{\Pi}^6 - i \widetilde{\Pi}^7}{2}} \sim i D^T C \gamma^5 D  \ , \\
\widetilde{\Pi}_{UD} & \equiv & {\displaystyle \frac{\widetilde{\Pi}^8 + i \widetilde{\Pi}^9}{\sqrt{2}}} \sim i U^T C \gamma^5 D \ ,
\end{array}
\label{TB-eigenstates}
\end{eqnarray}
for the technibaryons, where $U\equiv (U_L,U_R)^T$ and $D\equiv (D_L,D_R)^T$ are Dirac technifermions, and $C$ is the charge conjugation matrix, needed to form Lorentz-invariant objects. To these technibaryon charge eigenstates we must add the corresponding charge conjugate states ({\em e.g.} $\Pi_{UU}\rightarrow \Pi_{\overline{U}\overline{U}}$).

It is instructive to split the scalar matrix into four two by two blocks as follows:
\begin{equation}  
M=\begin{pmatrix} {\cal X}  & {\cal O}  \\ {\cal O}^T & {\cal Z} \end{pmatrix} \ ,
\end{equation}
with ${\cal X}$ and ${\cal Z}$ two complex symmetric matrices accounting for six independent degrees of freedom each and ${\cal O}$ is a generic complex two by two matrix featuring eight real bosonic fields. ${\cal O}$ accounts for the standard model like Higgs doublet and a second copy as well as for the three Goldstones which upon electroweak gauging will become the longitudinal components of the intermediate massive vector bosons. 

The electroweak subgroup can be embedded in SU(4), as explained in detail in \cite{Appelquist:1999dq}. Here SO(4) is the subgroup to which SU(4) is maximally broken. The generators $S^a$, with $a=1,2,3$, form an SU(2) subgroup of SU(4), which we denote by SU(2)$_{\rm V}$, while $S^4$ forms a U(1)$_{\rm V}$ subgroup. The $S^a$ generators, with $a=1,..,4$, together with the $X^a$ generators, with $a=1,2,3$, generate an SU(2)$_{\rm L}\times$SU(2)$_{\rm R}\times$U(1)$_{\rm V}$ algebra. This is easily seen by changing genarator basis from $(S^a,X^a)$ to $(L^a,R^a)$, where
\begin{eqnarray}
L^a \equiv \frac{S^a + X^a}{\sqrt{2}} = \begin{pmatrix}\frac{\tau^a}{2}\ \ \  & 0 \\ 0 & 0\end{pmatrix} \ , \ \
{-R^a}^T \equiv \frac{S^a-X^a}{\sqrt{2}}  = \begin{pmatrix}0 & 0 \\ 0 & -\frac{{\tau^a}^T}{2}\end{pmatrix} \ ,
\end{eqnarray}
with $a=1,2,3$. The electroweak gauge group is then obtained by gauging ${\rm SU(2)}_{\rm L}$, and the ${\rm U(1)}_Y$ subgroup of ${\rm SU(2)}_{\rm R}\times {\rm U(1)}_{\rm V}$, where
\begin{eqnarray}
Y =  -{R^3}^T + \sqrt{2}\ Y_{\rm V}\ S^4 \ ,
\end{eqnarray}
and $Y_{\rm V}$ is the U(1)$_{\rm V}$ charge. For example, from Eq.~(\ref{assign1}) and Eq.~(\ref{assign2}) we see that $Y_{\rm V}=y$ for the techniquarks, and $Y_{\rm V}=-3y$ for the new leptons. As SU(4) spontaneously breaks to SO(4), ${\rm SU(2)}_{\rm L}\times {\rm SU(2)}_{\rm R}$ breaks to ${\rm SU(2)}_{\rm V}$. As a consequence, the electroweak symmetry breaks to ${\rm U(1)}_Q$, where
\begin{eqnarray}
Q = \sqrt{2}\ S^3 + \sqrt{2}\ Y_{\rm V} \ S^4 \ .
\end{eqnarray}
Moreover the ${\rm SU(2)}_{\rm V}$ group, being entirely contained in the unbroken SO(4), acts as a custodial isospin, which insures that the $\rho$ parameter is equal to one at tree-level.

The electroweak covariant derivative for the $M$ matrix is
\begin{eqnarray}
D_{\mu}M =\partial_{\mu}M - i\,g \left[G_{\mu}(y)M + MG_{\mu}^T(y)\right]  \
, \label{covariantderivative}
\end{eqnarray}
where
\begin{eqnarray}
g\ G_{\mu}(Y_{\rm V}) & = & g\ W^a_\mu \ L^a + g^{\prime}\ B_\mu \ Y  \nonumber \\
& = & g\ W^a_\mu \ L^a + g^{\prime}\ B_\mu \left(-{R^3}^T+\sqrt{2}\ Y_{\rm V}\ S^4\right) \ .
\label{gaugefields}
\end{eqnarray}
Notice that in the last equation $G_\mu(Y_{\rm V})$ is written for a general U(1)$_{\rm V}$ charge $Y_{\rm V}$, while in Eq.~(\ref{covariantderivative}) we have to take the U(1)$_{\rm V}$ charge of the techniquarks, $Y_{\rm V}=y$, since these are the constituents of the matrix $M$, as explicitly shown in Eq.~(\ref{M-composite}).

Three of the nine Goldstone bosons associated with the broken generators become the longitudinal degrees of freedom of
the massive weak gauge bosons, while the extra six Goldstone bosons will acquire a mass due to extended technicolor interactions (ETC) as well as the
electroweak interactions per se. Using a bottom up approach we will not commit to a specific ETC theory but limit ourself to introduce the minimal low energy operators  needed to construct a phenomenologically viable theory. The new Higgs Lagrangian is
\begin{eqnarray}
{\cal L}_{\rm Higgs} &=& \frac{1}{2}{\rm Tr}\left[D_{\mu}M D^{\mu}M^{\dagger}\right] - {\cal V}(M) + {\cal L}_{\rm ETC} \ ,
\end{eqnarray}
where the potential reads
\begin{eqnarray}
{\cal V}(M) & = & - \frac{m^2}{2}{\rm Tr}[MM^{\dagger}] +\frac{\lambda}{4} {\rm Tr}\left[MM^{\dagger} \right]^2 
+ \lambda^\prime {\rm Tr}\left[M M^{\dagger} M M^{\dagger}\right] \nonumber \\
& - & 2\lambda^{\prime\prime} \left[{\rm Det}(M) + {\rm Det}(M^\dagger)\right] \ ,
\end{eqnarray}
and ${\cal L}_{\rm ETC}$ contains all terms which are generated by the ETC interactions, and not by the chiral symmetry breaking sector. Notice that the determinant terms (which are renormalizable) explicitly break the U(1)$_{\rm A}$ symmetry, and give mass to $\Theta$, which would otherwise be a massless Goldstone boson. While the potential has a (spontaneously broken) SU(4) global symmetry, the largest global symmetry of the kinetic term is SU(2)$_{\rm L}\times$U(1)$_{\rm R}\times$U(1)$_{\rm V}$ (where U(1)$_{\rm R}$ is the $\tau^3$ part of SU(2)$_{\rm R}$), and becomes SU(4) in the $g,g^\prime\rightarrow 0$ limit. Under electroweak gauge transformations, $M$ transforms like
\begin{eqnarray}
M(x) \rightarrow u(x;y) \ M(x) \ u^T(x;y) \ ,
\label{transf-M}
\end{eqnarray}
where
\begin{eqnarray}
u(x;Y_{\rm V}) = \exp{\left[i\alpha^a(x)L^a+i\beta(x)\left(-{R^3}^T+\sqrt{2}\ Y_{\rm V}\ S^4\right)\right]} \ ,
\label{u}
\end{eqnarray}
and $Y_{\rm V}=y$.
We explicitly break the SU(4) symmetry in order to provide mass to the Goldstone bosons which are not eaten by the weak gauge bosons. We, however, preserve the full 
SU(2)$_{\rm L}\times$SU(2)$_{\rm R}\times$U(1)$_{\rm V}$ subgroup of SU(4), since breaking SU(2)$_{\rm R}\times$U(1)$_{\rm V}$ to U(1)$_Y$ would result in a potentially dangerous violation of the custodial isospin symmetry. Assuming parity invariance we write: 
\begin{eqnarray}
{\cal L}_{\rm ETC} = \frac{m_{\rm ETC}^2}{4}\ {\rm Tr}\left[M B M^\dagger B + M M^\dagger \right] + \cdots \ ,
\end{eqnarray}
where the ellipses represent possible higher dimensional operators, and $B\equiv 2\sqrt{2}S^4$ commutes with the SU(2)$_{\rm L}\times$SU(2)$_{\rm R}\times$U(1)$_{\rm V}$ generators.

The potential ${\cal V}(M)$ is SU(4) invariant. It produces a VEV
which parameterizes the techniquark condensate, and spontaneously
breaks SU(4) to SO(4). In terms of the model parameters the VEV is
\begin{eqnarray}
v^2=\langle \sigma \rangle^2 = \frac{m^2}{\lambda + \lambda^\prime - \lambda^{\prime\prime} } \ ,
\label{VEV}
\end{eqnarray}
while the Higgs mass is
\begin{eqnarray}
M_H^2 = 2\ m^2 \ .
\end{eqnarray}
The linear combination $\lambda + \lambda^{\prime} -
\lambda^{\prime\prime}$ corresponds to the Higgs self coupling in
the SM. The three pseudoscalar mesons $\Pi^\pm$, $\Pi^0$ correspond
to the three massless Goldstone bosons which are absorbed by the
longitudinal degrees of freedom of the $W^\pm$ and $Z$ boson. The
remaining six uneaten Goldstone bosons are technibaryons, and all
acquire tree-level degenerate mass through, not yet specified, ETC interactions:
\begin{eqnarray}
M_{\Pi_{UU}}^2 = M_{\Pi_{UD}}^2 = M_{\Pi_{DD}}^2 = m_{\rm ETC}^2  \ .
\end{eqnarray}
The remaining scalar and pseudoscalar masses are
\begin{eqnarray}
M_{\Theta}^2 & = & 4 v^2 \lambda^{\prime\prime} \nonumber \\
M_{A^\pm}^2 = M_{A^0}^2 & = & 2 v^2 \left(\lambda^{\prime}+\lambda^{\prime\prime}\right)
\end{eqnarray}
for the technimesons, and
\begin{eqnarray}
M_{\widetilde{\Pi}_{UU}}^2 = M_{\widetilde{\Pi}_{UD}}^2 = M_{\widetilde{\Pi}_{DD}}^2 =
m_{\rm ETC}^2 + 2 v^2 \left(\lambda^{\prime} + \lambda^{\prime\prime }\right) \ ,
\end{eqnarray}
for the technibaryons. 
To gain insight on some of the mass relations one can use \cite{Hong:2004td}.

\subsubsection{Vector Bosons}
The composite vector bosons of a theory with a global SU(4) symmetry are conveniently described by the four-dimensional traceless Hermitian matrix
\begin{eqnarray}
A^\mu = A^{a\mu} \ T^a \ ,
\end{eqnarray}
where $T^a$ are the SU(4) generators: $T^a=S^a$, for $a=1, \dots ,6$, and $T^{a+6}=X^a$, for $a=1, \dots ,9$. Under an arbitrary SU(4) transformation, $A^\mu$ transforms like
\begin{equation}
A^\mu \ \rightarrow \ u\ A^\mu \ u^\dagger \ ,\ \ \ {\rm where} \ u\in {\rm SU(4)} \ .
\label{vector-transform}
\end{equation}
Eq.~(\ref{vector-transform}), together with the tracelessness of the matrix $A_\mu$, gives the connection with the techniquark bilinears:
\begin{equation}
A^{\mu,j}_{i}  \sim \ Q^{\alpha}_i  \sigma^{\mu}_{\alpha \dot{\beta}}  \bar{Q}^{\dot{\beta},j}
- \frac{1}{4} \delta_{i}^j Q^{\alpha}_k  \sigma^{\mu}_{\alpha \dot{\beta}} \bar{Q}^{\dot{\beta},k} \ .
\end{equation}
Then we find the following relations between the charge eigenstates and the wavefunctions of the composite objects:
\begin{eqnarray}
\begin{array}{rclcrcl}
v^{0\mu} & \equiv & A^{3\mu} \sim \bar{U} \gamma^\mu U - \bar{D} \gamma^\mu D & , & 
a^{0\mu} & \equiv & A^{9\mu} \sim \bar{U} \gamma^\mu \gamma^5 U - \bar{D} \gamma^\mu \gamma^5 D \\
v^{+\mu} & \equiv & {\displaystyle \frac{A^{1\mu}-i A^{2\mu}}{\sqrt{2}}} \sim \bar{D} \gamma^\mu U & , &
a^{+\mu} & \equiv & {\displaystyle \frac{A^{7\mu}-i A^{8\mu}}{\sqrt{2}}} \sim  \bar{D} \gamma^\mu  \gamma^5 U \\
v^{-\mu} & \equiv & {\displaystyle \frac{A^{1\mu}+i A^{2\mu}}{\sqrt{2}}} \sim  \bar{U} \gamma^\mu D & , &  
a^{-\mu} & \equiv & {\displaystyle \frac{A^{7\mu}+i A^{8\mu}}{\sqrt{2}}} \sim  \bar{U} \gamma^\mu  \gamma^5 D \\
v^{4\mu} & \equiv & A^{4\mu} \sim \bar{U} \gamma^\mu U + \bar{D} \gamma^\mu D  & , & & &
\end{array}
\label{TMV-eigenstates}
\end{eqnarray}
for the vector mesons, and
\begin{eqnarray}
\begin{array}{rcl}
x_{UU}^\mu & \equiv & {\displaystyle \frac{A^{10\mu}+i A^{11\mu}+A^{12\mu}+ i A^{13\mu}}{2}} \sim   U^T C \gamma^\mu \gamma^5 U \ , \\
x_{DD}^\mu & \equiv & {\displaystyle \frac{A^{10\mu}+i A^{11\mu}-A^{12\mu}- i A^{13\mu}}{2}} \sim   D^T C \gamma^\mu \gamma^5 D \ , \\
x_{UD}^\mu & \equiv & {\displaystyle \frac{A^{14\mu}+i A^{15\mu}}{\sqrt{2}}} \sim  D^T C \gamma^\mu \gamma^5 U \ , \\
s_{UD}^\mu & \equiv & {\displaystyle \frac{A^{6\mu}-i A^{5\mu}}{\sqrt{2}}} \sim U^T C \gamma^\mu  D \ ,
\end{array}
\label{TBV-eigenstates}
\end{eqnarray}
for the vector baryons.

There are different approaches on how to introduce vector mesons at the effective Lagrangian level. At the tree level they are all equivalent. The main differences emerge when exploring quantum corrections.

In the appendix we will show how to introduce the vector mesons in a way that renders the following Lagrangian amenable to loop computations.  
Based on these premises, the minimal kinetic Lagrangian is:
\begin{eqnarray}
{\cal L}_{\rm kinetic} = -\frac{1}{2}{\rm Tr}\Big[\widetilde{W}_{\mu\nu}\widetilde{W}^{\mu\nu}\Big] - \frac{1}{4}B_{\mu\nu}B^{\mu\nu}
-\frac{1}{2}{\rm Tr}\Big[F_{\mu\nu}F^{\mu\nu}\Big] + m_A^2 \ {\rm Tr}\Big[C_\mu C^\mu\Big] \ ,
\label{massterm}
\end{eqnarray}
where $\widetilde{W}_{\mu\nu}$ and $B_{\mu\nu}$ are the ordinary field strength tensors for the electroweak gauge fields. Strictly speaking the terms above are not only kinetic ones since the Lagrangian contains a mass term as well as self interactions. The tilde on $W^a$ indicates  that the associated states are not yet the standard model weak triplets: in fact these states mix with the composite vectors to form mass eigenstates corresponding to the ordinary $W$ and $Z$ bosons. $F_{\mu\nu}$ is the field strength tensor for the new SU(4) vector bosons,
\begin{eqnarray}
F_{\mu\nu} & = & \partial_\mu A_\nu - \partial_\nu A_\mu - i\tilde{g}\left[A_\mu,A_\nu\right]\ ,
\label{strength}
\end{eqnarray}
and the vector field $C_\mu$ is defined by
\begin{eqnarray}
C_\mu \ \equiv \ A_\mu \ - \ \frac{g}{\tilde{g}}\ G_\mu (y) \ .
\end{eqnarray}
As shown in the appendix this is the appropriate linear combination to take which transforms homogeneously under the electroweak symmetries:
\begin{eqnarray}
C_\mu(x) \ \rightarrow \ u(x;y)\ C_\mu(x) \ u(x;y)^\dagger \ ,
\label{transf-C}
\end{eqnarray}
where $u(x;Y_{\rm V})$ is given by Eq.~(\ref{u}). (Once again, the specific assignment $Y_{\rm V}=y$, due to the fact that the composite vectors are built out of techniquark bilinears.) The mass term in Eq.~(\ref{massterm}) is gauge invariant (see the appendix), and gives a degenerate mass to all composite vector bosons, while leaving the actual gauge bosons massless. (The latter acquire mass as usual from the covariant derivative term of the scalar matrix $M$, after spontaneous symmetry breaking.)

The $C_\mu$ fields couple with $M$ via gauge invariant operators. Up
to dimension four operators the Lagrangian is (see the appendix for a more general treatment):
\begin{eqnarray}
{\cal L}_{\rm M-C} & = & \tilde{g}^2\ r_1 \ {\rm Tr}\left[C_\mu C^\mu M M^\dagger\right]
+ \tilde{g}^2\ r_2 \ {\rm Tr}\left[C_\mu M {C^\mu}^T M^\dagger \right] \nonumber \\
& + & i \ \tilde{g}\ r_3 \ {\rm Tr}\left[C_\mu \left(M (D^\mu M)^\dagger - (D^\mu M) M^\dagger \right) \right]
+ \tilde{g}^2\ s \ {\rm Tr}\left[C_\mu C^\mu \right] {\rm Tr}\left[M M^\dagger \right] \ . \nonumber \\
\end{eqnarray}
The dimensionless parameters $r_1$, $r_2$, $r_3$, $s$ parameterize
the strength of the interactions between the composite scalars and
vectors in units of $\tilde{g}$, and are therefore naturally
expected to be of order one. However, notice that for
$r_1=r_2=r_3=0$ the overall Lagrangian possesses two independent
SU(2)$_{\rm L}\times$U(1)$_{\rm R}\times$U(1)$_{\rm V}$ global
symmetries. One for the terms involving $M$ and one for the terms
involving $C_\mu$~\footnote{The gauge fields explicitly break the
original SU(4) global symmetry to SU(2)$_{\rm L}\times$U(1)$_{\rm
R}\times$ U(1)$_{\rm V}$, where U(1)$_{\rm R}$ is the $T^3$ part of
SU(2)$_{\rm R}$, in the SU(2)$_{\rm L}\times$SU(2)$_{\rm
R}\times$U(1)$_{\rm V}$ subgroup of SU(4).}. The Higgs potential
only breaks the symmetry associated with $M$, while leaving the
symmetry in the vector sector unbroken. This {\em enhanced symmetry}
guarantees that all $r$-terms are still zero after loop corrections.
Moreover if one chooses $r_1$, $r_2$, $r_3$ to be small the near enhanced symmetry will protect these values against large corrections \cite{Casalbuoni:1995qt,Appelquist:1999dq}.

We can also construct dimension four operators including only
$C_{\mu}$ fields. These new operators will not affect our analysis
but will be relevant when investigating corrections to the trilinear
and quadrilinear gauge bosons interactions. We will include these terms in appendix~\ref{app:lagr}.

\subsubsection{Fermions and Yukawa Interactions}
The fermionic content of the effective theory consists of the standard model quarks and leptons, the new lepton doublet $L=(N,E)$ introduced to cure the Witten anomaly, and a composite techniquark-technigluon doublet. 

We now consider the limit according to which the SU(4) symmetry is, at first, extended to ordinary quarks and leptons. Of course, we will need to break this symmetry to accommodate the standard model phenomenology. We start by arranging the SU(2) doublets in SU(4) multiplets as we did for the techniquarks in Eq.~(\ref{SU(4)multiplet}). We therefore introduce the four component vectors $q^i$ and $l^i$,
\begin{eqnarray} 
q^i= \begin{pmatrix}
u^i_L \\
d^i_L \\
-i\sigma^2 {u^i_R}^* \\
-i\sigma^2 {d^i_R}^*
\end{pmatrix}\ , \quad
l^i= \begin{pmatrix}
\nu^i_L \\
e^i_L \\
-i\sigma^2 {\nu^i_R}^* \\
-i\sigma^2 {e^i_R}^*
\end{pmatrix}\ ,
\end{eqnarray}
where $i$ is the generation index. Note that such an extended SU(4) symmetry automatically predicts the presence of a right handed neutrino for each generation. In addition to the standard model fields there is an SU(4) multiplet for the new leptons,
\begin{eqnarray}
L = \begin{pmatrix}
N_L \\
E_L \\
-i\sigma^2 {N_R}^* \\
-i\sigma^2 {E_R}^*
\end{pmatrix}\ ,
\end{eqnarray}
and a multiplet for the techniquark-technigluon bound state,
\begin{eqnarray} 
\widetilde{Q}= \begin{pmatrix}
\widetilde{U}_L \\
\widetilde{D}_L \\
-i\sigma^2 {\widetilde{U}_R}^* \\
-i\sigma^2 {\widetilde{D}_R}^*
\end{pmatrix}\ .
\end{eqnarray}
With this arrangement, the electroweak covariant derivative for the fermion fields can be written
\begin{eqnarray}
D_\mu \ = \  \partial_\mu \  - \  i \ g \ G_\mu (Y_{\rm V})  \ , 
\end{eqnarray}
where $Y_{\rm V}=1/3$ for the quarks, $Y_{\rm V}=-1$ for the leptons, $Y_{\rm V}=-3y$ for the new lepton doublet, and $Y_{\rm V}=y$ for the techniquark-technigluon bound state. One can check that these charge assignments give the correct electroweak quantum numbers for the standard model fermions. In addition to the covariant derivative terms, we should add a term coupling $\widetilde{Q}$ to the vector field $C_\mu$, which transforms globally under electroweak gauge transformations. Such a term naturally couples the composite fermions to the composite vector bosons which otherwise would only feel the week interactions. Based on this, we write the following gauge part of the fermion Lagrangian:
\begin{eqnarray}
{\cal L}_{\rm fermion} & = & i\ \overline{q}^i_{\dot{\alpha}}  \overline{\sigma}^{\mu,\dot{\alpha} \beta} D_\mu  q^i_\beta 
+ i\ \overline{l}^i_{\dot{\alpha}}  \overline{\sigma}^{\mu,\dot{\alpha} \beta} D_\mu  l^i_\beta 
+ i\ \overline{L}_{\dot{\alpha}}  \overline{\sigma}^{\mu,\dot{\alpha} \beta} D_\mu  L_\beta 
+ i\ \overline{\widetilde{Q}}_{\dot{\alpha}}  \overline{\sigma}^{\mu,\dot{\alpha} \beta} D_\mu  \widetilde{Q}_\beta  \nonumber \\
& + & x\ \overline{\widetilde{Q}}_{\dot{\alpha}}  \overline{\sigma}^{\mu,\dot{\alpha} \beta} C_\mu  \widetilde{Q}_\beta
\label{fermion-kinetic}
\end{eqnarray}
The terms coupling the standard model fermions or the new leptons to $C_\mu$ are in general not allowed. In fact under electroweak gauge transformations any four-component fermion doublet $\psi$ transforms like
\begin{eqnarray}
\psi \rightarrow  u(x;Y_{\rm V}) \ \psi \ ,
\label{transf-psi}
\end{eqnarray}
and from Eq.~(\ref{transf-C}) we see that a term like $\psi^\alpha  \sigma^{\mu}_{\alpha \dot{\beta}} C_\mu  \overline{\psi}^{\dot{\beta}}$ is only invariant if $Y_{\rm V}=y$. Then we can distinguish two cases. First, we can have
$y\neq 1/3$ and $y\neq -1$, in which case $\psi^\alpha  \sigma^{\mu}_{\alpha \dot{\beta}} C_\mu  \overline{\psi}^{\dot{\beta}}$ is only invariant for $\psi=\widetilde{Q}$. Interaction terms of the standard model fermions with components of $C_\mu$ are still possible, but these would break the SU(4) chiral simmetry even in the limit in which the electroweak gauge interactions are switched off. Second, we can have $y=1/3$ or $y=-1$. Then $\psi^\alpha  \sigma^{\mu}_{\alpha \dot{\beta}} C_\mu  \overline{\psi}^{\dot{\beta}}$ is not only invariant for $\psi=\widetilde{Q}$, but also for either $\psi=q^i$ or $\psi=l^i$, respectively. In the last two cases, however, the corresponding interactions are highly suppressed, since these give rise to anomalous couplings of the light fermions with the standard model gauge bosons, which are tightly constrained by experiments.

We now turn to the issue of providing masses to ordinary fermions.  In the first chapter the simplest ETC model has been briefly reviewed. Many extensions of technicolor have been suggested in the literature to address this problem. Some of the extensions use another strongly coupled gauge dynamics,  others introduce fundamental scalars. Many variants of the schemes presented above exist and a review of the major models is the one by Hill and Simmons \cite{Hill:2002ap}. At the moment there is not yet a consensus on which is the correct ETC. To keep the number of fields minimal we make the most economical ansatz, i.e. we parameterize our ignorance about a complete ETC theory by simply coupling the fermions to our low energy effective Higgs. This simple construction minimizes the flavor changing neutral currents problem. It is worth mentioning that it is possible to engineer a schematic ETC model proposed first by Randall in \cite{Randall:1992vt} and adapted for the MWT in \cite{Evans:2005pu} for which the effective theory presented in the main text can be considered a minimal description. Another non minimal way to give masses to the ordinary fermions is to (re)introduce a new Higgs doublet as already done many times in the literature \cite{Simmons:1988fu,Dine:1990jd,Kagan:1990az,Kagan:1991gh,Carone:1992rh,Carone:1993xc,Gudnason:2006mk}. 

Depending on the value of $y$ for the techniquarks, we can write different Yukawa interactions which couple the standard model fermions to the matrix $M$. Let $\psi$ denote either $q^i$ or $l^i$. If $\psi$ and the techniquark multiplets $Q^a$ have the same U(1)$_{\rm V}$ charge, then the Yukawa term
\begin{eqnarray}
- \psi^T M^* \psi + {\rm h.c.} \ ,
\label{yukawa1}
\end{eqnarray}
is gauge invariant, as one can check explicitly from Eq.~(\ref{transf-M}) and Eq.~(\ref{transf-psi}). Otherwise, if $\psi$ and $Q^a$ have different U(1)$_{\rm V}$ charges, we can only write a gauge invariant Lagrangian with the off-diagonal terms of $M$, which contain the Higgs and the Goldstone bosons:
\begin{eqnarray}
- \psi^T M_{\rm off}^*\ \psi + {\rm h.c.} \ .
\label{yukawa2}
\end{eqnarray}
In fact $M_{\rm off}$ has no U(1)$_{\rm V}$ charge, since
\begin{eqnarray}
S^4 M_{\rm off} + M_{\rm off} {S^4}^T = 0 \ ,
\end{eqnarray}
The last equation implies that the U(1)$_{\rm V}$ charges of $\psi^T$ and $\psi$ cancel in Eq.~(\ref{yukawa2}). The latter is actually the only viable Yukawa Lagrangian for the new leptons, since the corresponding U(1)$_{\rm V}$ charge is $Y_{\rm V}=-3y \neq y$, and for the ordinary quarks, since Eq.~(\ref{yukawa1}) contains $qq$ terms which are not color singlets. 

We notice however that neither Eq.~(\ref{yukawa1}) nor Eq.~(\ref{yukawa2}) are phenomenologically viable yet, since they leave the SU(2)$_{\rm R}$ subgroup of SU(4) unbroken, and the corresponding Yukawa interactions do not distinguish between the up-type and the down-type fermions. In order to prevent this feature, and recover agreement with the experimental input, we break the SU(2)$_{\rm R}$ symmetry to U(1)$_{\rm R}$ by using the projection operators $P_U$ and $P_D$, where
\begin{eqnarray}
P_U = \begin{pmatrix} 1 & 0 \\ 0 & \frac{1+\tau^3}{2} \end{pmatrix} \ , \quad
P_D = \begin{pmatrix} 1 & 0 \\ 0 & \frac{1-\tau^3}{2} \end{pmatrix} \ .
\end{eqnarray}
Then, for example, Eq.~(\ref{yukawa1}) should be replaced by
\begin{eqnarray}
- \psi^T \left(P_U M^* P_U\right) \psi - \psi^T \left(P_D M^* P_D\right) \psi + {\rm h.c.} \ .
\label{yukawa3}
\end{eqnarray}

 {}For illustration we distinguish two different cases for our analysis, $y\neq -1$ and $y= -1$, and write the corresponding Yukawa interactions:
 \newline
(i) $y\neq -1$. In this case we can only form gauge invariant terms with the standard model fermions by using the off-diagonal $M$ matrix. Allowing for both $N-E$ and $\widetilde{U}-\widetilde{D}$ mass splitting, we write
\begin{eqnarray}
{\cal L}_{\rm Yukawa} &=& -\ y_u^{ij}\ q^{i T} \left(P_U M_{\rm off}^* P_U\right) q^j
- y_d^{ij}\ q^{i T} \left(P_D M_{\rm off}^* P_D\right) q^j \nonumber \\
&& -\ y_\nu^{ij}\ l^{i T} \left(P_U M_{\rm off}^* P_U\right) l^j
- y_e^{ij}\ l^{i T} \left(P_D M_{\rm off}^* P_D\right) l^j \nonumber \\
&& -\ y_N\ L^T \left(P_U M_{\rm off}^* P_U\right) L
-\ y_E\ L^T \left(P_D M_{\rm off}^* P_D\right) L \nonumber \\
&& -\ \ y_{\widetilde{U}} \widetilde{Q}^T \left(P_U M^* P_U\right) \widetilde{Q} 
- \ y_{\widetilde{D}} \widetilde{Q}^T \left(P_D M^* P_D\right) \widetilde{Q} \ + {\rm h.c.} \ ,
\label{yukawa-1}
\end{eqnarray}
where $y_u^{ij}$, $y_d^{ij}$, $y_\nu^{ij}$, $y_e^{ij}$ are arbitrary complex matrices, and $y_N$, $y_E$, $y_{\widetilde{U}}$, $y_{\widetilde{D}}$ are complex numbers.
\newline
Note that the underlying strong dynamics already provides a dynamically generated mass term for $\widetilde{Q}$ of the type: 
\begin{equation}
{k}\,  \widetilde{Q}^T  M^* \widetilde{Q} + {\rm h.c.} \ ,
\end{equation}
with ${k}$ a dimensionless coefficient of order one and entirely fixed within the underlying theory.  The splitting between the up and down type techniquarks is due to physics beyond the technicolor interactions \footnote{Small splittings with respect to the electroweak scale will be induced by the standard model corrections per se.}. Hence the Yukawa interactions for $\widetilde{Q}$ must be interpreted as already containing the dynamical generated mass term.  

(ii) $y= -1$. In this case we can form gauge invariant terms with the standard model leptons and the full $M$ matrix:
\begin{eqnarray}
{\cal L}_{\rm Yukawa} &=& -\ y_u^{ij}\ q^{i T} \left(P_U M_{\rm off}^* P_U\right) q^j
- y_d^{ij}\ q^{i T} \left(P_D M_{\rm off}^* P_D\right) q^j \nonumber \\
&& -\ y_\nu^{ij}\ l^{i T} \left(P_U M^* P_U\right) l^j
- y_e^{ij}\ l^{i T} \left(P_D M^* P_D\right) l^j \nonumber \\
&& -\ y_N\ L^T \left(P_U M_{\rm off}^* P_U\right) L
-\ y_E\ L^T \left(P_D M_{\rm off}^* P_D\right) L \nonumber \\
&& -\ \ y_{\widetilde{U}} \widetilde{Q}^T \left(P_U M^* P_U\right) \widetilde{Q} 
- \ y_{\widetilde{D}} \widetilde{Q}^T \left(P_D M^* P_D\right) \widetilde{Q} \ + {\rm h.c.} \ .
\label{yukawa-2}
\end{eqnarray}
Here we are assuming Dirac masses for the neutrinos, but we can easily add also Majorana mass terms. At this point one can exploit the symmetries of the kinetic terms to induce a GIM mechanism, which works out exactly like in the standard model. Therefore, in both Eq.~(\ref{yukawa-1}) and Eq.~(\ref{yukawa-2}) we can assume $y_u^{ij}$, $y_d^{ij}$, $y_\nu^{ij}$, $y_e^{ij}$ to be diagonal matrices, and replace the $d^i_L$ and $\nu^i_L$ fields, in the kinetic terms, with $V_q^{ij} d^j_L$ and $V_l^{ij} \nu^j_L$, respectively, where $V_q$ and $V_l$ are the mixing matrices.  

When $y=-1$ $\widetilde{Q}$ has the same quantum numbers of the ordinary leptons, except for the technibaryon number. If the technibaryon number is violated they can mix with the ordinary leptons behaving effectively as a  fourth generation leptons (see Eq.~(\ref{yukawa-2})). However this will reintroduce, in general, anomalous couplings with intermediate gauge bosons for the ordinary fermions and hence we assume the mixing to be small.

\subsection{Constraining the MWT effective Lagrangian via WSRs  \& $S$ parameter }

In our effective theory the $S$ parameter is directly proportional to the parameter $r_3$ via:
\begin{eqnarray}
S= \frac{8\pi}{\tilde{g}^2}\, \chi \,(2- \chi)   \ , \quad {\rm with} \quad \chi = \frac{v^2\tilde{g}^2}{2M^2_A} \, r_3 \ ,
\label{S}
\end{eqnarray}
where we have expanded in $g/\tilde{g}$ and kept only the leading order. The full expression can be found in the appendix D of \cite{Foadi:2007ue} and it is also reported in the appendix here.
We can now use the  sum rules to relate $r_3$ to other parameters in the theory for the running and the walking case.  Within the effective theory we deduce:
\begin{eqnarray}
F^2_V = \left(1 - \, \chi \frac{r_2}{r_3}\right)\, \frac{2M^2_A}{\tilde{g}^2} = \frac{2M^2_V}{\tilde{g}^2}\ , \quad F^2_A = 2\frac{M^2_A}{\tilde{g}^2}(1 - \chi)^2 \ , \quad F^2_{\pi} = v^2 ( 1 -\chi \, r_3 ) \ .\end{eqnarray}
Hence the first WSR reads:
\begin{eqnarray}
1+r_2 - 2 r_3 = 0 \ ,
\end{eqnarray}
while the second:
\begin{eqnarray}
 (r_2 - r_3) (v^2\tilde{g}^2 (r_2 + r_3) - 4 M^2_A) = a \frac{16\pi^2}{d(R)}  v^2\left( 1 - \chi \, r_3\right)^2 \ .
\end{eqnarray}

To gain analytical insight we consider the limit in which $\tilde{g} $ is small while $g/\tilde{g}$ is still much smaller than one. To leading order in $\tilde{g}$ the second sum rule simplifies to:
\begin{eqnarray}
r_3 - r_2  = a \frac{4\pi^2}{d(R)}\frac{v^2}{M^2_A} \ ,
\end{eqnarray}
Together with the first sum rule we find:
\begin{eqnarray}
r_2 = 1 - 2 t \ , \qquad r_3 = 1 - t \ ,
\end{eqnarray}
with
\begin{eqnarray}
t= a \frac{4\pi^2}{d(R)}\frac{v^2}{M^2_A} \ .
\end{eqnarray}
The approximate $S$ parameter reads.
\begin{eqnarray}
S= 8\pi\, \frac{v^2}{M^2_A} (1-t)   \ . \end{eqnarray}
A positive $a$  renders $S$ smaller than expected in a running theory for a given value of the axial mass. In the next subsection we will make a similar analysis without taking the limit of small $\tilde{g}$.

 \subsubsection{Axial-Vector Spectrum via WSRs}

It is is interesting to determine the relative vector to axial
spectrum as function of one of the two masses, say the axial one,
for a fixed value of the $S$ parameter meant to be associated to a given underlying gauge theory. 

\begin{figure}[!t]
\centering
\resizebox{10cm}{!}{\includegraphics{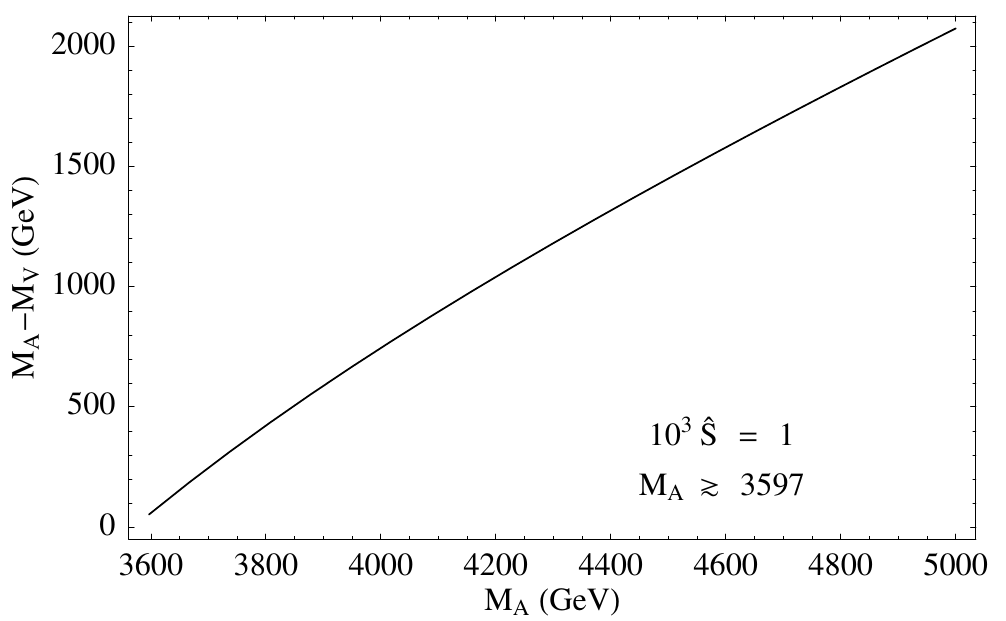}}
\caption{In the picture above we  have set $10^3 \hat{S}=1$, corresponding to $S\simeq 0.11$.  In the appendix we have provided the relation between $\hat{S}$ and the traditional S. Here we have imposed the first and the second WSR for $a=0$. This corresponds to an underlying gauge theory with a standard running behavior of the coupling constant. }
\label{runningwsr}
\end{figure}

For a running type dynamics (i.e. $a=0$) the two WSRs force the vector mesons to be quite heavy (above 3 TeV) in order to have a relatively low $S$ parameter ($S\simeq$ 0.1). This can be seen directly from Eq.~(\ref{S}) in the running regime, where $r_2=r_3=1$. This leads to
\begin{eqnarray}
M_A^2 \ \gtrsim \ \frac{8\pi v^2}{S} \ ,
\label{lowMA}
\end{eqnarray}
which corresponds to $M_A\gtrsim$ 3.6 TeV, for $S\simeq$ 0.11. Perhaps a more physical way to express this is to say that it is hard to have an intrinsically small $S$ parameter for running type theories. By small we mean smaller than the scaled up technicolor version of QCD with two techniflavors, in which $S\simeq$ 0.3. In Figure \ref{runningwsr} we plot the difference between the axial and vector mass as function of the axial mass, for $S\simeq$0.11. Since Eq.~(\ref{lowMA}) provides a lower bound for $M_A$, this plot shows that in the running regime the axial mass is always heavier than the vector mass. In fact the $M_A^2-M_V^2$ difference is proportional to $r_2$, with a positive proportionality factor (see the appendix), and $r_2=1$ in the running regime.

\begin{figure}[!t]
\centering
\begin{tabular}{cc}
\hskip -1.2cm\resizebox{8.5cm}{!}{\includegraphics{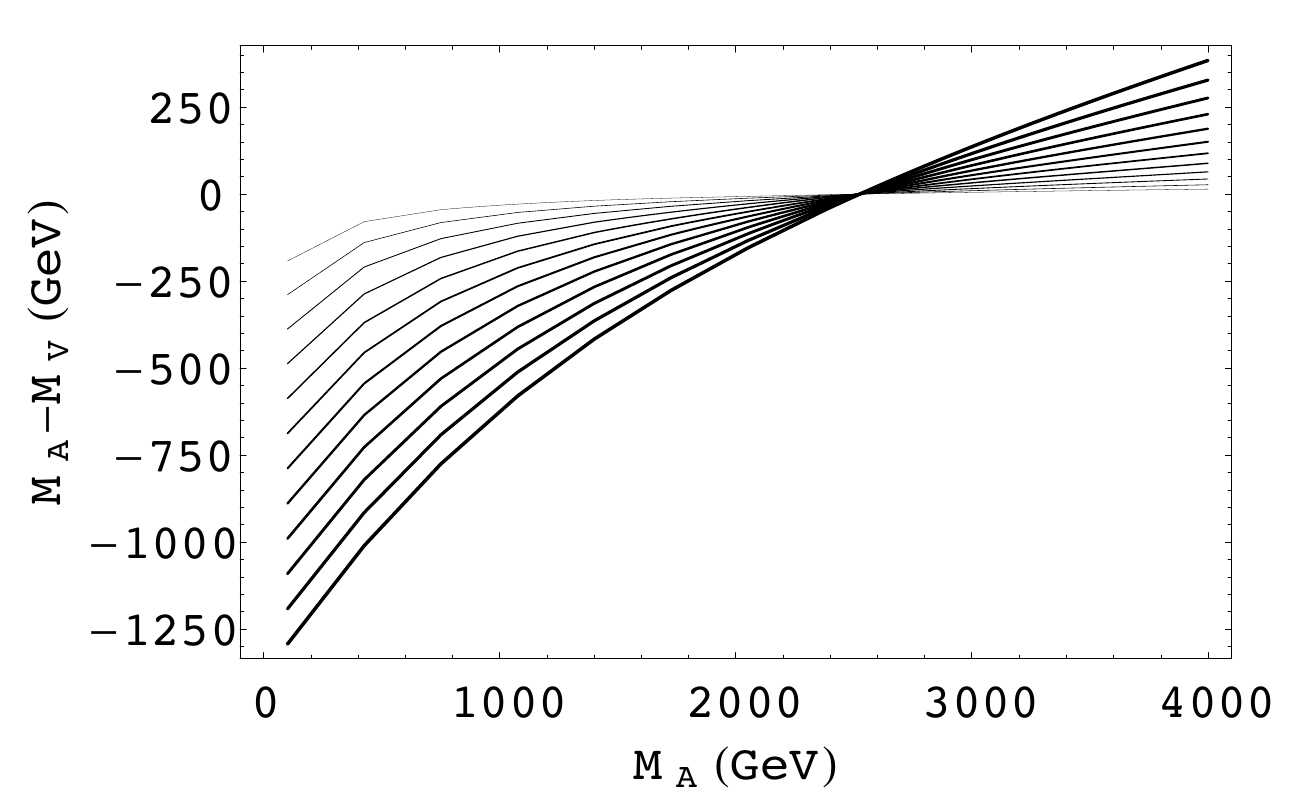}} & \resizebox{8.5cm}{!}{\includegraphics{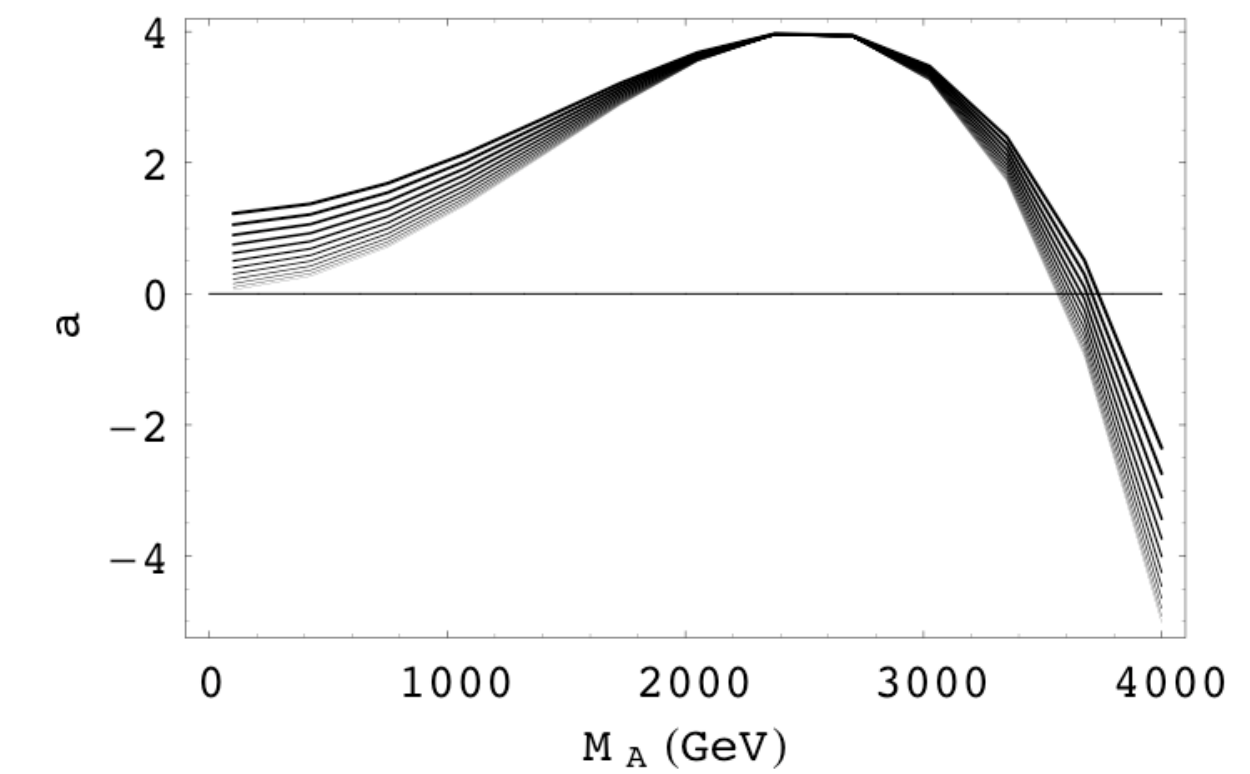}}
\end{tabular}
\caption{In the two pictures above we have set $10^3 \hat{S}=1$, corresponding to $S\simeq 0.11$, and the different curves are obtained by varying $\tilde{g}$ from one (the thinnest curve)  to eight  (the thickest curve). We have imposed the first WSR. {\it Left Panel}: We plot the allowed values of $M_A-M_V$ as function of $M_A$. {\it Right Panel}: We compute the value that $a$ should assume as function of $M_A$ in order for the second WSR to be satisfied in the walking regime. Note that $a$ is expected to be positive or zero. }
\label{walkwsr}
\end{figure}

When considering the second WSR modified by the walking dynamics, we observe that it is possible to have quite light spin one  vector mesons compatible with a small $S$ parameter. We numerically solve the first and second WSR in presence of the contribution due to walking in the second sum rule. The results are summarized in Figure \ref{walkwsr}. As for the running case we set again $S\simeq 0.11$. This value is close to the estimate in the underlying MWT \footnote{For the MWT we separate the contribution due to the new leptonic sector (which will be dealt with later in the main text) and the one due to the underlying strongly coupled gauge theory which is expected to be well represented by the perturbative contribution and is of the order of $1/2\pi$. When comparing with the $S$ parameter from the vector meson sector of the effective theory we should subtract from the underlying $S$ the one due to the new fermionic composite states $\tilde{U}$ and $\tilde{D}$. This contribution is very small since it is $1/6\pi$ in the limit when these states are degenerate and heavier than the $Z$ gauge boson.}. The different curves are obtained by varying $\tilde{g}$ from one (the thinnest curve)  to eight (the thickest curve). We plot the allowed values of $M_A-M_V$ as function of $M_A$ in the left panel, having imposed only the first sum rule. In the right panel we compute the corresponding value that $a$ should assume as function of $M_A$ in order for the second WSR to be satisfied in the walking regime as well.
Our computation shows that it is possible to have walking theories with light vector mesons and a small $S$ parameter. Such a scenario, however, needs a positive value of $a$, together with an exotic vector spectrum according to which the axial is lighter than its associated vector meson\footnote{The sum rules do not contain sufficient information to predict the relative spectral ordering of the vector and axial vector mesons. We have hence studied both orderings.}. However for spin one fields heavier than roughly 2.5 TeV and with still a positive $a$ one has an axial meson heavier than the vector one.  A degenerate spectrum allows for a small $S$ but with  relatively large values of $a$ and spin one masses around 2.5 TeV. We observe that $a$ becomes zero (and eventually negative) when the vector spectrum becomes sufficiently heavy. In other words we recover the running behavior for large masses of spin-one fields. Although in the plot we show negative values of $a$ one should stop the analysis after having reached a zero value of $a$. In fact, for masses heavier than roughly 3.5 TeV the second WSR for the running behavior, i.e. $a=0$, is enforced.  

Our results are general and help elucidating how different underlying dynamics will manifest itself at LHC. Any four dimensional strongly interacting theory replacing the Higgs mechanism, with two Dirac techniflavors gauged under the electroweak theory, is expected to have a spectrum of the low lying vector resonances like the one presented above. We note that the resonance inversion does not imply a negative or zero S parameter. 

\subsubsection{Reducing $S$ via New Leptons}
\begin{figure}[!t]
\centering
\begin{tabular}{cc}
\resizebox{6.5cm}{!}{\includegraphics{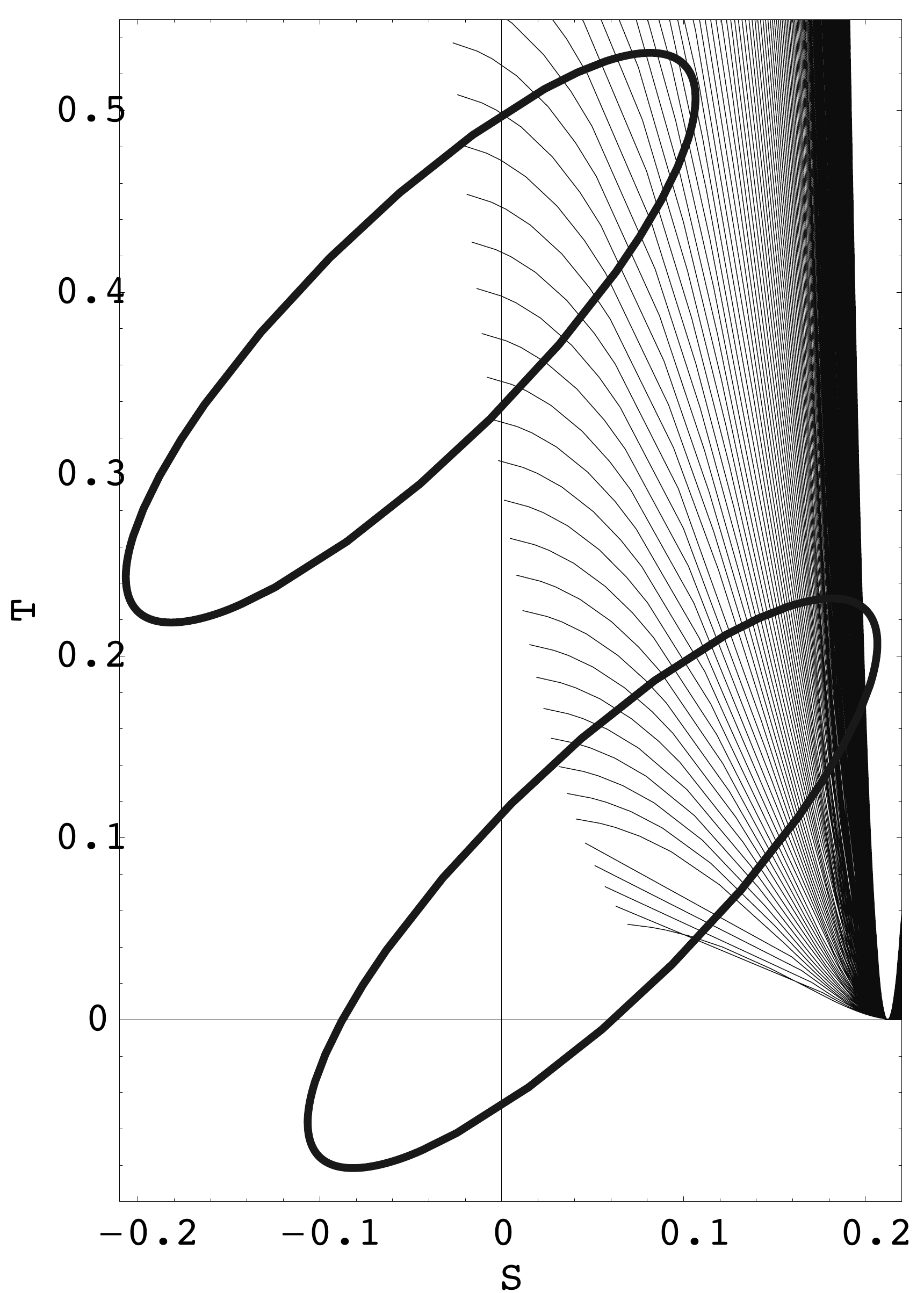}}
~~&~~~~
\resizebox{6.39cm}{!}{\includegraphics{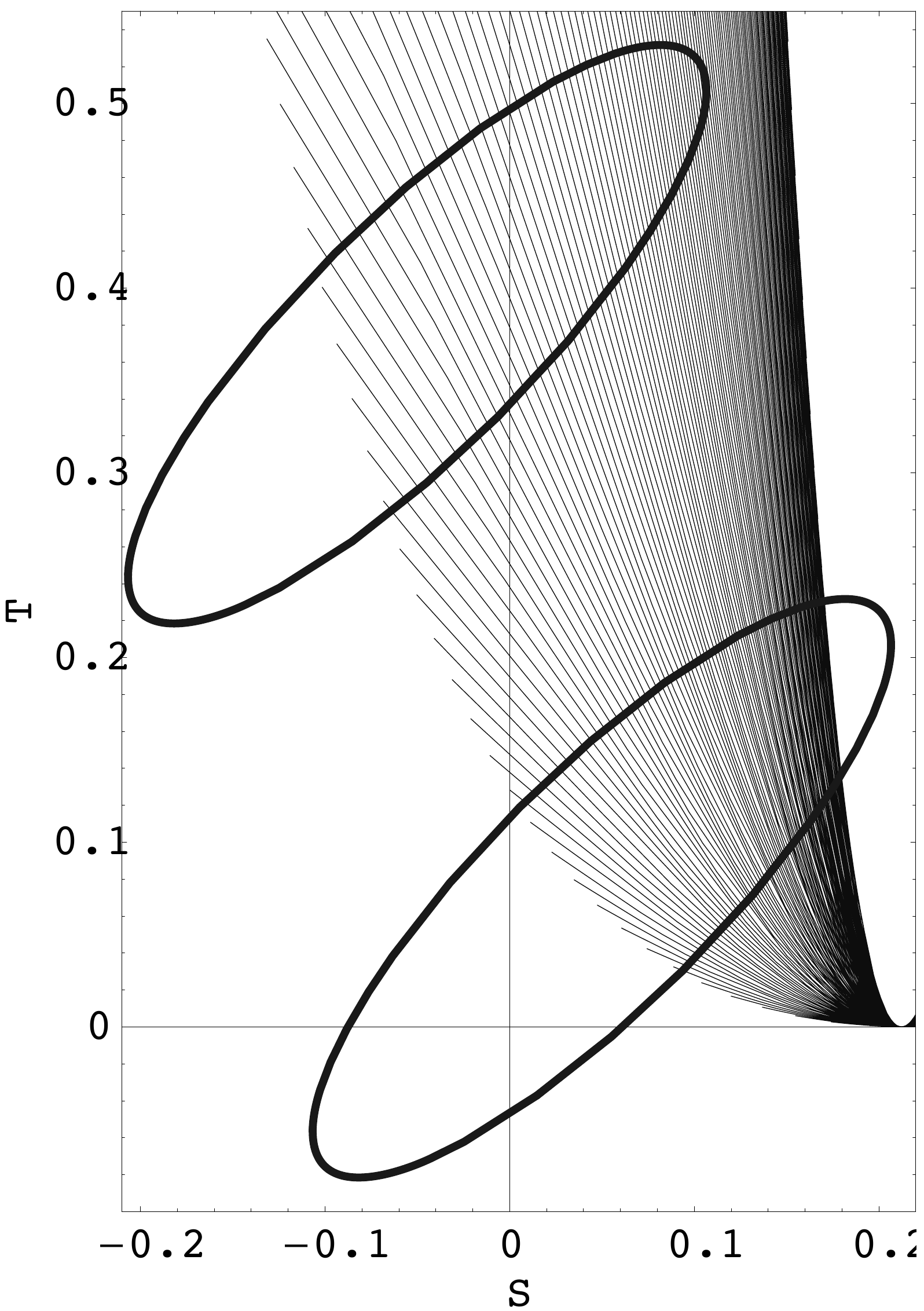}}
\end{tabular}
\caption{The ellipses represent the 68\% confidence region for the $S$ and $T$ parameters. The upper ellipse is for  a reference Higgs mass of the order of a TeV, the lower curve is for a light Higgs with mass around 114 GeV. The contribution from the MWT theory per se and of the leptons as function of the new lepton masses is expressed by the dark grey region. The left panel has been obtained using a SM type hypercharge assignment while the right hand graph is for $y=1$. }
\label{ST}
\end{figure}

We have studied the effects of the lepton family on the
electroweak parameters in~\cite{Dietrich:2005jn}, we summarize here the main results
in Figure \ref{ST}.  The ellipses represent the one standard deviation
confidence region for the $S$ and $T$ parameters. The upper ellipse
is for  a reference Higgs mass of the order of 1 TeV while the lower
curve is for a light Higgs with mass around 114 GeV. The
contribution from the MWT theory per se and of the leptons as
function of the new lepton masses is expressed by the dark grey
region. The left panel has been obtained using a SM type hypercharge assignment while the right hand graph is for $y=1$. In both pictures the regions of overlap between the theory and the precision contours are achieved when the upper component of the weak isospin doublet is lighter than the lower component. The opposite case leads to a total $S$ which is larger than the one predicted within the new strongly coupled dynamics per se.  This is due to the sign of the hypercharge for the new leptons. The mass range used in the plots, in the case of the SM hypercharge assignment is $100-1000$~GeV for the new electron and $50-800$~GeV for the new Dirac neutrino, while it is $100-800$ and $100-1000$~GeV respectively for the $y=1$ case. The plots have been obtained assuming a Dirac mass for the new neutral lepton (in the case of a SM hypercharge assignment).  {The range of the masses for which the theory is in the ellipses, for a reference Higgs mass of a TeV, is $100-400$~GeV for the new neutrino and about twice the mass of the neutrino for the new electron.}

The analysis for the Majorana mass case has been performed in \cite{Kainulainen:2006wq} where one can again show that it is possible to be within the 68\% contours. 

The contour plots we have drawn take into account the values of the top
mass which has dropped dramatically since our first comparison of our model 
theory in \cite{Dietrich:2005jn} to the experimental data \cite{EWWG}. 

We have provided a comprehensive extension of the standard model which embodies (minimal) walking technicolor theories and their interplay with the standard model particles. Our extension of the standard model features all of the relevant low energy effective degrees of freedom linked to our underlying minimal walking theory. These include scalars, pseudoscalars as well as spin one fields. The bulk of the Lagrangian has been spelled out in \cite{Foadi:2007ue}. The link with underlying strongly coupled gauge theories has been achieved via the time-honored Weinberg sum rules. The modification of the latter according to walking has been taken into account. We have also analyzed the case in which the underlying theory behaves like QCD rather than being near an infrared fixed point. This has allowed us to gain insight on the spectrum of the spin one fields which is an issue of phenomenological interest. In the appendix we have: i) provided the explicit construction of all of the SU(4) generators, ii) shown how to construct the effective Lagrangian in a way which is amenable to quantum corrections,  iii) shown the explicit form of the mass matrices for all of the particles, iv) provided a summary of all of the relevant electroweak parameters and their explicit dependence on the coefficients of our effective theory.

We have introduced the model in a format which is useful for collider phenomenology, as well as for computing corrections due to (walking) technicolor for different observable of the standard model, even in the flavor sector. 

\subsubsection{ Beyond $S$ and $T$: New Constraints for Walking Technicolor}
In \cite{Foadi:2007se} we investigated the effects of the electroweak precision measurements beyond the time honored $S$ and $T$ ones. 
Once the hypercharge of the underlying technifermions is fixed all of the derived precision parameters defined in \cite{Barbieri:2004qk} are function solely of the gauge couplings, masses of the gauge bosons and the first excited spin-one states and one more parameter $\chi$:
\begin{eqnarray}
\hat{S} &=& \frac{(2 - \chi )\chi g^2}{2\tilde{g}^2}
      \ , \\ 
W &=& \frac{g^2}{2\tilde{g}^2}\frac{M_W^2 }{M_A^2M_V^2}{(M_A^2+(\chi-1)^2M_V^2)}
\ ,  \\
Y &=&  \frac{g'^2}{2\tilde{g}^2}\frac{M_W^2}{M_A^2M_V^2} {((1+4y^2)M_A^2+(\chi -1)^2M_V^2)}
 \ , \\
X &=& \frac{g\,g'}{2\tilde{g}^2} \, \frac{M_W^2}{M^2_A M^2_V}{ (M_A^2 - (\chi -1)^2M_V^2 )} \ .
\end{eqnarray}
$\hat{T}=\hat{U}=V=0$.  $g$ and $g^{\prime}$ are the weak and hypercharge couplings, $M_W$ the gauge boson mass, $y$ the coefficient parameterizing different hypercharge choices of the underlying technifermions \cite{Foadi:2007ue}, $\tilde{g}$ the technistrong vector mesons coupling to the Goldstones in the technicolor limit, i.e. $a=0$. It was realized in \cite{Appelquist:1999dq, Duan:2000dy} and further explored in \cite{Foadi:2007ue} that for walking theories, i.e. $a\neq 0$, the WSRs allow for a new parameter $\chi$ which in the technicolor limit reduces to $\chi_0=\tilde{g}^2 v^2/2M^2_A$ 
with $F^2_{\pi}=v^2(1-\chi^2/\chi_0)$ the electroweak vacuum expectation value and $M_{A(V)}$ the mass of the axial(vector) lightest spin-one field. To make direct contact with the WSRs and for the reader's convenience we recall the relations:\begin{eqnarray}
F^2_V =  \frac{2M^2_V}{\tilde{g}^2}\ , \quad F^2_A = 2\frac{M^2_A}{\tilde{g}^2}(1 - \chi)^2 \ .\end{eqnarray}
We have kept the leading order in the electroweak couplings over the technistrong coupling $\tilde{g}$ in the expressions above while we used the full expressions \cite{Foadi:2007ue} in making the plots.

How do we study the constraints? From the expressions above we have four independent parameters, $\tilde{g}$, $\chi$, $M_V$ and $M_A$ at the effective Lagrangian level. Imposing the first WSR and assuming a fixed value of $\hat{S}$ leaves two independent parameters which we choose to be $\tilde{g}$ and $M_A$. From the modified second WSR we read off the value of $a/d(R)$.

In \cite{Foadi:2007se} we first constrained the spectrum and couplings of theories of  WT with a positive value of the $\hat{S}$ parameter compatible with the associated precision measurements at the one sigma level. More specifically we will take $\hat{S}\simeq 0.0004$ which is the highest possible value compatible with precision data for a very heavy Higgs \cite{Barbieri:2004qk}. Of course the possible presence of another sector can allow for a larger intrinsic $\hat{S}$. We are interested in the constraints coming from W and Y after having fixed $\hat{S}$.  The analysis can easily be extended to take into account sectors not included in the new strongly coupled dynamics.
\begin{figure*}[htb] 
%\begin{center}
\begin{tabular}{ccc}
%\begin{array}{cc} 
{\includegraphics[height=9cm,width=5cm]{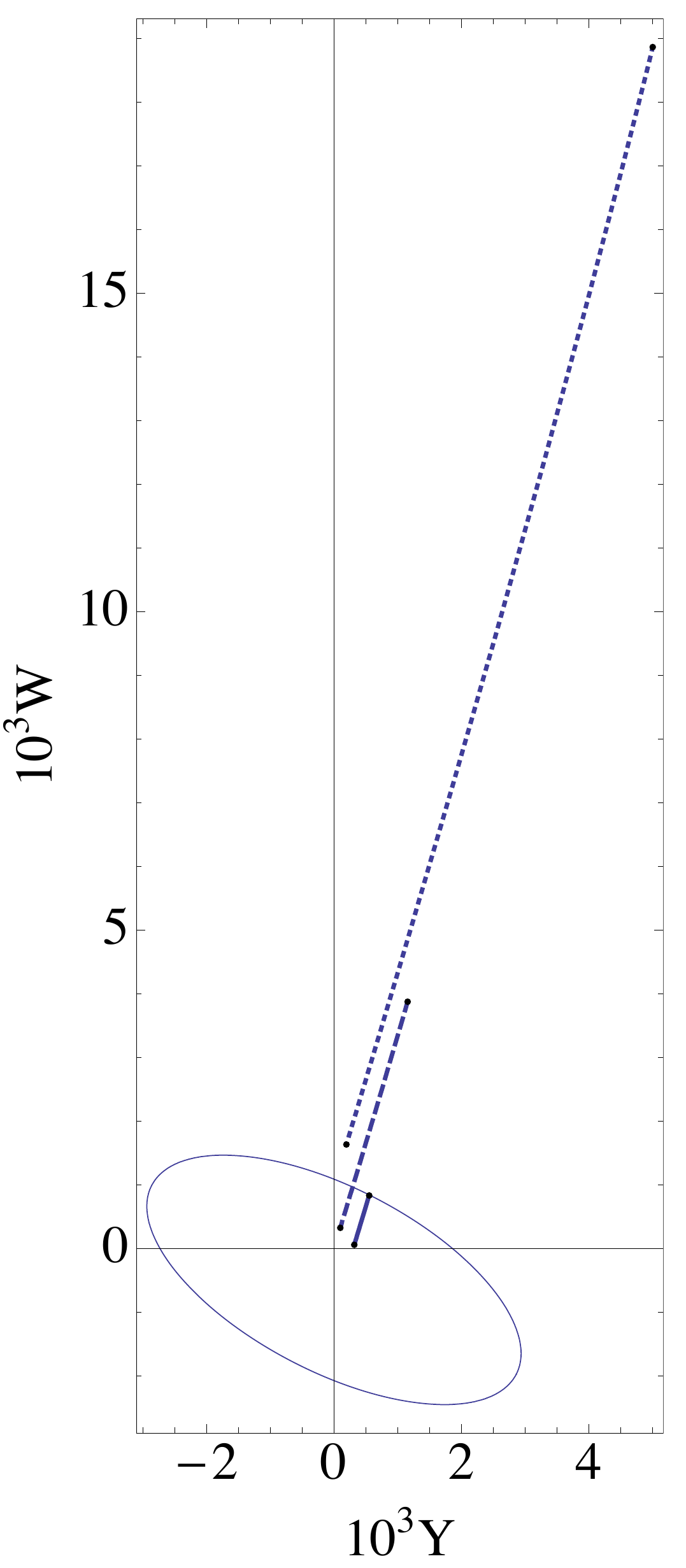}} ~~~&~~~ 
{\includegraphics[height=9cm,width=5cm]{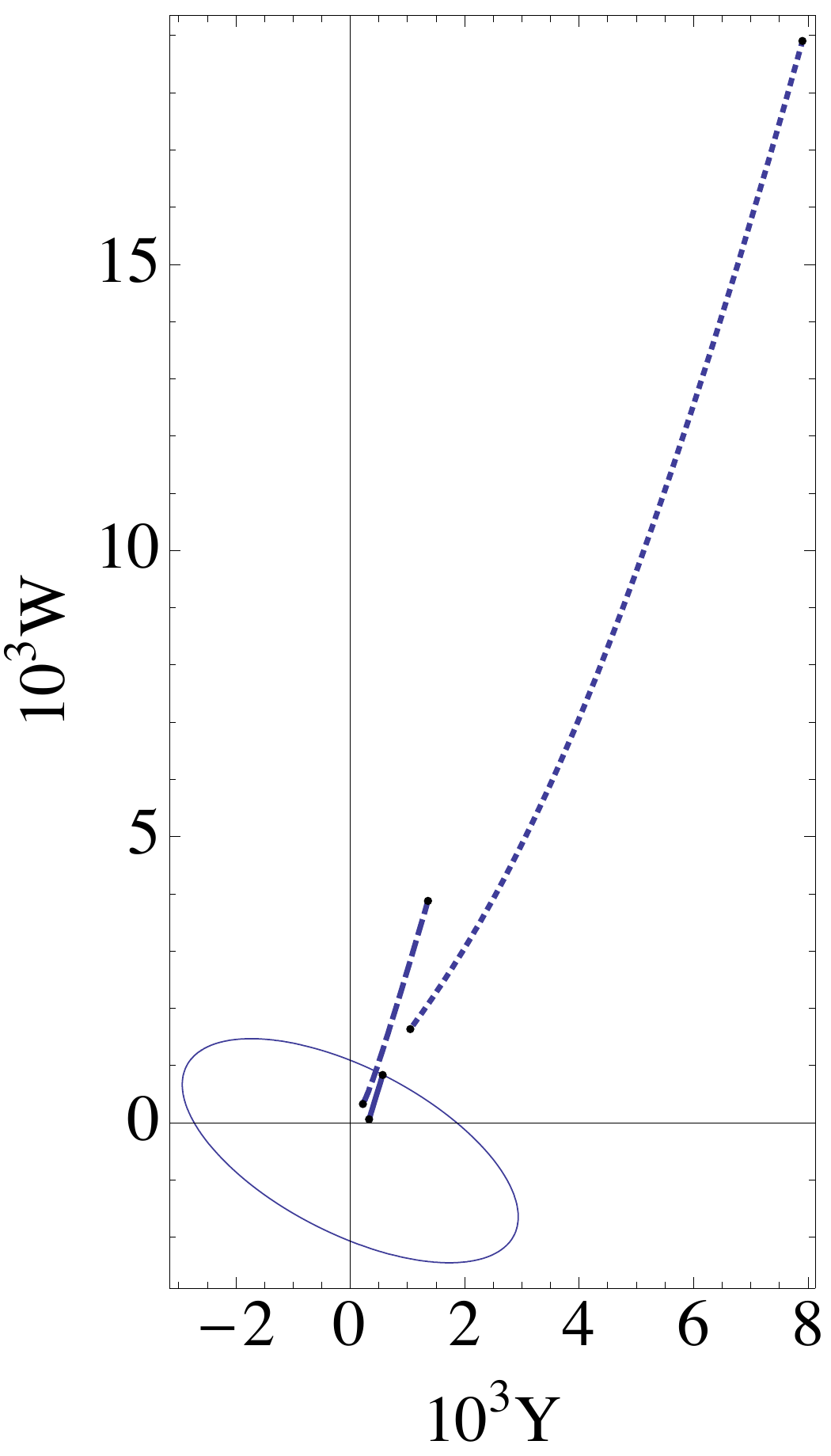}}~~~&~~~{\includegraphics[height=9cm,width=5cm]{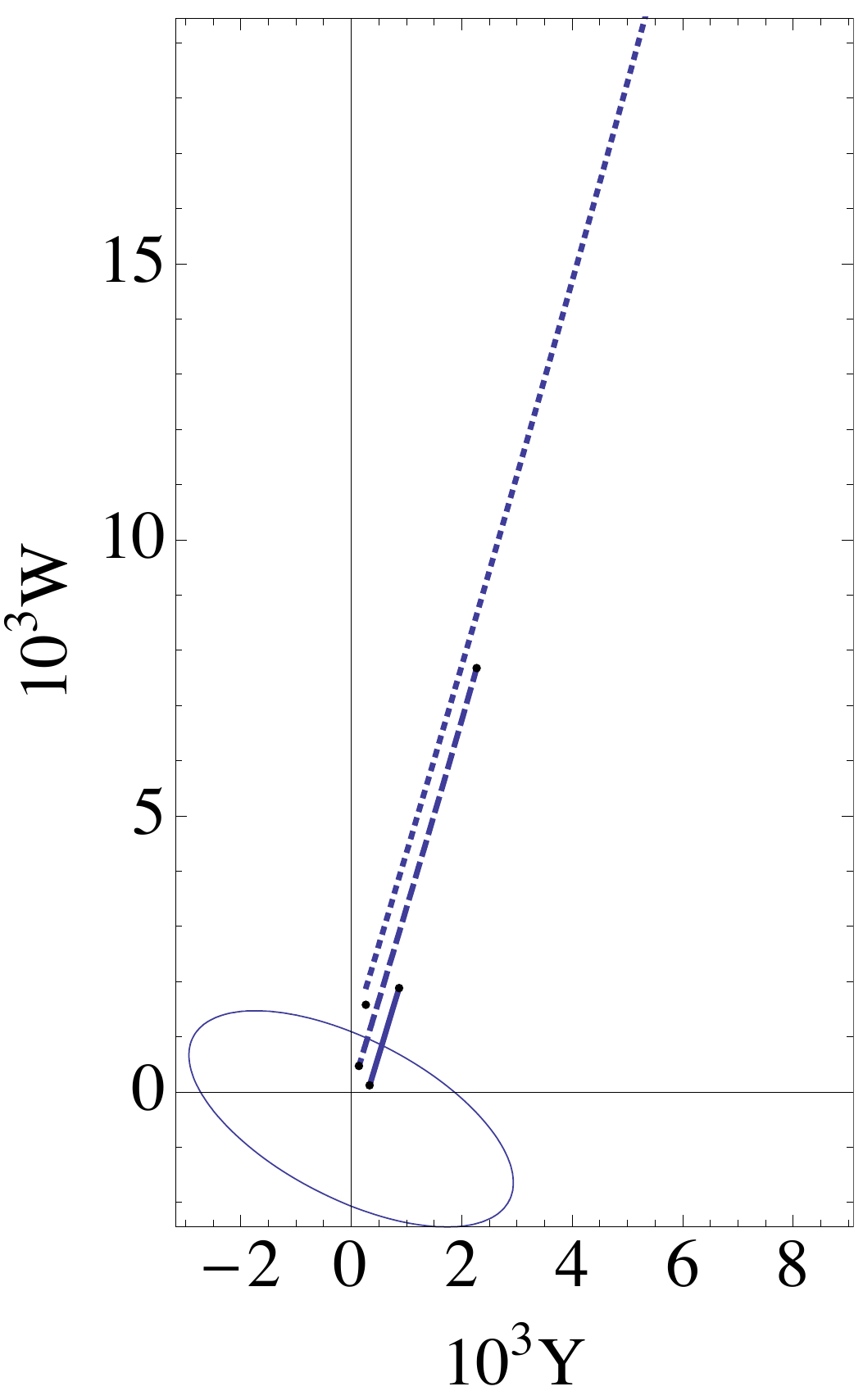}}
\\ 
~~~{WT with $y=0$}&~~~{WT with $y=1$ }&~~~CT with $y=0$
%\end{array}
\end{tabular}
%\end{center}
\caption{\label{fig:label1}  The ellipses in the WY plane corresponds to the 95\% confidence level obtained scaling the standard error ellipse axis by a 2.447 factor. The three segments, meant to be all on the top of each other,  in each plot correspond to different values of $\tilde{g}$. The  solid line corresponds to $\tilde{g}=8$, the dashed line to $\tilde{g}=4$ and the dotted one to $\tilde{g}=2$. The lines are drawn as function of $M_A$ with the point closest to the origin obtained for $M_A = 600$ GeV while the further away corresponds to $M_A=150$ GeV. We assumed $\hat{S} = 0.0004$ for WT while $\hat{S}$ is $0$ in CT by construction.  
}
\end{figure*}
A light spin-one spectrum can be achieved only if the axial is much lighter than the associated vector meson. The second is that WT models, even with small $\hat{S}$, are sensitive to the  W-Y constraints as can be seen from the plots in Fig.~\ref{fig:label1}. Since X is a higher derivative of $\hat{S}$ it is not constraining. We find that WT dynamics with a small $\tilde{g}$ coupling and a light axial vector boson is not preferred by electroweak data. Only for values of $\tilde{g}$ larger than or about $8$ the axial vector meson can be light, i.e. of the order of $200$ GeV. However WT dynamics with a small intrinsic S parameter does not allow the spin-one vector partner to be degenerate with the light axial but  predicts it to be much heavier Fig.~\ref{fig:label2}. If the spin-one masses are very heavy then the spectrum has a standard ordering pattern, i.e. the vector meson lighter than the axial meson. We also show in Fig.~\ref{fig:label2} the associated value of $a$. We were the first to make the prediction of a very light axial vector mesons in \cite{Foadi:2007ue} on the base of the modified WSRs, even lighter than the associated vector mesons.  Eichten and Lane put forward a similar suggestion in \cite{Eichten:2007sx}.  We find that a WT dynamics alone compatible with precision electroweak data can accommodate a light spin-one axial resonance only if the associated vector partner is much heavier and in the regime of a strong $\tilde{g}$ coupling.   $a$.  We find tension with the data at a level superior to the 95\% confidence level for: a) WT models featuring $M_A \simeq M_V$ spectrum with a common and very light mass; b) WT models with an axial vector meson lighter than $300$ GeV and $\tilde{g}$ smaller than $4$, an axial vector meson with a mass lighter than or around $600$ GeV and $\tilde{g}$ smaller than $2$.  

\subsubsection
{Introducing and constraining Custodial Technicolor}
We now constrain also models proposed in \cite{Appelquist:1999dq, Duan:2000dy} which, at the effective Lagrangian level, possess an explicit {\it custodial} symmetry for the S parameter. We will refer to this class of models as custodial technicolor (CT) \cite{Foadi:2007se} . The new custodial symmetry is  present in the BESS 
models \cite{Casalbuoni:1988xm,Casalbuoni:1995yb,Casalbuoni:1995qt} which will therefore be constrained as well. In this case we expect our constraints to be similar to the ones also discussed in \cite{Casalbuoni:2007dk}.

Custodial technicolor corresponds to the case for which $M_A=M_V=M$ and $\chi=0$. The effective Lagrangian acquires a new symmetry, relating a vector and an axial field, which can be interpreted as a custodial symmetry for the S parameter \cite{Appelquist:1999dq, Duan:2000dy}.  The only non-zero parameters are now:
\begin{eqnarray}
W &=& \frac{g^2}{\tilde{g}^2}\frac{M_W^2 }{M^2}
%+(2+(\chi-2)\chi)g^2
\ ,  \\
Y &=&  \frac{g'^2}{2\tilde{g}^2}\frac{M_W^2}{M^2} {(2+4y^2)} \ .
\end{eqnarray}
A CT model cannot be achieved in walking dynamics and must be interpreted as a new framework. In other words CT does not respect the WSRs and hence it can only be considered as a phenomenological type model in search of a fundamental strongly coupled theory. To make our point clearer note that a degenerate spectrum of light spin-one resonances (i.e. $M<4\pi F_{\pi}$) leads to  a very large $\hat{S}=g^2 F^2_{\pi}/4M^2$. We needed only the first sum rule together with the statement of degeneracy of the spectrum to derive this $\hat{S}$ parameter. This statement is universal and it is true for WT and ordinary technicolor. 
The Eichten and Lane \cite{Eichten:2007sx} scenario of almost degenerate and very light spin-one states can only be achieved within a near CT models. A very light vector meson with a small number of techniflavors fully gauged under the electroweak can  be accommodated in CT. This scenario was considered in \cite{ Zerwekh:2005wh,Zerwekh:2007pw} and our constraints apply here.

We find that in CT it is possible to have a very light and degenerate spin-one spectrum  if $\tilde{g}$ is sufficiently large, of the order say of 8 or larger as in the WT case.

We constrained the electroweak parameters intrinsic to WT or CT, however, in general other sectors may contribute to the electroweak observables, an explicit example is the new heavy lepton family introduced above \cite{Dietrich:2005jn}. 

To summarize we have suggested in \cite{Foadi:2007se} a way to constrain WT theories with any given S parameter. We have further constrained relevant models featuring a custodial symmetry protecting the S parameter.  When increasing the value of the S parameter while reducing the amount of walking we recover the technicolor constraints \cite{Peskin:1990zt}. We found bounds on the lightest spectrum of WT and CT theories with an intrinsically small S parameter. Our results are applicable to {\it any} dynamical model of electroweak symmetry breaking featuring near conformal dynamics \'a la walking technicolor.

\begin{figure*}[htb] 
\begin{center}
\begin{tabular}{cc}
%\begin{array}{cc} 
{\includegraphics[height=5cm,width=7cm]{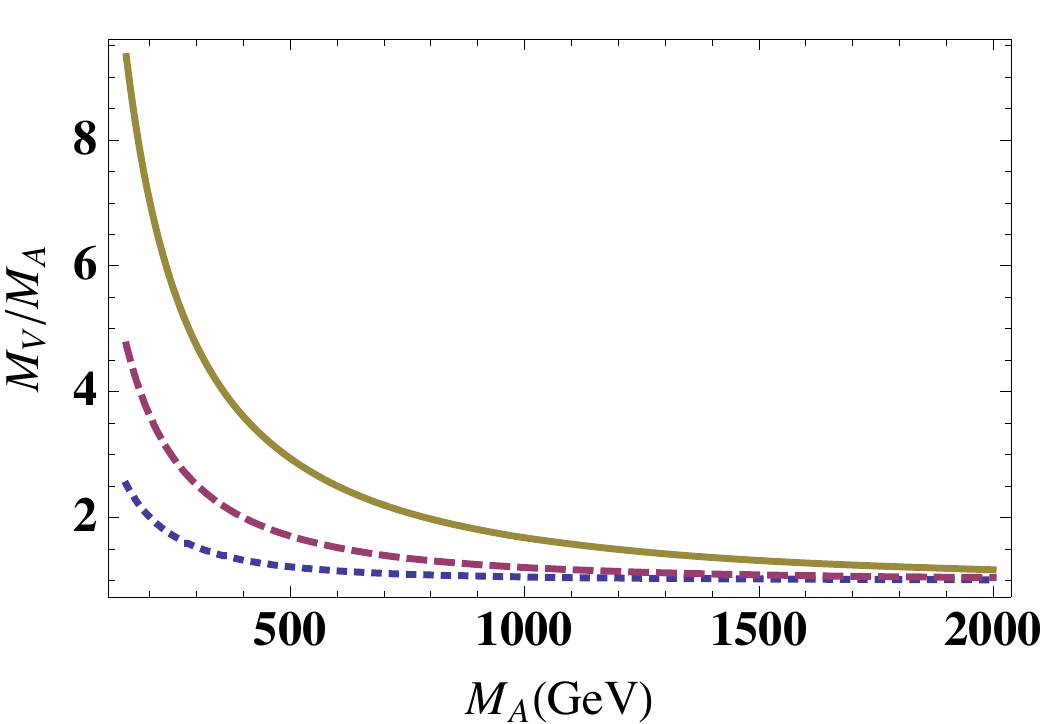}}~~~~&~~~~ 
{\includegraphics[height=4.8cm,width=7.2cm]{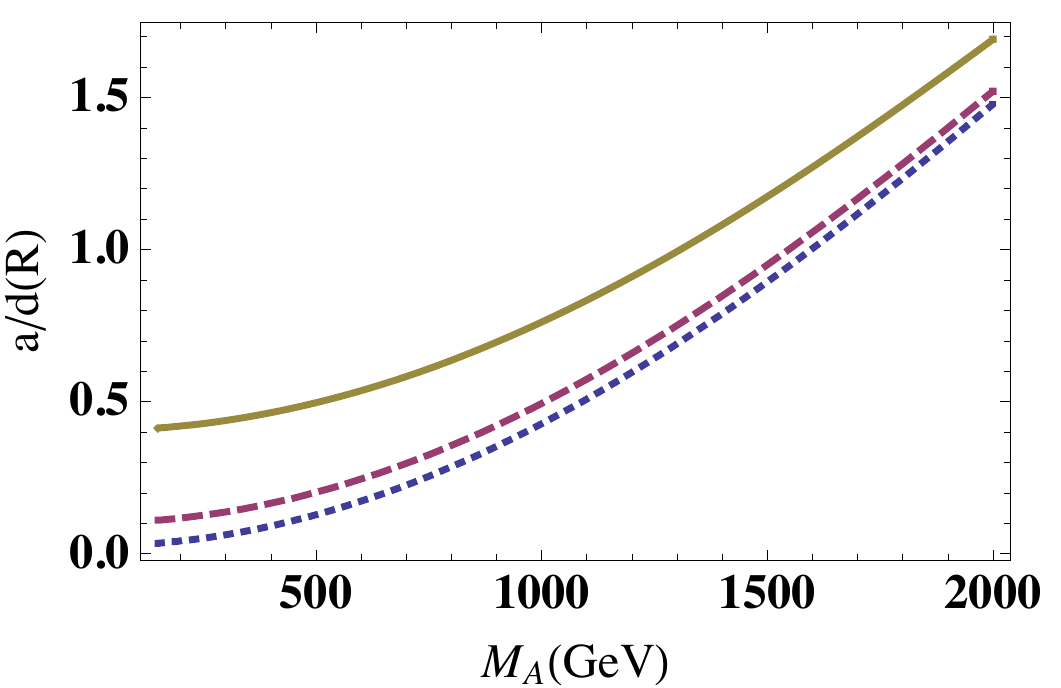}}
%\end{array}
\end{tabular}
\end{center}
\caption{\label{fig:label2}  
In the left panel we plot the ratio of the vector over axial mass as function of the axial mass for a WT theory with an intrinsic small S parameter. The vector and axial spectrum is close only when their masses are of the order of the TeV scale and around 2 TeV and onwards the vector is lighter than the axial. The right panel shows the value $a/d(R)$ as function of the axial mass. In both plots the solid, dashed and dotted lines corresponds respectively to  $\tilde{g}=8,4,2$.
}
\end{figure*}%

\subsubsection{An ETC example for MWT and The Top Mass}

It is instructive to present a simple model \cite{Evans:2005pu} which shows how one can embed MWT in an extended technicolor model
capable of generating the top quark mass.  

When the techni-quarks are not in the fundamental representation
of the technicolor group it can be hard to feed
down the electroweak symmetry breaking condensate to generate the
standard model fermion masses
\cite{Lane:1989ej,Christensen:2005cb}.  Here, following \cite{Evans:2005pu}, we
wish to highlight that the minimal model can be recast as an SO(3)
theory with fundamental representation techni-quarks. The model
can therefore rather easily be enlarged to an extended technicolor
theory \cite{Lane:1989ej} in the spirit of many examples in the
literature. We will concentrate on the top quark sector - the ETC
gauge bosons in this sector violate weak isospin and one must be
careful to compute their contribution to the T
parameter \cite{Chivukula:1995dc}.

We start by recognizing that adjoint multiplets of SU(2) can be written as
fundamental representations of SO(3). This trick will now allow us
to enact a standard ETC pattern from the literature - it is
particularly interesting that for this model of higher dimensional
representation techniquarks there is a simple ETC model. We will
follow the path proposed in \cite{Randall:1992vt} where we gauge
the full flavor symmetry of the fermions. 

If we were simply interested in the fourth family then the
enlarged ETC symmetry is a Pati-Salam type
unification. We stack the doublets

\beq \left[\left( \begin{array}{c} U^a \\ D^a \end{array}
\right)_L \ , \quad \left( \begin{array}{c} N \\ E \end{array}
\right)_L \right],\hspace{1.5cm} \left[U_R^a\ ,~N_R \right],
\hspace{1.5cm} \left[D_R^a \ ,~E_R \right]\eeq into 4 dimensional
multiplets of SU(4). One then invokes some symmetry breaking
mechanism at an ETC scale (we will not speculate on the mechanism
here though see figure 1) 

\beq SU(4)_{ETC} \rightarrow SO(3)_{TC} \times U(1)_Y \eeq

The technicolor dynamics then proceeds to generate a techniquark
condensate $\langle \bar{U} U \rangle = \langle \bar{D} D \rangle
\neq 0$. The massive gauge bosons associated with the broken ETC
generators can then feed the symmetry breaking condensate down to
generate fourth family lepton masses

\beq m_N = m_E \simeq {\langle \bar{U}U \rangle \over
\Lambda_{ETC}^2} \eeq 

One could now naturally proceed to include the third (second,
first) family by raising the ETC symmetry group to SU(8) (SU(12),
SU(16)) and a series of appropriate symmetry breakings. This would
generate masses for all the standard model fermions but no isospin
breaking mass contributions within fermion doublets. The simplest
route to generate such splitting is to make the ETC group chiral
so that different ETC couplings determine the isospin +1/2 and
-1/2 masses. Let us only enforce such a pattern for the top quark
and fourth family here since the higher ETC scales are far beyond
experimental probing.

We can, for example, have the SU(7) multiplets
\beq \left[\left( \begin{array}{c} U^a \\ D^a \end{array}
\right)_L\ , \quad \left( \begin{array}{c} N \\ E \end{array}
\right)_L\ , \quad \left( \begin{array}{c} t^c \\ b^c \end{array}
\right)_L \right], \hspace{1.5cm} \left[U_R^a\ ,~ N_R\ ,~t^c_R
\right] \eeq here $a$ will become the technicolor index and $c$
the QCD index. We also have a right handed SU(4) ETC group that
only acts on \beq \left[D_R^a\ , \quad E_R \right]\eeq

The right handed bottom quark is left out of the ETC dynamics and
only has proto-QCD SU(3) dynamics. The bottom quark will thus be
left massless. The symmetry breaking scheme at, for example, a
single ETC scale would then be

\beq SU(7) \times SU(4) \times SU(3) \rightarrow SO(3)_{TC} \times SU(3)_{QCD}  \eeq

The top quark now also acquires a mass from the broken gauge
generators naively equal to the fourth family lepton multiplet.
%Traditional estimates of the electroweak vev and ETC generated
%masses based on one loop diagrams are given by
%\beq v^2_{weak} = {d(R_{\rm TC}) \over 4 \pi^2} \Sigma(0)^2, \hspace{1cm} m = {d(R_{\rm TC})
%\over 4 \pi^2} {\Sigma(0)^2 \over \Lambda_{ETC}^2} \Sigma(0) \eeq
%here $\Sigma(0)$ is the techniquark self energy at zero momentum
%which must lie around the TeV scale. Such estimates suggest that a top
%quark mass beyond a few 10s of GeV would be hard to achieve even
%with an ETC scale of 1 TeV.
Walking dynamics has many features though that one would expect to
overcome the traditional small size of the top mass in ETC models.
Firstly the enhancement of the techniquark self energy at high
momentum enhances the ETC generated masses by a factor potentially
as large as  $\Lambda_{ETC}/ \Sigma(0)$. In
\cite{Appelquist:2003hn} it is argued that this effect alone may
be sufficient to push the ETC scale to 4 TeV and still maintain
the physical top mass.

The technicolor coupling is near conformal and strong
so the ETC dynamics will itself be quite strong at its breaking
scale which will tend to enhance light fermion masses
\cite{Evans:1994fb}. In this ETC model the top quark will also feel
the effects of the extra massive octet of axial gluon-like gauge
fields that may induce a degree of top condensation a l\`{a} top
color models \cite{Miransky:1989nu}. We conclude that a 4-8 TeV ETC scale for generating the top mass is possible. In this model the fourth family lepton would then have a mass of the same order and well in excess of the current search limit $M_Z/2$.

\subsubsection{The Next to Minimal Walking Technicolor Theory (NMWT)}
\label{4}

The theory with three technicolors contains an even number of electroweak doublets, and hence
it is not subject to a Witten anomaly.  
The doublet of technifermions, is then represented again as:
\bear
Q_L^{\{C_1,C_2 \}}
=
\left(\begin{array}{l}U^{\{C_1,C_2 \}}\\ D^{\{C_1,C_2 \}}\end{array}\right)_L \ ,
\qquad 
Q_R^{\{C_1,C_2\}}&=&\left(U_R^{\{C_1,C_2\}},~ D_R^{\{C_1,C_2\}}\right) \ .
\nn
\eear
Here $C_i=1,2,3$ is the technicolor index and $Q_{L(R)}$ is a doublet (singlet) with respect 
to the weak interactions. 
Since the two-index symmetric representation of $SU(3)$ is complex the flavor symmetry is $SU(2)_L\times SU(2)_R\times U(1)$. 
Only three Goldstones emerge and
are absorbed in the longitudinal components of the weak vector bosons.

Gauge anomalies are absent with the choice $Y=0$ for the hypercharge of the left-handed technifermions:
\bear
Q_L^{(Q)}
=
\left(\begin{array}{l} U^{(+1/2)}\\ D^{(-1/2)}\end{array}\right)_L \ .
\eear
Consistency requires for the right-handed technifermions (isospin singlets):
\bear
Q_R^{(Q)}&=&\left(U_R^{(+1/2)},~D_R^{-1/2}\right)
\nn
Y&=&~~+1/2,~-1/2 \ .
\eear 
All of these states will be bound into hadrons. There is no need for an associated fourth family of leptons, and hence it is not expected to be observed in the experiments.

Here the low-lying technibaryons are fermions constructed with three techniquarks in the following way:
\begin{eqnarray}
B_{f_1,f_2,f_3;\alpha} = Q^{\{C_1,C_2 \}}_{L;\alpha,
f_1}Q^{\{ C_3,C_4 \}}_{L;\beta,f_2} Q^{\{ C_5,C_6 \}}_{L;\gamma,f_3}\epsilon^{\beta \gamma}
\epsilon_{C_1 C_3 C_5}\epsilon_{C_2 C_4 C_6} \ .
\end{eqnarray}
where $f_i=1,2$ corresponds to $U$ and $D$ flavors, and we are not specifying the flavor symmetrization which in any 
event will have to be such that the full technibaryon wave function is fully antisymmetrized in technicolor, flavor and spin. 
$\alpha$, $\beta$, and $\gamma$ assume the values of one or 
two and represent the ordinary spin. Similarly we can construct different technibaryons using only right fields or a mixture of left and right.
%\newpage

\subsection{Beyond Minimal Walking Technicolor (BMWT)}

When going beyond MWT one finds new and interesting theories able to break the electroweak symmetry while featuring a walking dynamics and yet not at odds with precision measurements, at least when comparing with the naive $S$ parameter.  A compendium of these theories can be found in \cite{Dietrich:2006cm}. Here we will review only the principla type of models one can construct.

%%%%%%%%%%%%%%%%%%%%%%%%%%%%%%%%%%%%%%%%%%%%%%%%%%%%%%%%%%%%%%%%%%%%%

\subsubsection{Partially gauged technicolor\label{pgt}}

A small modification of the traditional technicolor approach, which neither
involves additional particle species nor more complicated gauge groups, 
allows constructing several other viable candidates. It consists in letting 
only one doublet of techniquarks transform non-trivially under the electroweak
symmetries with the rest being electroweak singlets, as first suggested in 
\cite{Dietrich:2005jn} and later also used in \cite{Christensen:2005cb}.
Still, all techniquarks transform under the technicolor gauge group. Thereby only one techniquark doublet contributes directly\footnote{Via Technicolor interactions all of the matter content of the theory will affect physical observables associated to the sector coupled to the electroweak symmetry.} to the oblique 
parameter which is thus kept to a minimum for theories which need more
than one family of techniquarks to be quasi-conformal. It is the condensation 
of that first electroweakly charged family that breaks the electroweak 
symmetry. The techniquarks which are uncharged under the electroweak gauge group are
natural building blocks for components of dark matter.

Among the partially gauged cases one of the interesting candidates \cite{Dietrich:2006cm} is the theory with eight techniflavors
in the two-index antisymmetric representation of SU(4). The techniquarks of 
one of the four families carry electroweak charges while the others are 
electroweak singlets. Gauge anomalies are avoided if the two electrically 
charged techniquarks possess half-integer charges.
The technihadron spectrum contains technibaryons made of only two
techniquarks because in the two-index
antisymmetric representation of SU(4) a singlet can already be formed in
that case.
Otherwise technisinglets can also be formed from four techniquarks. All 
technihadrons formed from techniquarks without electrical charges
can contribute to dark matter.
Due to the special charge assignment of the electrically charged particles
(opposite half-integer charges) certain combinations of those can also be 
contained in electrically uncharged technibaryons. For instance we can 
construct the following completely neutral technibaryon:
\bear
\epsilon_{t_1t_2t_3t_4}Q_L^{{t_1t_2},f}Q_L^{{t_3t_4},f^{\prime}}
{\epsilon_{ff^{\prime}}} \ ,
\eear
where $\epsilon_{ff^{\prime}}$ saturates the SU(2)$_L$ indices of the two
gauged techniquarks and the first antisymmetric tensor $\epsilon$ is summed 
over the technicolor indices. We have suppressed the spin indices. This particle is an interesting candidate for 
dark matter and it is hardly detectable in any earth based experiment
\cite{Gudnason:2006yj}.

Since the two index antisymmetric representation of $SU(4)$ is real the model's flavor symmetry is enhanced
to SU(2$N_f$=16)\footnote{It is slightly explicitly broken by the
electroweak interactions.  Here, additionally, there is a difference, on
the electroweak level, between gauged and ungauged techniquark families.}, 
which, when it breaks to SO(16), induces 135 Goldstone
bosons\footnote{Obviously the Goldstone bosons must receive a sufficiently large
mass. This is usually achieved in extended technicolor. Still, they could be
copious at LHC.}. 

It is worth recalling that the centre group symmetry left invariant
by the fermionic matter is a Z$_2$
symmetry. Hence there is a well defined order parameter for confinement
\cite{Sannino:2005sk} which can play a
role in the early Universe.

\subsubsection{Split technicolor}

We summarize here also another
possibility \cite{Dietrich:2005jn} according to which we keep the technifermions gauged under the electroweak theory in the fundamental 
representation of the $SU(N)$ technicolor group while still reducing the number of techniflavors needed to be
near the conformal window. Like for the partially gauged case described above 
this can be achieved by adding matter uncharged under the weak interactions. 
The difference to section \ref{pgt} is that this part of matter transforms 
under a different representation of the technicolor gauge group than the 
part coupled directly to the electroweak sector. {}For example, for definiteness let's choose it to be a 
massless Weyl fermion in the adjoint representation of the technicolor gauge 
group. The resulting theory has the same matter content as $N_f$-flavor super QCD but without the scalars;
hence the name "split technicolor." We expect the critical number of flavors above which one enters the
conformal window $N_f^\mathrm{II}$ to lie within the range
\bear
\frac{3}{2}<\frac{N^\mathrm{II}_f}{N}<\frac{11}{2} \ .
\eear
The lower bound is the exact supersymmetric value for a non-perturbative
conformal fixed point \cite{Intriligator:1995au}, while the upper bound is
the one expected in the theory without a technigluino. The matter content of
"split technicolor" lies between that of super QCD and QCD-like theories with 
matter in the fundamental representation.

For two colors the number of (techni)flavors needed to be near the
conformal window in the split case is at least three, while for three
colors more than five flavors are required. These values are still larger
than the ones for theories with fermions in the two-index
symmetric representation. It is useful to remind the reader that in
supersymmetric theories the critical number of flavors needed to enter the
conformal window does not coincide with the critical number of flavors
required to restore chiral symmetry. The scalars in supersymmetric theories
play an important role from this point of view. We note that a split
technicolor-like theory has been used in \cite{Hsu:2004mf}, to
investigate the strong CP problem.

Split technicolor shares some features with theories of split
supersymmetry advocated and studied in
\cite{ArkaniHamed:2004fb,Giudice:2004tc} as possible extensions of the
standard model. Clearly, we have introduced split technicolor---differently
from split supersymmetry---to address the hierarchy problem. This is why we
do not expect new scalars to appear at energy scales higher than the one of
the electroweak theory.

By using the beta function for different representations it is now possible to constrain the phase diagram of gauge theories with different fermionic matter representation and construct explicit split technicolor models \cite{RS-Progress}.

\newpage
\section{Exploring Unification within Technicolor}
Unification of the standard model couplings is an attractive feature of grand unified models. Unification, however, it is not a fundamental prerequisite of any specific extension of the SM nevertheless it is instructive to investigate what happens to the SM couplings when the Higgs sector is replaced with a new strongly coupled theory a la' technicolor \cite{Gudnason:2006mk}.  

 We start by investigating the  one-loop evolution of the SM couplings
 once the SM Higgs is replaced by the MWT model. Quite surprisingly we
 find that the SM coupling constants unify much better than with the
 standard model Higgs being
 present. We compare our results with
 different time-honored technicolor models and show that either they
 are not competitive unification wise or they are not
 a prime candidate
 for walking technicolor theories according to our results
 \cite{Dietrich:2006cm}. With a small modification of the technicolor
 gauge interactions we show how we can envision unification also
 with the technicolor coupling constant at the same energy scale.

Technicolor requires some other mechanism to provide the standard model fermion masses. This mechanism will, in general, have an effect on our results. We have estimated these corrections by providing a simple/minimal model which consists in adding a new Higgs field on the top of the minimal walking theory whose main purpose is to provide mass to standard model fermions. This construction has already been used in the literature
\cite{Simmons:1988fu,Dine:1990jd,Kagan:1990az,Kagan:1991gh,Carone:1992rh,Carone:1993xc}.
 In a more natural theory this field will be replaced, perhaps, by some new strong dynamics. The model parametrizes our ignorance of a more fundamental extended technicolor theory.  

A general feature of a unified theory of the SM interactions is the prediction of the proton decay. A unification energy scale of the order of, or larger than, $10^{15}$ GeV leads, typically, to phenomenologically acceptable proton decay rates.
Despite the good, but yet not
  perfect, degree of
  unification - when compared, for example, to the minimal
    supersymmetric standard model result for unification - we discover
  that the proton decays too fast since the unification scale is
quite low.
To cure the proton decay problem we then add a QCD colored Weyl fermion
  transforming according to the adjoint representation of $SU(3)$ and
  one Weyl fermion transforming according to the adjoint of
  $SU_L(2)$. These fermions are known in supersymmetric extensions of
  the standard model as the gluino and the wino.  They will, however have, more general couplings to the SM fermions.

  We then compare with
  the supersymmetric predictions for unification at the same order in
  perturbation theory and discover that the present theory unifies
  better. 

The reader may consider adding matter transforming according to even
higher dimensional representations than the adjoint one or more
generally higher than the two-index type matter. We know
\cite{Dietrich:2006cm}, however, that a very limited number of
theories with fermions in higher dimensional representations remain
asymptotically free when the rank of the gauge group increases.  The
unified gauge group must necessarily have quite a large rank
constraining matter to have at most two indices to insure asymptotic
freedom. Recall, that to avoid low energy fine tuning of the
coupling constants the unified gauge theory must be asymptotically
free. This requirement, {\it de facto}, limits the maximum allowed
representation in the theory.

The phenomenology, both for collider experiments and cosmology, of this
novel extension of the standard model is very rich with many features
common to both supersymmetry and technicolor.

The evolution of the coupling constant $\alpha_n$, at the
one-loop level, of a gauge theory
is controlled by
\begin{equation}\label{running}
{\alpha_{n}^{-1}(\mu) = \alpha_{n}^{-1}(M_Z) - \frac{b_n}{2\pi}\ln
\left(\frac{\mu}{M_Z}\right) \ ,}
\end{equation}
where $n$ refers to the gauge group being $SU(n),$ for $n\geq 2$ or $U(1),$
 for $n=1\ $.

The first coefficient of the beta function $b_n$ is
\begin{eqnarray}
b_n =  \frac{2}{3}T(r) N_{wf} + \frac{1}{3} T(r')N_{cb} -
\frac{11}{3}C_2(G) \ ,
\end{eqnarray}
where $T(r)$ is the Casimir of the
representation $r$ to which the
fermions belong, $T(r')$ is the
Casimir of the representation $R'$
to which the bosons belong. $N_{wf}$ and $N_{cb}$ are respectively the
number of Weyl fermions and the number of complex scalar bosons. $C_2(G)$ is the
quadratic Casimir of the adjoint representation of the gauge group.

The SM gauge group is $SU(3)\times SU(2)\times U(1)$. We have three
associated coupling constants which one can imagine to unify at some
very high energy scale $M_{GUT}$. This means that the three
couplings are all equal at the scale $M_{GUT}$, i.e.
$\alpha_3(M_{GUT})= \alpha_2(M_{GUT})=\alpha_1(M_{GUT})$ with
$\alpha_1= \alpha/(c^2\cos^2 \theta_w)$ and $\alpha_2 = \alpha/
\sin^2\theta_w$, where $c$ is a normalization constant to be
determined shortly.

Assuming one-loop unification using
Eq.~(\ref{running}) for $n=1,2,3,$ one
finds the following relation
\begin{eqnarray} \label{unification}
\frac{b_3-b_2}{b_2-b_1} & = & \frac{\alpha_3^{-1} -\alpha^{-1}\sin^2
\theta_w}{(1+c^2)\alpha^{-1} \label{1}
\sin^2\theta_w-c^2\alpha^{-1}} \ .
\end{eqnarray}
In the above expressions the
Weinberg angle
$\theta_w$, the electromagnetic coupling constant $\alpha$ and the
strong coupling constant $\alpha_3$ are all evaluated at the
$Z$ mass. For a
given particle content we shall denote the LHS of
Eq.~(\ref{unification}) by $B_{\rm theory}$ and the RHS by
$B_{\rm
  exp}$. Whether $B_{\rm theory}$ and $B_{\rm exp}$ agree is a simple way to
check if
the coupling constants unify. We shall use the experimental
values $\sin^2 \theta_w (M_Z) = 0.23150\pm 0.00016$,
$\alpha^{-1}(M_Z) = 128.936 \pm 0.0049$, $\alpha_3(M_Z) =
0.119\pm 0.003$ and $M_Z = 91.1876(21)$ GeV \cite{Yao:2006px}.
The unification scale is given by the expression
\begin{eqnarray}
{M_{GUT} = M_Z
\exp
\left[{{2\pi}\frac{\alpha_2^{-1}(M_Z)-\alpha_1^{-1}(M_Z)}{b_2-b_1}}\right]
\ . }
\end{eqnarray}

While the normalizations of the
coupling constants of the two
non-Abelian gauge groups are fixed by
the appropriately normalized
generators of the gauge groups, the
normalization of the Abelian
coupling constant is a priori arbitrary. The normalization of the
Abelian coupling constant can be
fixed by a rescaling of the
hypercharge $Y \rightarrow cY$ along with
$g\to g/c\ $. The normalization
constant $c$ is
chosen by imposing that all three coupling constants have a common
normalization
\begin{eqnarray}
\text{Tr}\,(c^2Y^2) = \text{Tr}\,(T_3^2) \ ,
\end{eqnarray}
where $T_3$ is the generator of the weak
  isospin and
the trace is over all the relevant fermionic particles on which the
generators act. It is sufficient to fix it for a given fermion
  generation (in a complete multiplet of the unification group).

The previous normalization is consistent with
an $SU(5)$-type normalization for the generators of $U(1)$
of hypercharge,
$SU(2)_L$ and $SU(3)_c\ $.

As well explained in the paper by Li
and Wu \cite{Li:2003zh}:
{\it  At one-loop a contribution to $b_3 - b_2$ or $b_2 - b_1$ emerges
  only from particles not forming complete representations
  \footnote{Such as the five and the ten dimensional representation of
    the unifying gauge group $SU(5)$.} of the unified gauge
  group}. {}For example the gluons, the weak gauge bosons and the
Higgs particle of the SM do not form complete
representations of $SU(5)$ but ordinary quarks and leptons do. Here we
mean that these particles form complete representations of $SU(5)$, all
the way from the unification scale
    down to the electroweak scale.
The particles not forming complete representations will presumably
join at the unification scale with new particles and together then
form complete representations of the unified gauge group. Note, that
although there is no contribution to the unification point of the
particles forming complete
representations, the running of each
coupling constant is affected by all of the particles present at low
energy.

\subsection{Dis-unification in SM}
As a warm up, we consider the SM with
$N_g$ generations. In
this case we find $c=\sqrt{3/5}$, which is the same value one finds
when the hypercharge is upgraded to one of the generators of $SU(5)$,
and therefore the beta function
coefficients are

\begin{eqnarray}
b_3 & = & \frac{4}{3}N_g -11 \ ,\\
b_2 & = & {\frac{4}{3} N_g - \frac{22}{3} +
\underbrace{\frac{1}{6}}_{\rm Higgs} \ , } \\
b_1 & = & {\frac{3}{5}\left( \frac{20}{9}N_g +\frac{1}{6} \right) =
\frac{4}{3}N_g +\underbrace{\frac{1}{10}}_{\rm Higgs} \ .}
\end{eqnarray}

Here $N_g$ is the number of
generations. It is clear that the
SM does not unify since $B_{\rm theory} \sim 0.53$ while
$B_{\rm exp} \sim 0.72\ $.

\begin{figure}[htbp]
\begin{center}
\includegraphics[width=0.6\linewidth]{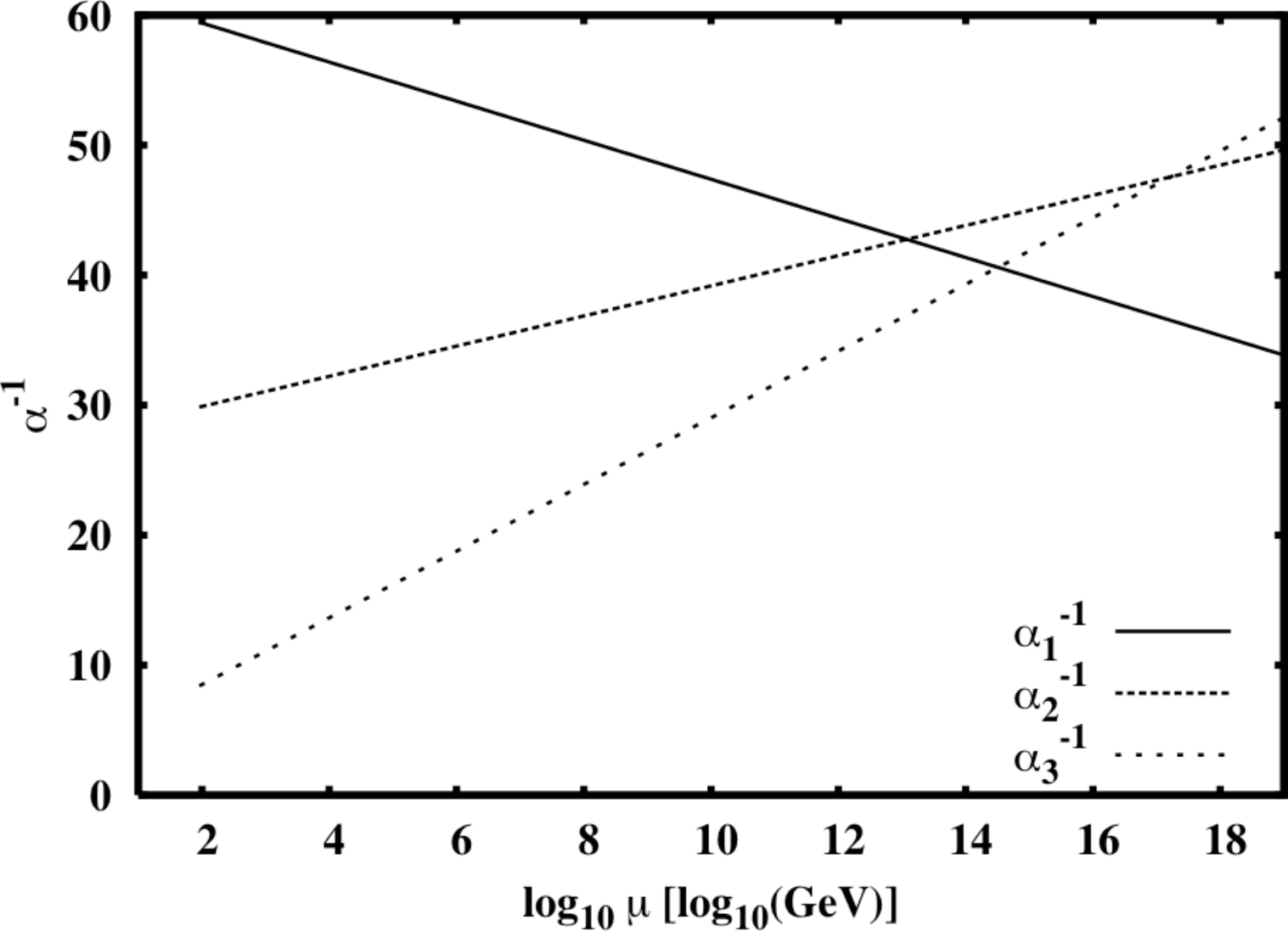}
\end{center}
\caption{The running of the three standard model gauge couplings.}
\label{SM}
\end{figure}

Note that the spectrum relevant for computing
$B_{\rm theory}$  is constituted by the gauge bosons and the standard
model
Higgs. The contribution of quarks and leptons drops out in agreement
with the fact that they form complete representations of the unifying
gauge group which, given the present normalization for $c$, is at
least $SU(5)$. Hence the predicted value of $B_{\rm theory}$ is
independent of the number of
generations. However the overall running
for the three couplings is dependent on the number of
  generations and in
Fig.~\ref{SM} we show the behavior of the three couplings with $N_g
= 3$.

\subsection{Studying SU(3)$\times$SU(2)$\times$U(1) Unification in Technicolor}
Here we compare a few examples in which the standard model Higgs is
replaced by a technicolor-like theory. A similar analysis was performed in \cite{Christensen:2005bt}. In this section we press on phenomenological successful technicolor models with technimatter in
higher dimensional representations and demonstrate that the simplest
model helps unifying the SM couplings while other more
traditional approaches are less successful. We also show that by a
small modification of the technicolor
dynamics, all of the four
couplings can unify \footnote{Since the technicolor dynamics is
  strongly coupled at the electroweak scale the last point on the
  unification of all of the couplings is meant to be only
  illustrative.}.

\subsubsection{Minimal Walking Technicolor (MWT)}
\begin{figure}[!tbp]
\begin{center}
\includegraphics[scale=0.6]{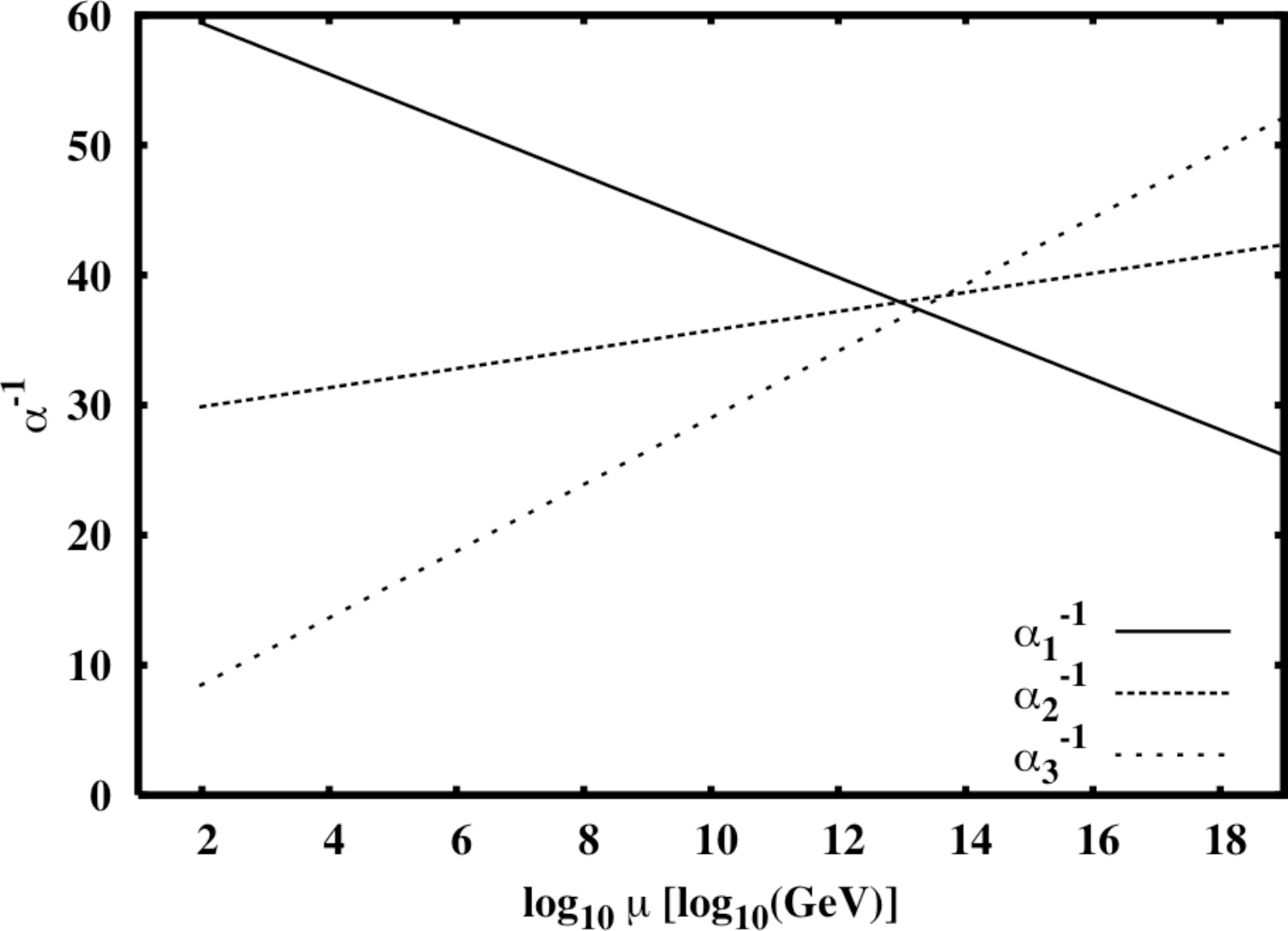}
\end{center}
\caption{The running of the SM gauge couplings in the
    presence of adjoint technifermions (the technicolor
coupling is not included here).}\label{TC}
\end{figure}

We examine what happens to the running of the SM couplings when the Higgs sector is replaced by the
MWT theory introduced earlier. This model has
technicolor group $SU(2)$ with two techniflavors in the two-index
symmetric representation of the technicolor group. As already
mentioned to avoid Witten's $SU(2)$
anomaly, the minimal solution is to
add a
new lepton family. We still assume an $SU(5)$-type
unification leading to $c^2=3/5$.
The beta function
coefficients will be those of
  the SM minus the Higgs plus the extra
contributions from the techniparticles, ergo
\begin{eqnarray}
b_3 & = & \frac{4}{3}N_g -11 \ ,\\
b_2 & = & \frac{4}{3} N_g - \frac{22}{3} + \frac{2}{3}
\frac{1}{2}\left( \frac{2(2+1)}{2} + 1 \right) = \frac{4}{3} \left(
N_g + 1 \right) - \frac{22}{3} \ ,\\
b_1 & = & \frac{3}{5} \left( \frac{20}{9} N_g + \frac{20}{9} \right)
= \frac{4}{3}\left( N_g+1 \right) \ ,
\end{eqnarray}
where $N_g$ is the number of ordinary SM generations. From this we
see that $B_{\rm theory}=0.68$ and $B_{\rm exp} = 0.72\ $ and hence argue that we have a better
unification than in the standard model
with an elementary Higgs. The running of the SM couplings is shown in
Fig.~\ref{TC} for three ordinary standard model generations.
Note that the increase in $B_{\rm theory}$ with respect to the
SM is due to the fact that,
typically, bosonic contributions
are numerically suppressed with respect to fermionic
ones and
that,
while $b_1-b_2 = 22/3$ receives only a contribution from the gauge
sector, $b_2-b_3 = 11/3 + 4/3$ has
two contributions, a gauge one and a
fermionic one. These results are a
direct consequence of the fact that we have no
ordinary quarks related to the new leptonic family.

\subsubsection{Traditional Walking and Non-Walking One Family Model}
Here the technicolor particles also carry ordinary color and the
technifermions constitute complete representations of $SU(5)$, hence
the SM coupling unification receives no improvement with respect to
the SM case. This is so since the numerical effect of the Higgs on
the unification is small. {}For a one-family $SU(N)$ theory, we
  have $B_{\rm theory} = 1/2 \ $.

\subsubsection{Partially Electroweak-Gauged Technicolor}

This approach consists in letting
only one doublet of techniquarks transform non-trivially under the electroweak
symmetries with the rest of the matter remaining in electroweak singlets, as first suggested in
\cite{Dietrich:2005jn}. In this case all techniquarks transform
still under the technicolor gauge group and hence contribute to
rendering
the technicolor dynamics quasiconformal without affecting the leading
perturbative contribution  to the electroweak precision
parameters. The reader can find a rather exhaustive investigation of
walking technicolor theories with fermions in different
representations of the technicolor gauge group in
\cite{Dietrich:2006cm}.

Denoting by ${\rm
  dim}[R_T]$ the dimension of the technicolor representation under
which all the techniquarks transform we find the following result for
$B_{\rm theory}$
\begin{eqnarray}
B_{\rm theory} = \frac{1}{2}\left[\frac{11+{\rm dim}[R_T]}{11-{\rm dim}[R_T]/ {5}}\right] \ .
\end{eqnarray}
Since we have gauged only one technidoublet with respect to the weak interactions to avoid Witten's global anomaly with respect to the weak interactions we take ${\rm dim}[R_T]$ to be even. We have used the following hypercharge assignment free from gauge anomalies
\begin{align}
Y(Q^a_L)=& \; 0 \ ,&\quad Y(U^a_R,D^a_R)&=\left(\frac{1}{2},-\frac{1}{2}\right) \ ,&\quad a=1,\dots,{\rm dim}[R_T]
\ ,
\end{align}
for the technidoublet charged under
the electroweak interactions.

We find that $B_{\rm theory}$ is 0.73 for ${\rm dim}[R_T]=4$ and it
increases with larger values of ${\rm dim}[R_T]\ $. We can also consider
the case ${\rm dim}[R_T]$ odd while solving Witten's anomaly by adding
for example a new weak doublet uncharged under technicolor. We see
that from the unification point of view partially electroweak gauged
technicolor models are comparable with the MWT model presented
earlier.

However, for the model to be phenomenologically viable the new
technicolor theory should pass the electroweak precision
constraints. A complete list of
walking-type technicolor theories
passing the precision tests can be found in
\cite{Dietrich:2006cm}. The simplest unification condition requires
the technicolor representation, in this case, to be
four dimensional. This can only be achieved when the technifermions
are arranged in the fundamental representation of the
$SU(4)$-technicolor gauge group. According to Table III in
\cite{Dietrich:2006cm} one needs, at least, fifteen techniflavors for
the theory to have a walking behavior with a reasonable
$S$
parameter. In \cite{Dietrich:2006cm},
this theory has not been listed
as a prime candidate and hence will
not be considered further here.

\subsubsection{The Technicolor Coupling Constant}

Until now we have not discussed the technicolor coupling
constant $\alpha_{TC}$.
It is possible that the technicolor
interaction does not unify with the other three forces or unifies
later. A single step unification is though esthetically more appealing to
us. Here we focus on the minimal walking theory which has already
shown to be a promising theory for the unification of the standard
model couplings.
Remembering that the Casimir of the
two-index symmetric representation of $SU(N_{TC})$ is
$(N_{TC}+2)/2$ the first coefficient
of the beta function
$b_{TC}$ is easily found to be
\begin{eqnarray}
b_{TC} = \frac{2}{3} (N_{TC}+2)  N_{f} - \frac{11}{3}N_{TC} \ ,
\end{eqnarray}
where $N_{TC}$ is the number of
technicolors and $N_f$ is the
number of techniflavors. For two
colors and two flavors we find
$b_{TC}=-2$. Observing that, somewhat accidentally, also $b_2=-2$ for
three ordinary SM
generations, we
conclude that the technicolor coupling constant cannot unify with the
other three couplings at the same point. We are assuming, quite
naturally, that the low energy starting points of $\alpha_2$ and
$\alpha_{TC}$ are different.

Insisting that the technicolor coupling constant must
unify with the other coupling constants at
$M_{GUT}$, we need to modify
at a given scale
$ {X}<M_{GUT}$ either the
overall running of the SM couplings or the one of technicolor. To make less
steep the running of the SM couplings one could add new
generations. To avoid the loss of asymptotic freedom for the week
coupling we find that at most only one entire new SM like
generation can be added at an intermediate
scale.
If we, however, choose not to modify the running of the SM coupling constants, the
running of the technicolor coupling constant must at some point $X < M_{GUT}$ become
steeper. This can be achieved by enhancing the number of technigluons
and lowering the contribution due to the
techniquarks at the
scale $X$. An elegant way to implement this idea is to imagine
that the techniquarks - belonging to
the three dimensional
two-index symmetric representation of $SU(2)$ - are
embedded in the
fundamental representation of $SU(3)$ at the scale $X$. At energies
below $X$ we have $b^{<
  X}_{TC}=-2$ and for energies larger than
$X$ we have $b_{TC}^{> X} = -29/3$. If we take the technicolor
coupling to start running at the electroweak scale $M_{EW}\sim 246 \
\text{GeV}$ and unifying with the three SM couplings at the
unification scale we find an expression for the intermediate scale
$X$
\begin{equation}
\ln X = \frac{1}{b_{TC}^{< X} - b_{TC}^{>
 X}}\bigg\{2\pi \bigg(\alpha_{TC}^{-1}(M_{EW}) - \alpha_{TC}^{-1}(M_{GUT})\bigg) +
 b_{TC}^{< X} \ln M_{EW} - b_{TC}^{> X}\ln M_{GUT} \bigg\} \ .
 \label{XX}
\end{equation}
If we take the starting point of the running of the technicolor
coupling to be the critical coupling close to the conformal
window
we have ${\alpha_{TC}(M_{EW}) =
\pi/(3C_2(\square\hspace{-1pt}\square)) = \pi/6}\ $. Also
using the numbers
$\alpha_{TC}(M_{GUT}) = {\alpha_i(M_{GUT}) \sim 0.026\ ,
  i=1,2,3\ }$, $M_{GUT} \sim 9.45 \times 10^{12}
\,\text{GeV}$ we find the intermediate scale
to be $X \sim 830$ GeV.

\subsubsection{Proton Decay}
Grand Unified Theories lead, generally, to proton decay. Gauge bosons
of mass ${M_{V} < M_{GUT}}$ are
responsible for the decay of the proton into $\pi^0$ and $e^+$. The
lifetime of the proton is estimated to be \cite{Giudice:2004tc}
\begin{eqnarray}
\tau &=& {\frac{4f^2_{\pi}M^4_{V}}{\pi m_p \alpha_{GUT}^2 \left(1+D+F\right)^2
\alpha_N^2\left[A_R^2+\left(1+|V_{ud}|^2 \right)^2A_L^2\right]}} \\
& = &{\left( \frac{M_{GUT}}{10^{16}\ \text{GeV}} \right)^4 \left(
\frac{\alpha_{GUT}^{-1}}{35} \right)^2 \left( \frac{0.015\
\text{GeV}^3}{\alpha_N} \right)^2 \left(\frac{2}{\mathcal{A}} \right)^2 2.7
\times 10^{35}\ \text{yr} \ , }
\end{eqnarray}
where we have used $f_{\pi} = 0.131\ \text{GeV}$, the chiral
Lagrangian factor $1+D+F = 2.25$, the operator renormalization
factors $\mathcal{A} \equiv A_L=A_R$ and the hadronic matrix element
is taken from lattice results \cite{Aoki:1999tw} to be $\alpha_N
=-0.015\ \text{GeV}^3$. Following Ross \cite{Ross:1985ai}, we have
estimated $\mathcal{A} \sim 2$ but a larger value $\sim 5$ is quoted in \cite{Giudice:2004tc} .
The lower bound on the unification scale comes from
the Super-Kamiokande limit $\tau > 5.3 \times 10^{33}$
{yr}
\cite{Suzuki:2001rb}
\begin{eqnarray}
M_{GUT} > M_{V} & > & {\left( \frac{35}{\alpha_{GUT}^{-1}} \right)^{1/2}
\left( \frac{\alpha_N}{-0.015\ \text{GeV}^3}\right)^{1/2} \left(
\frac{\mathcal{A}}{2} \right)^{1/2}\ 3.7\times 10^{15}\ \text{GeV} \ .}
\end{eqnarray}

In the MWT model extension of the SM we find ${\alpha^{-1}_{GUT} \sim
  37.5}$ and $M_{GUT} \sim  10^{13} \,\text{GeV}$ yielding too fast
proton decay.

\subsubsection{Constructing a Simple Unifying Group}
We provide a simple embedding of our matter content into a unifying gauge group. To construct this group we first summarize the charge assignments in table \ref{single}.
\begin{table}[h]
\caption{Quantum Numbers of the MWT + One SM Family}
\begin{center}
\begin{tabular}{c|c|c|c|c}
&$SO_{TC}(3)$&$SU_c(3)$&$SU_L(2)$&$U_Y(1)$ \\
\hline \hline
$q_L$ &1&3&2&1/6 \\
$u_R$ & 1 &3 &1 & 2/3 \\
$d_R$ & 1&3 &1&-1/3\\
$L $& 1&1&2&-1/2 \\
$e_R$ & 1 &1&1&-1 \\
\hline
$Q_L$ &3&1&2&1/6 \\
$U_R$ & 3 &1 &1 & 2/3 \\
$D_R$ & 3&1 &1&-1/3\\
${\cal L}_L $& 1&1&2&-1/2 \\
$\zeta_R$ & 1 &1&1&-1 \\
\end{tabular}
\end{center}
\label{single}
\end{table}
For simplicity we have considered right transforming leptons only
for the charged ones. Also, the techniquarks are classified as being
fundamentals of $SO(3)$ rather than adjoint of $SU(2)$. Except for
topological differences, linked to the center group of the two
groups, there is no other difference. This choice allows us to show
the resemblance of the technicolor fermions with ordinary quarks. We
can now immediately arrange each SM family within an ordinary
$SU(5)$ gauge theory. The relevant question is how to incorporate
the technicolor sector (here we mean also the new Lepton family). An
easy way out is to double the weak and hypercharge gauge groups as
described in table \ref{tabledouble}.
\begin{table}[h]
\caption{MWT + One SM Family enlarged gauge group}
\begin{center}
\begin{tabular}{c|c|c|c|c|c|c}
&$SO_{TC}(3)$&$SU_1(2)$&$U_1(1)$&$SU_c(3)$&$SU_2(2)$&$U_{2}(1)$ \\
\hline \hline
$q_L$ &1&1&0&3&2&1/6 \\
$u_R$& 1&1&0 &3 &1 & 2/3 \\
$d_R$ &1&1& 0&3 &1&-1/3\\
$L $&1&1 &0&1&2&-1/2 \\
$e_R$ &1&1& 0 &1&1&-1 \\
\hline
$Q_L$ &3&2&1/6&1&1&0 \\
$U_R$ & 3 &1 & 2/3&1&1&0 \\
$D_R$ & 3&1 &-1/3&1&1&0\\
${\cal L}_L $& 1&2&-1/2&1&1&0 \\
$\zeta_R$ & 1 &1&-1&1&1&0 \\
\end{tabular}
\end{center}
\label{tabledouble}
\end{table}
This assignment allows us to arrange the low energy matter  fields
into complete representations of $SU(5)\times SU(5)$. To recover the
low energy assignment one invokes a spontaneous breaking of the
group down to $SO(3)_{TC}\times SU_c(3)\times SU_L(2) \times U_Y(1)$
\footnote{To achieve such as a spontaneous breaking of the gauge
group one needs new matter fields around or slightly above the grand
unified scale transforming with respect to both the gauge groups. }.
We summarize in table \ref{GUT} the technicolor and SM fermions
transformation properties with respect to the grand unified group.
\begin{table}[h]
\caption{GUT}
\begin{center}
\begin{tabular}{c|c|c}
&$SU(5)$&$SU(5)$ \\
\hline \hline
$\bar{A}_{SM}$ &1&$\overline{10}$\\
$F_{SM}$ & 1&5 \\
\hline
$\bar{A}_{MWT}$ &$\overline{10}$&1\\
$F_{MWT}$& 5&1 \\
\end{tabular}
\end{center}
\label{GUT}
\end{table}
Here the fields $A$ and $F$ are standard Weyl fermions and the gauge
couplings of the two $SU(5)$ groups need to be the same. We have
shown here that it is easy to accommodate all of the matter fields
in a single semi-simple gauge group. This is a minimal embedding and
others can be envisioned. New fields must be present at the grand
unified scale (and hence will not affect the running at low energy)
guaranteeing the desired symmetry breaking pattern.

\subsubsection{Providing Mass to the fermions}
We have not yet considered the problem of how the ordinary fermions
acquire mass. Many  extensions of technicolor have been suggested in
the literature to address this issue. Some of the extensions make
use of yet another strongly coupled gauge dynamics,  others
introduce fundamental scalars. It is even possible to marry
supersymmetry and technicolor.  Many variants of the schemes
presented above exist. A nice review of the major models is the one
by Hill and Simmons \cite{Hill:2002ap}. It is fair to say that at
the moment there is not yet a consensus on which is the correct ETC.
Although it is beyond the scope of this initial investigation to
provide a complete working scheme for mass generation we find it
instructive to construct the simplest model able to provide mass to
all of the fermions and which does not affect our results, but
rather improves them.

We parametrize our ETC, or better our ignorance about a complete ETC theory, with the (re)introduction of a single Higgs type doublet on the top of the minimal walking theory whose main purpose is to give mass to the ordinary fermions. This simple construction leads to no flavor changing neutral currents and does not upset the agreement with the precision tests which our MWT theory already passes brilliantly. We are able to give mass to all of the fermions and the contribution to the beta functions reads:
\begin{eqnarray}
b_3 & = & \frac{4}{3}N_g -11 \ ,\\
b_2 & = & \frac{4}{3} \left(
N_g + 1 \right) - \frac{22}{3}  + \frac{1}{6}\ ,\\
b_1 & = &  \frac{4}{3}\left( N_g+1 \right) +\frac{1}{10} \ ,
\end{eqnarray}
leading to
\begin{equation}
B_{theory}=0.71 \ ,
\end{equation}
a value which, at the one loop level, is even closer to the
experimental value of $0.72$ than the original MWT theory alone. The
unification scale is also slightly higher than in MWT alone and it
is of the order of $1.2 \times 10^{13}$~ GeV.  The ETC construction
presented above has already been used many times in the literature
\cite{Simmons:1988fu,Dine:1990jd,Kagan:1990az,Kagan:1991gh,Carone:1992rh,Carone:1993xc}.
We find the results very encouraging.  We wish to add that the need
for walking dynamics in the gauge sector is important since it helps
reducing the value of the S-parameter which is typically large even
before taking into account the problems due to the introduction of
an ETC sector.

\subsection{Adding Adjoint SM Matter: A novelty in Composite Higgs Models}
We wish to improve on the unification point (before taking into account of possible ETC type corrections) and delay it, energy-wise, to
avoid the experimental bounds on the proton decay.

We hence need a minimal modification of our extension of the SM with
the following properties: i) it is natural, i.e. it does not
reintroduce the hierarchy problem, ii) it does not affect the working
technicolor sector, iii) it allows for a straightforward unification
with a resulting theory which is asymptotically free, iv) it yields a
phenomenologically viable proton decay rate and possibly leads also
to dark matter candidates.

Point i) forces us to add new fermionic-type matter while ii) can be
satisfied by modifying the matter content of the SM per se. A simple
thing to do is to explore the case in which we consider adjoint
fermionic matter for the strong and weak interactions. We will show
that this is sufficient to greatly improve the proton decay problem
while also improving unification with respect to the MWT theory. To be
more specific, we add one colored
Weyl fermion transforming solely
according to the adjoint representation of $SU(3)$ and a Weyl fermion
transforming according to the adjoint
representation of $SU_L(2)$. These
fermions can
be identified with the gluino and wino in supersymmetric extensions of
the SM.  The big
hierarchy is still under control in the present model.

Since our theory is not supersymmetric the introduced fermions need
not be degenerate with the associated gauge bosons. Their masses can
be of the order of, or larger than, the electroweak scale. Finally,
naturality does not forbid the presence of a fermion associated to
the hypercharge gauge boson and hence this degree of freedom may
occur in the theory. Imagining a unification of the value of the
masses at the unification scale also requires the presence of such a
${U(1)}$ bino-type fermion.

In this case
the one-loop beta function coefficients are
\begin{eqnarray}
b_3 &=& \frac{4}{3}N_g -11 + 2 \ , \\
b_2 &=& \frac{4}{3} \big( N_g +1 \big) - \frac{22}{3} + \frac{4}{3} \ ,
\\
b_1 &=& \frac{4}{3} \big( N_g +1 \big) \ .
\end{eqnarray}
This gives $B_{\rm theory} = 13/18 \sim
0.72{(2)}$ which is in
excellent agreement with the experimental value. Note also that the
unification scale is ${M_{GUT} \sim 2.65 \times 10^{15} \text{GeV}}$
which brings the proton decay within the correct order of magnitude set by
experiments.
\begin{figure}[ht]
\begin{center}
\includegraphics[width=0.6\linewidth]{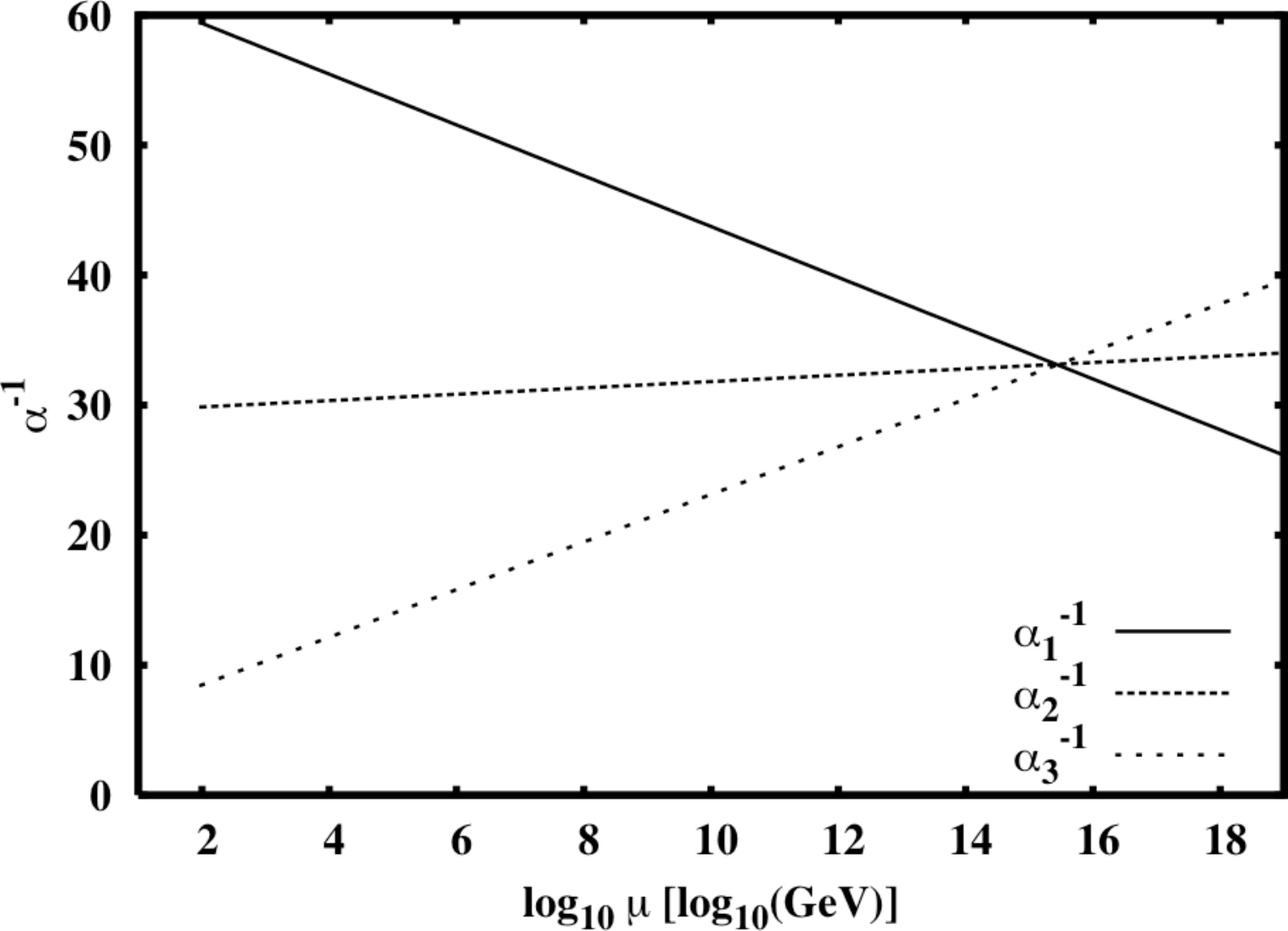}
\end{center}
\caption{The running of the three SM gauge couplings in the new model
  with also adjoint fermionic matter for the SM gauge groups.}
\label{SM-Improved}
\end{figure}

\begin{figure}[ht]
\begin{center}
\mbox{\subfigure{\resizebox{!}{5.3cm}{\includegraphics{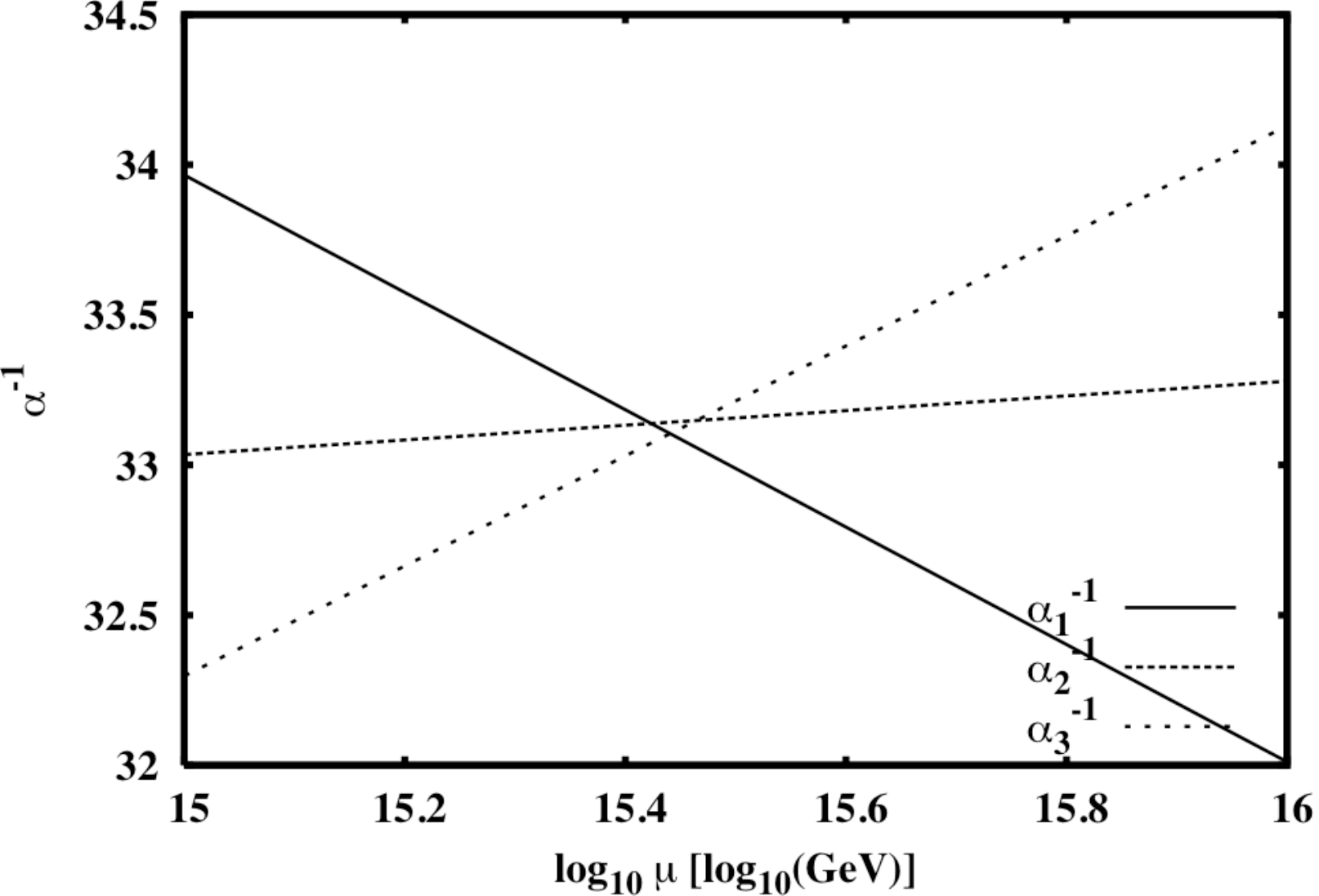}}}
\quad
\subfigure{\resizebox{!}{5.3cm}{\includegraphics{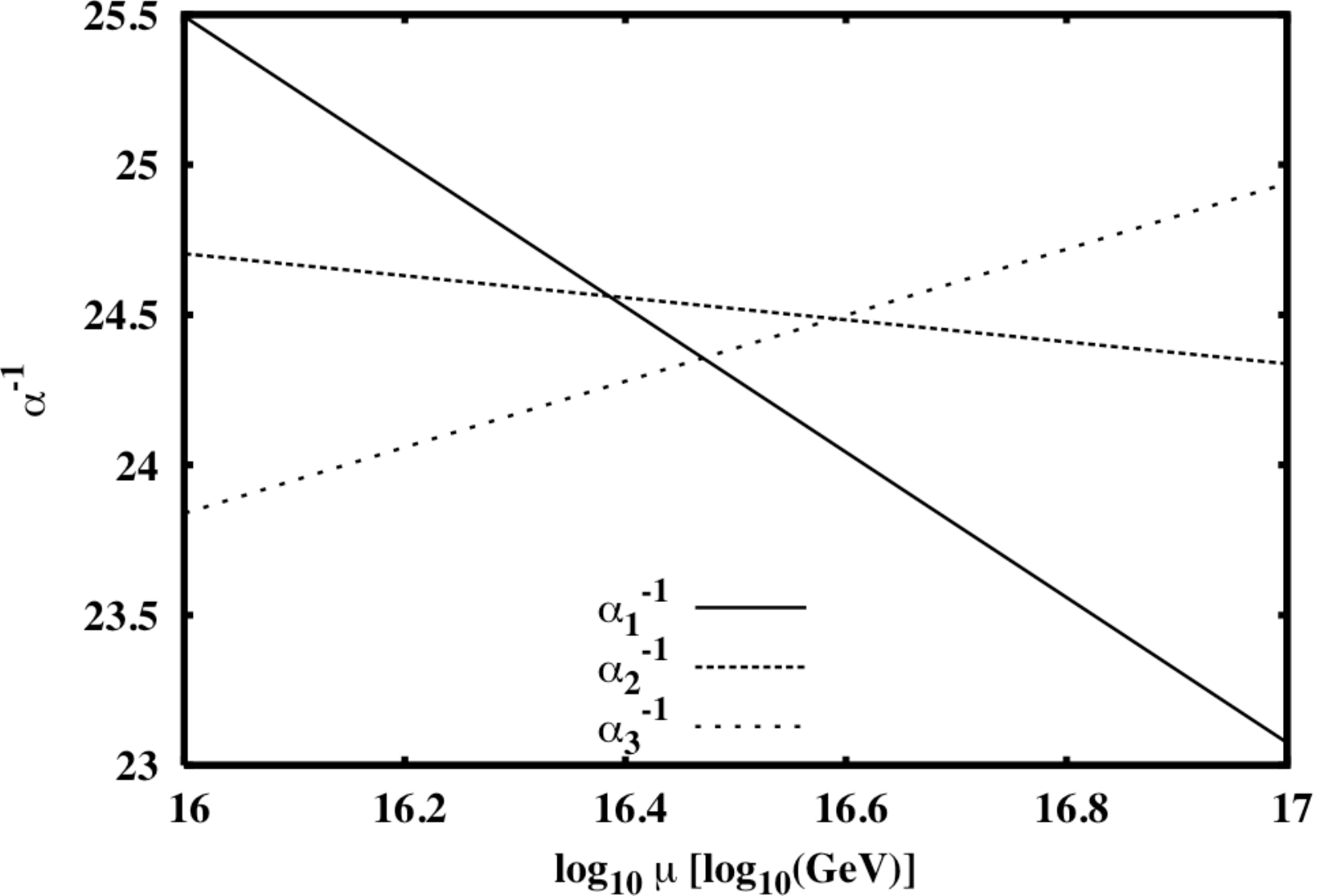}}}}
\end{center}
\caption{\textit{Left Panel}: A zoom around the unification point of the running of the
  three SM gauge couplings in the new model with extra fermionic adjoint matter
  for the SM gauge groups. \textit{Right Panel}: A zoom around the
  unification point for the couplings in the MSSM. }
\label{ZoomCompare}
\end{figure}

We can make the technicolor coupling unify with the SM couplings, as
done in the MWT section.  Using
Eq.~(\ref{XX}), we find now $X\sim
10^8$ GeV. We recall here that $X$ is the scale above which our
technicolor theory becomes an ${SU(3)}$ gauge theory with the fermions
transforming according to the fundamental representation.

It is phenomenologically appealing that the scale $X$ is much higher than the electroweak scale. This allows our technicolor coupling to walk for a sufficiently large range of energy to allow for the introduction of extended technicolor interactions needed to give masses to the SM particles.
\begin{figure}[htbp]
\begin{center}
\includegraphics[width=0.6\linewidth]{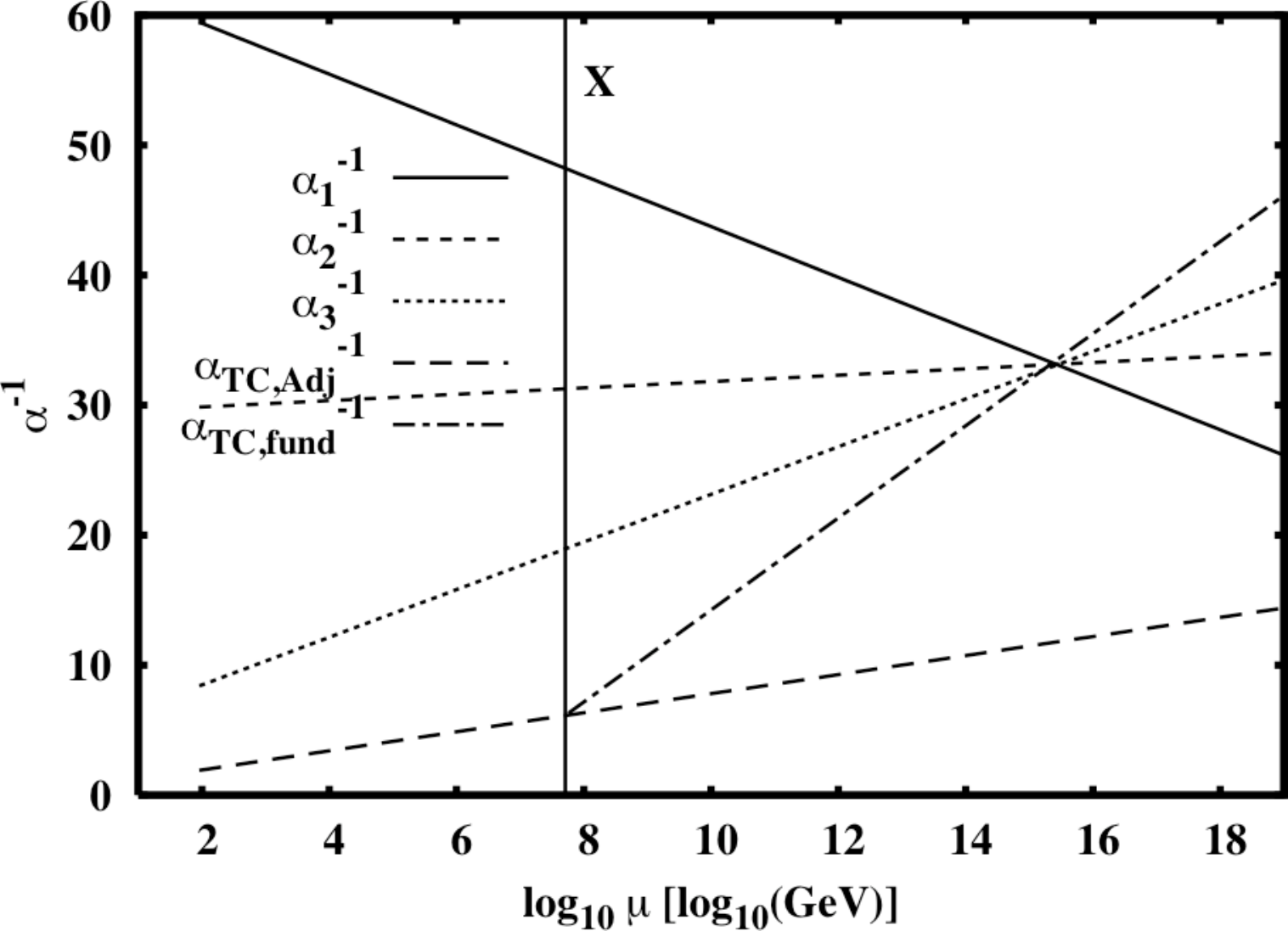}
\end{center}
\caption{The running of the three SM gauge couplings as well as the
  technicolor one. MWT is made to unify with the other three couplings
  by enhancing the gauge group from ${SU(2)}$ of technicolor
  to ${SU(3)}$ while
  keeping the same fermionic matter content. We see that the scale
  where this enhancement of the gauge group should dynamically occur
  to obtain complete unification is around $10^8$ GeV. }
\label{SM-ImprovedX}
\end{figure}
%%%%
%%%%
\subsubsection{Comparing with the MSSM}
Unification of the SM gauge couplings is considered one of the
strongest points in favor of a supersymmetric extension of the SM
and hence it is reasonable to compare our results with the {SUSY}
ones. In {SUSY}, one finds $B_{\rm theory}=0.714$ which is
remarkably close to the experimental value $B_{\rm exp}\sim 0.72$
but it is not better than the value predicted in the present model
which is $0.72(2)$. Obviously this comparison must be taken with a
grain of salt since we still need to provide mass to the SM fermions
and take care of the threshold corrections. {}For example according
to the model introduced in section III, subsection G, to give masses
to all of the fermions yields a theoretical value for the
unification which is around $0.76$.

There are three possible candidates for dark matter here, depending on
which one is the lightest one and on the extended technicolor
interactions which we have not yet specified but that we will
explore in the future: The chargeless fermion in the adjoint
representation of $SU(2)_L$, i.e. the wino-like object as well as the
bino-type one.  The third possibility is the heavy neutrino-like fermion whose dark matter potential features are being currently investigated \cite{Kainulainen:2006wq}.

We have introduced a technicolor model which leads to the unification of the SM gauge couplings. At the one-loop level the model provides a higher degree of unification when compared to other technicolor models and to the minimal supersymmetric extension of the SM.

The phenomenology, both for collider experiments and cosmology, of the present extension of the SM is very
rich and needs to be explored in much detail.

The model has many
features in common with split and non-split supersymmetry \cite{ArkaniHamed:2004fb,Giudice:2004tc} and also with models proposed in \cite{Bajc:2006ia,Dorsner:2006fx} while others in common with
technicolor.

\section{Conclusion}

We  summarized some of the features associated to dynamical breaking of the electroweak symmetry. The status of the phase diagram for $SU(N)$ gauge theories with fermionic matter has been summarized.  As a relevant example for breaking the electroweak symmetry dynamically we  introduced Minimal Walking Technicolor. Other extensions, such as partially gauged and split technicolor have also been reviewed.  Unification of the standard model gauge couplings has been discussed within old and new technicolor models. A number of appendices have been added to support the reader with group theoretical results, notation and some needed technical information.  In one of the appendices we  have also discussed how to gain information on the spectrum of new strongly coupled theories relevant for the dynamical breaking of the electroweak theory using alternative large N limits. 

 \section{Acknowledgments}
I am very happy to thank S. Catterall, S. Chivukula, L. Del Debbio, D. D. Dietrich, R. Foadi, M. T. Frandsen, M. J\"{a}rvinen, M. P. Lombardo, I. Masina, T. Ryttov, J. Schechter, E. H. Simmons, M. Svensson, K. Tuominen and R. Zwicky, for pleasant fruitful collaborations on the various topics presented in this review, comments and/or careful reading of the manuscript. This review started during my staying at CERN Theory Division in 2007.  I am partially supported by the Marie Curie Excellence Grant under contract MEXT-CT-2004-013510.

\newpage
\appendix

\section{Basic group theory relations \label{eleven}}

The Dynkin indices label the highest weight
of an irreducible representation and uniquely characterise the
representations. The Dynkin indices for some of the most common
representations are given in Table \ref{dynkin}. For details on the concept
of Dynkin indices see, for example \cite{Dynkin:1957um,Slansky:1981yr}.

%----------------------------------------------------------------------------
\begin{table}
\begin{center}
\begin{tabular}{r|c}
representation&Dynkin indices\\
\hline
singlet&(000\dots 00)\\
fundamental (F)&(100\dots 00)\\
antifundamental ($\bar{\mathrm{F}}$)&(000\dots 01)\\
adjoint (G)&(100\dots 01)\\
$n$-index symmetric (S$_n$)&(n00\dots 00)\\
2-index antisymmetric (A$_2$)&(010\dots 00)\\
\end{tabular}
\caption{Examples for Dynkin indices for some common representations.}
\label{dynkin}
\end{center}
\end{table}
%----------------------------------------------------------------------------

For a representation, R, with the Dynkin indices
$(a_1,a_2,\dots,a_{N-2},a_{N-1})$ the
quadratic Casimir operator reads \cite{White:1992aa}
\bear 2N \,C_2(\mathrm{R}) &=& \sum_{m=1}^{N-1}[ N(N-m)ma_m + m(N-m){a_m}^2 +\nonumber \\&&
\sum_{n=0}^{m-1}2n(N-m)a_na_m ] \label{C} \eear
and the dimension of R is given by
\bear
d(\mathrm{R})
=
\prod_{p=1}^{N-1}
\left\{
\frac{1}{p!}
\prod_{q=p}^{N-1}
\left[
\sum_{r=q-p+1}^p(1+a_r)
\right]
\right\},
\label{d}
\eear
which gives rise to the following structure
\bear d(\mathrm{R}) &=& (1+a_1)(1+a_2)\dots(1+a_{N-1}) \times
\nn && \times (1+\sfr{a_1+a_2}{2}) \dots
(1+\sfr{a_{N-2}+a_{N-1}}{2}) \times
\nn && \times (1+\sfr{a_1+a_2+a_3}{3}) \dots
(1+\sfr{a_{N-3}+a_{N-2}+a_{N-1}}{3}) \times
\nn && \times \dots \times
\nn && \times
(1+\sfr{a_1+\dots+a_{N-1}}{N-1}). \eear
The Young tableau associated to a given Dynkin index
$(a_1,a_2,\dots,a_{N-2},a_{N-1})$ is easily constructed. The length of
row $i$ (that is the number of boxes per row) is given in terms of the Dynkin
indices by the expression $r_i = \sum_{i}^{N-1} a_{i}$. The length of
each column is indicated by $c_k$; $k$ can assume any positive integer value.
Indicating the total number of boxes associated to a given Young tableau with 
$b$ one has another compact expression for $C_2(\mathrm{R})$,
\bear
2N \, C_2(\mathrm{R})
= 
N\left[ bN +\sum_i r^2_i  -\sum_ic_i^2 - \frac{b^2}{N} \right]\ ,
\eear
and the sums run over each column and row.

\section{Realization of the generators for MWT \label{appgen}}

It is convenient to use the following representation of SU(4)
\beq S^a = \begin{pmatrix} \bf A & \bf B \\ {\bf B}^\dag & -{\bf A}^T
\end{pmatrix} \ , \qquad X^i = \begin{pmatrix} \bf C & \bf D \\ {\bf
    D}^\dag & {\bf C}^T \end{pmatrix} \ , \eeq
where $A$ is hermitian, $C$ is hermitian and traceless, $B = -B^T$ and
$D = D^T$. The ${S}$ are also a representation of the $SO(4)$
generators, and thus leave the vacuum invariant $S^aE + E {S^a}^T = 0\ $.
Explicitly, the generators read
\beq S^a = \frac{1}{2\sqrt{2}}\begin{pmatrix} \tau^a & \bf 0 \\ \bf 0 &
  -\tau^{aT} \end{pmatrix} \ , \quad a = 1,\ldots,4 \ , \eeq
where $a = 1,2,3$ are the Pauli matrices and $\tau^4 =
\mathbbm{1}$. These are the generators of SU$_V$(2)$\times$ U$_V$(1).
\beq S^a = \frac{1}{2\sqrt{2}}\begin{pmatrix} \bf 0 & {\bf B}^a \\
{\bf B}^{a\dag} & \bf 0 \end{pmatrix} \ , \quad a = 5,6 \ , \eeq
with
\beq B^5 = \tau^2 \ , \quad B^6 = i\tau^2 \ . \eeq
The rest of the generators which do not leave the vacuum invariant are
\beq X^i = \frac{1}{2\sqrt{2}}\begin{pmatrix} \tau^i & \bf 0 \\
\bf 0 & \tau^{iT} \end{pmatrix} \ , \quad i = 1,2,3 \ , \eeq
and
\beq X^i = \frac{1}{2\sqrt{2}}\begin{pmatrix} \bf 0 & {\bf D}^i \\
{\bf D}^{i\dag} & \bf 0 \end{pmatrix} \ , \quad i = 4,\ldots,9 \ ,
\eeq
with
\beq\begin{array}{r@{\;}c@{\;}lr@{\;}c@{\;}lr@{\;}c@{\;}l}
D^4 &=& \mathbbm{1} \ , & \quad D^6 &=& \tau^3 \ , & \quad D^8 &=& \tau^1 \ , \\
D^5 &=& i\mathbbm{1} \ , & \quad D^7 &=& i\tau^3 \ , & \quad D^9 &=& i\tau^1
\ .
\end{array}\eeq

The generators are normalized as follows
\beq {\rm Tr}\left[S^aS^b\right] =\frac{1}{2}\delta^{ab}\ , \qquad \ , {\rm Tr}\left[X^iX^j\right] =
\frac{1}{2}\delta^{ij} \ , \qquad {\rm Tr}\left[X^iS^a\right] = 0 \ . \eeq

\section{Vector Mesons as Gauge Fields} \label{sec:hidden}
We show how to rewrite the vector meson Lagrangian in a gauge invariant way.  We assume the scalar sector to transform according to a given but otherwise arbitrary representation of the flavor symmetry group $G$. This is a straightforward generalization of the Hidden Local Gauge symmetry idea \cite{Bando:1984ej,Bando:1987br}, used in a similar context for the BESS models \cite{Casalbuoni:1995qt}. At the tree approximation this approach is identical to the one introduced first in \cite{Kaymakcalan:1984bz,Kaymakcalan:1983qq}.

\subsection{Introducing Vector Mesons}
Let us start with a generic flavor symmetry group $G$ under which a scalar field $M$ transforms globally in a given, but generic, irreducible representation $R$.  We also introduce an algebra valued  one-form $A=A^{\mu}dx_{\mu}$ taking values in a copy of the algebra of the group $G$, call it $G^{\prime}$, i.e.
\begin{eqnarray}
 A_{\mu}=A^{a}_{\mu} T^a \ , \qquad {\rm with } \qquad T^a \in {\cal{A}}(G^{\prime}) \ .
 \end{eqnarray}
At this point the full group structure is the semisimple group $G\times G^{\prime}$. $M$ does not transform under $G^{\prime}$. Given that $M$ and $A$ belong to two different groups we need another field to connect the two. We henceforth introduce a new scalar field $N$ transforming according to the fundamental of $G$ and to the antifundamental of $G^{\prime}$. We then upgrade $A$ to a gauge field over $G^{\prime}$.
\begin{table}[t]
\caption{Field content}
\begin{center}
\begin{tabular}{c||cc}
&\,\,\,$G$\,\,\,&\,\,\,$G^{\prime}$\,\,\, \\
\hline
&&\\
$M$\,\,\,&\,\,\,$R$\,\,\,&\,\,\,${\mathbf 1}$\,\,\, \\
&&\\
$N$\,\,\,&\,\,\,$\fund$\,\,\,&\,\,\,$\overline{\fund}$\,\,\, \\
&&\\
$A_{\mu}$\,\,\,&\,\,\,${\mathbf 1}$\,\,\,&\,\,\,${\rm Adj}$\,\,\, \\
\end{tabular}
\end{center}
\label{default}
\end{table}%
The covariant derivative for $N$ is:
\begin{eqnarray}
D_{\mu}N=\partial_{\mu} N + i\,\tilde{g} \, N\,A_{\mu} \ .
\end{eqnarray}
We now force $N$ to acquire the following vacuum expectation value
\begin{eqnarray}
\langle N^i_j \rangle = \delta^i_j \, v^{\prime} \ ,
\end{eqnarray}
which leaves the diagonal subgroup - denoted with $G_{V}$ - of $G \times G^{\prime}$ invariant. Clearly $G_V$ is a copy of $G$. Note that it is always possible to arrange a suitable potential term for $N$ leading to the previous pattern of symmetry breaking. $v/v^{\prime}$ is expected to be much less than one and the {\it unphysical} massive degrees of freedom associated to the fluctuations of $N$ will have to be integrated out. The would-be Goldstone bosons associated to $N$ will become the longitudinal components of the massive vector mesons.

To connect $A$ to $M$ we define the one-form transforming only under $G$ via $N$ which - in the deeply spontaneously broken phase of $N$ - reads:
\begin{eqnarray}
\frac{ {\rm Tr} [N N^{\dagger}]}{{\rm dim}(F)} \, P_{\mu} =\frac{D_{\mu}N N^{\dagger} - N D_{\mu}N^{\dagger}} {2\, i \tilde{g}} \ , \qquad P_{\mu} \rightarrow u P_{\mu} u^{\dagger} \ ,
\end{eqnarray}
with $u$ being an element of $G$ and dim($F$) the dimension of the fundamental representation of $G$. When evaluating $P_{\mu}$ on the vacuum expectation value for $N$ we recover $A_{\mu}$:
\begin{eqnarray}
\langle P_{\mu } \rangle =  A_{\mu} \ .
\end{eqnarray}
  At this point it is straightforward to write the Lagrangian containing $N$, $M$ and $A$ and their self-interactions.
 Being in the deeply broken phase of $G\times G^{\prime}$ down to $G_{V}$ we count $N$ as a dimension zero field. This is consistent with the normalization for $P_{\mu}$.

The simplest\footnote{Another nonminimal term is ${\rm Tr} \left[NFN^{\dagger}M (NFN^{\dagger})^{T} M^{\dagger}\right]$.} kinetic term of the Lagrangian is:
\begin{eqnarray}
L_{kinetic} = -\frac{1}{2}{\rm Tr} \left[F_{\mu\nu}F^{\mu\nu}\right]  + \frac{1}{2}{\rm Tr} \left[DN DN^{\dagger}\right] +\frac{1}{2}{\rm Tr} \left[ \partial M \partial M^{\dagger} \right] \ .
\end{eqnarray}
The second kinetic term will provide a mass to the vector mesons.
Besides the potential terms for $M$ and $N$ there is another part of the Lagrangian which is of interest to us. This is the one mixing $P$ and $M$.  Up to dimension four and containing at most two powers of $P$ and $M$ this is:
\begin{eqnarray}
L_{P-M} & = & \tilde{g}^2\ r_1 \ {\rm Tr}\left[P_\mu P^\mu M M^\dagger\right]
+ \tilde{g}^2\ r_2 \ {\rm Tr}\left[P_\mu M {P^\mu}^T M^\dagger \right] \nonumber \\
& + & i \ \tilde{g}\ r_3 \ {\rm Tr}\left[P_\mu \left(M (D^\mu M)^\dagger - (D^\mu M) M^\dagger \right) \right]
+ \tilde{g}^2\ s \ {\rm Tr}\left[P_\mu P^\mu \right] {\rm Tr}\left[M M^\dagger \right] \ . \nonumber \\
\end{eqnarray}
The dimensionless parameters $r_1$, $r_2$, $r_3$, $s$ parameterize the strength of the interactions between the composite scalars and vectors in units of $\tilde{g}$, and are therefore expected to be of order one. We have assumed $M$ to belong to the two index symmetric representation of a generic G= SU(N). It is straightforward to generalize the previous terms to the case of an arbitrary representation $R$ with respect to any group G.
Further higher derivative interactions including $N$ can be included systematically.

\subsection{Further Gauging of G}
In this case we add another gauge field $G_{\mu}$ taking values in the algebra of $G$. We then define the correct covariant derivatives for $M$ and $N$. {}For $N$, for example, we have:
\begin{eqnarray}
D_{\mu}N=\partial_{\mu} N -i\,g\,G_{\mu}\,N+ i\,\tilde{g} \, N\,A_{\mu} \ .
\label{covariant-N}
\end{eqnarray}
Evaluating the previous expression on the vacuum expectation value of $N$ we recover the field $C_{\mu}$ introduced in the text. To be more precise we need to use $P_{\mu}$ again but with the covariant derivative for $N$ replaced by the one in the equation above.
\begin{table}[hb]
\caption{Field content}
\begin{center}
\begin{tabular}{c||cc}
&\,\,\,$G$\,\,\,&\,\,\,$G^{\prime}$\,\,\, \\
\hline
&&\\
$M$\,\,\,&\,\,\,$R$\,\,\,&\,\,\,${\mathbf 1}$\,\,\, \\
&&\\
$N$\,\,\,&\,\,\,$\fund$\,\,\,&\,\,\,$\overline{\fund}$\,\,\, \\
&&\\
$A_{\mu}$\,\,\,&\,\,\,${\mathbf 1}$\,\,\,&\,\,\,${\rm Adj}$\,\,\, \\
&&\\
$G_{\mu}$\,\,\,&\,\,\,${\rm Adj}$\,\,\,&\,\,\,${\mathbf 1}$\,\,\, \\
\end{tabular}
\end{center}
\label{default2}
\end{table}%

\subsection{Effective Lagrangian and Mass Matrices} \label{app:lagr}
In this section we summarize and generalize the effective Lagrangians for the scalar and vector sectors, and include the explicit mass matrices for the mixings of the composite vectors with the fundamental gauge fields. 
\subsubsection{Scalar Sector}
The composite scalars are assembled in the matrix $M$ of Eq.~(\ref{M}). In terms of the mass eigenstates this reads
\small
\begin{eqnarray}
M=
\begin{pmatrix}
i \Pi_{UU} + \widetilde{\Pi}_{UU} & {\displaystyle \frac{i \Pi_{UD} + \widetilde{\Pi}_{UD}}{\sqrt{2}}} &
{\displaystyle \frac{\sigma+i\Theta + i \Pi^0 + A^0}{2}} & {\displaystyle \frac{i\Pi^+ + A^+}{\sqrt{2}}} \\
\smallskip \\
{\displaystyle \frac{i \Pi_{UD} + \widetilde{\Pi}_{UD}}{\sqrt{2}}} & i \Pi_{DD} + \widetilde{\Pi}_{DD} &
{\displaystyle \frac{i\Pi^- + A^-}{\sqrt{2}}} & {\displaystyle \frac{\sigma+i\Theta - i \Pi^0 - A^0}{{2}}} \\
\smallskip \\
{\displaystyle \frac{\sigma+i\Theta + i \Pi^0 + A^0}{2}} & {\displaystyle \frac{i\Pi^- + A^-}{\sqrt{2}}} &
i \Pi_{\overline{UU}} + \widetilde{\Pi}_{\overline{UU}} & {\displaystyle \frac{i \Pi_{\overline{UD}} + \widetilde{\Pi}_{\overline{UD}}}{\sqrt{2}}} \\
\smallskip \\
{\displaystyle \frac{i\Pi^+ + A^+}{\sqrt{2}}} & {\displaystyle \frac{\sigma+i\Theta - i \Pi^0 - A^0}{2}} &
{\displaystyle \frac{i \Pi_{\overline{UD}} + \widetilde{\Pi}_{\overline{UD}}}{\sqrt{2}}} & i \Pi_{\overline{DD}} + \widetilde{\Pi}_{\overline{DD}}
\end{pmatrix} \ , \nonumber \\
\end{eqnarray}
\normalsize
where $\sigma=v+H$. The Lagrangian for the Higgs sector, including the spontaneously broken potential, and the ETC mass term for the uneaten Goldstone bosons, is
\begin{eqnarray}
{\cal L}_{\rm Higgs} &=& \frac{1}{2}{\rm Tr}\left[D_{\mu}M D^{\mu}M^{\dagger}\right] + \frac{m^2}{2}{\rm Tr}[MM^{\dagger}] \nonumber \\
& - & \frac{\lambda}{4} {\rm Tr}\left[MM^{\dagger} \right]^2 - \lambda^\prime {\rm Tr}\left[M M^{\dagger} M M^{\dagger}\right]
+  2\lambda^{\prime\prime} \left[{\rm Det}(M) + {\rm Det}(M^\dagger)\right] \nonumber \\
& + & \frac{m_{\rm ETC}^2}{4}\ {\rm Tr}\left[M B M^\dagger B + M M^\dagger \right] \ ,
\end{eqnarray}
where the covariant derivative is given by Eq.~(\ref{covariantderivative}).

\subsubsection{Vector Sector}
In terms of the charge eigenstates the matrix $A^\mu$ is
\begin{eqnarray}
A^\mu =
\begin{pmatrix}
{\displaystyle \frac{a^{0\mu}+v^{0\mu}+v^{4\mu}}{2\sqrt{2}}} & 
{\displaystyle \frac{a^{+\mu}+v^{+\mu}}{2}} &  
{\displaystyle \frac{x_{UU}^\mu}{\sqrt{2}}} & 
{\displaystyle \frac{x_{UD}^\mu+s_{UD}^\mu}{2}} \\
\smallskip \\
{\displaystyle \frac{a^{-\mu}+v^{-\mu}}{2}} & 
{\displaystyle \frac{-a^{0\mu}-v^{0\mu}+v^{4\mu}}{2\sqrt{2}}} & 
{\displaystyle \frac{x_{UD}^\mu-s_{UD}^\mu}{2}} & 
{\displaystyle \frac{x_{DD}^\mu}{\sqrt{2}}} \\
\smallskip\\
{\displaystyle \frac{x_{\overline{UU}}^\mu}{\sqrt{2}}} & 
{\displaystyle \frac{x_{\overline{UD}}^\mu-s_{\overline{UD}}^\mu}{2}} & 
{\displaystyle \frac{a^{0\mu}-v^{0\mu}-v^{4\mu}}{2\sqrt{2}}} & 
{\displaystyle \frac{a^{-\mu}-v^{-\mu}}{2}} \\
\smallskip \\
{\displaystyle \frac{x_{\overline{UD}}^\mu+s_{\overline{UD}}^\mu}{2}} & 
{\displaystyle \frac{x_{\overline{DD}}^\mu}{\sqrt{2}}} & 
{\displaystyle \frac{a^{+\mu}-v^{+\mu}}{2}} & 
{\displaystyle \frac{-a^{0\mu}+v^{0\mu}-v^{4\mu}}{2\sqrt{2}}}
\end{pmatrix} \ .
\end{eqnarray}
The most general Lagrangian for the gauge and vector fields can be conveniently written using the $N$ and $P_\mu$ fields of appendix~\ref{sec:hidden}. Demanding $CP$ invariance, and including terms up to dimension four, we have 
\begin{eqnarray}
{\cal L}_{\rm vector} =  & - & \frac{1}{2}{\rm Tr} \Big[\widetilde{W}_{\mu\nu}\widetilde{W}^{\mu\nu}\Big]
-\frac{1}{4}{\rm Tr}\Big[B_{\mu\nu}B^{\mu\nu}\Big]
-\frac{1}{2}{\rm Tr} \Big[F_{\mu\nu}F^{\mu\nu}\Big] \nonumber \\
&+& \frac{1}{2}{\rm Tr} \left[D_\mu N \left(D^\mu N\right)^{\dagger}\right] 
+ \frac{1}{2}{\rm Tr} \left[D_\mu M \left(D^\mu M\right)^\dagger \right] \nonumber \\
& + &  \tilde{g}^2 \ a_1 \ {\rm Tr}\Big[P_\mu P^\mu \Big]^2 + \tilde{g}^2  \ a_2 \ {\rm Tr}\Big[P_\mu P^\mu P_\nu P^\nu\Big]
+ \tilde{g}^2 \ a_3 \ {\rm Tr}\Big[P_\mu P_\nu P^\mu P^\nu \Big]  \nonumber \\
& - & i\ \tilde{g} \ b \ {\rm Tr}\Big[[P_\mu,P_\nu] N F^{\mu\nu} N^\dagger \Big] \nonumber \\
& + & \tilde{g}^2\ r_1 \ {\rm Tr}\Big[P_\mu P^\mu M M^\dagger\Big]
+ \tilde{g}^2\ r_2 \ {\rm Tr}\Big[P_\mu M {P^\mu}^T M^\dagger \Big] \nonumber \\
& + & i \ \tilde{g}\ r_3 \ {\rm Tr}\Big[P_\mu \left(M (D^\mu M)^\dagger - (D^\mu M) M^\dagger \right) \Big]
+ \tilde{g}^2\ s \ {\rm Tr}\Big[P_\mu P^\mu \Big] {\rm Tr}\Big[M M^\dagger \Big] \ , \nonumber \\
\end{eqnarray}
where the field strength tensor $F^{\mu\nu}$ is given by Eq.~(\ref{strength}), and the covariant derivatives of $M$ and $N$ are respectively given by Eq.~(\ref{covariantderivative}) and Eq.~(\ref{covariant-N}). Notice that we have excluded the  terms $i\, {\rm Tr}\left[[P_\mu,P_\nu]G^{\mu\nu}\right]$ and ${\rm Tr}\left[N F_{\mu\nu} N^\dagger G^{\mu\nu}\right]$ with order one couplings.  There terms in the limit of no weak interactions are reserved solely to technicolor interactions.  Here $G^{\mu
\nu}$ contains $\widetilde{W}^{\mu\nu}$ and $B^{\mu\nu}$. 

The covariant derivative terms give rise to mass terms for the charged and neutral vector bosons:
\begin{eqnarray}
{\cal L}_{\rm mass} =
\begin{pmatrix} \widetilde{W}^-_\mu & v^-_\mu & a^-_\mu  \end{pmatrix} {\bf {\cal M}_{\rm C}^2}
\begin{pmatrix} \widetilde{W}^{+\mu} \\ v^{+\mu} \\ a^{+\mu}  \end{pmatrix} +
\frac{1}{2}\begin{pmatrix} B_\mu & \widetilde{W}^3_\mu & v^0_\mu & a^0_\mu & v^4_\mu \end{pmatrix} {\bf {\cal M}_{\rm N}^2}
\begin{pmatrix} B^\mu \\ \widetilde{W}^{3\mu} \\ v^{0\mu} \\ a^{0\mu} \\ v^{4\mu} \end{pmatrix} \ ,
\end{eqnarray}
where
\begin{eqnarray}
{\bf {\cal M}_{\rm C}^2} =
\begin{pmatrix}
{\displaystyle \frac{g^2 M_V^2 (1+\omega)}{\tilde{g}^2}} & 
-{\displaystyle \frac{g M_V^2}{\sqrt{2}\tilde{g}}} & 
-{\displaystyle \frac{g M_A^2 (1-\chi)}{\sqrt{2}\tilde{g}}} \\
\smallskip \\
-{\displaystyle \frac{g M_V^2}{\sqrt{2}\tilde{g}}} & 
M_V^2 &
0 \\
\smallskip \\
-{\displaystyle \frac{g M_A^2 (1-\chi)}{\sqrt{2}\tilde{g}}} & 
0 & 
M_A^2
\end{pmatrix} \ ,
\end{eqnarray}
\begin{eqnarray}
{\bf {\cal M}_{\rm N}^2} =
\begin{pmatrix}
{\displaystyle \frac{g^{\prime 2}M_V^2 (1+2y^2+\omega)}{\tilde{g}^2}} & 
-{\displaystyle \frac{g g^\prime M_V^2 \omega}{\tilde{g}^2}} & 
-{\displaystyle \frac{g^\prime M_V^2}{\sqrt{2}\tilde{g}}} &
{\displaystyle \frac{g^\prime M_A^2 (1-\chi)}{\sqrt{2}\tilde{g}}} & 
-{\displaystyle \frac{g^\prime M_V^2 (2y)}{\sqrt{2}\tilde{g}}} \\
\smallskip \\
-{\displaystyle \frac{g g^\prime M_V^2\omega}{\tilde{g}^2}}  &
{\displaystyle \frac{g^2 M_V^2 (1+\omega)}{\widetilde{g}^2}} &
-{\displaystyle \frac{g M_V^2}{\sqrt{2}\tilde{g}}} &
-{\displaystyle \frac{g M_A^2 (1-\chi)}{\sqrt{2}\tilde{g}}} &
0 \\
\smallskip \\
-{\displaystyle \frac{g^\prime M_V^2}{\sqrt{2}\tilde{g}}} &
-{\displaystyle \frac{g M_V^2}{\sqrt{2}\tilde{g}}} &
M_V^2 &
0 &
0 \\
\smallskip \\
{\displaystyle \frac{g^\prime M_A^2 (1-\chi)}{\sqrt{2}\tilde{g}}} &
-{\displaystyle \frac{g M_A^2 (1-\chi)}{\sqrt{2}\tilde{g}}} &
0 &
M_A^2 &
0 \\
\smallskip \\
-{\displaystyle \frac{g^\prime M_V^2 (2y)}{\sqrt{2}\tilde{g}}} &
0 &
0 &
0 &
M_V^2
\end{pmatrix}  \ . \nonumber \\
\end{eqnarray}
Here $M_V$ and $M_A$ are the masses of the vector and axial vector bosons in absence of electroweak interactions, and are related by
\begin{eqnarray}
M_A^2 = M_V^2 + \frac{1}{2}v^2\tilde{g}^2 r_2 \ .
\end{eqnarray}
The parameters $\omega$ and $\chi$ are defined by
\begin{eqnarray}
\omega & \equiv & \frac{v^2\tilde{g}^2}{4 M_V^2} (1+r_2-2r_3) \nonumber \\
\chi & \equiv & \frac{v^2\tilde{g}^2}{2 M_A^2} r_3 \ ,
\end{eqnarray}
where $\chi$ has been used already in Eq.~(\ref{S}).

The vector baryons do not mix with the fundamental gauge fields and thus their masses do not receive tree-level electroweak corrections. Therefore, $x_{UU}$, $x_{UD}$, and $x_{DD}$ are axial mass eigenstates, and $s_{UD}$ is a vector mass eigenstate:
\begin{eqnarray}
& M_{x_{UU}} & = M_{x_{UD}} = M_{x_{DD}} = M_A  \ , \nonumber \\
& M_{s_{UD}} & = M_V \ . 
\end{eqnarray}

\section{Rainbow Approximation}
\label{ra}
{}For nonsupersymmetric theories one way to get quantitative estimates is to use the
\emph{rainbow} approximation
to the Schwinger-Dyson equation
\cite{Pagels:1974se}, see Fig.~\ref{rainbowselfenergy}. 
\ctfig{0.3}{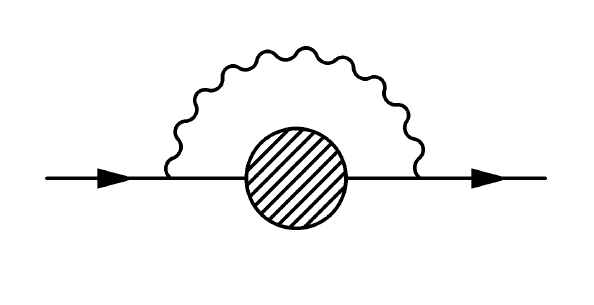}{Rainbow approximation for the
fermion self energy function. The boson is a gluon.} 

\noindent
In this case the full nonperturbative fermion propagator in momentum space reads
\beq iS^{-1}(p) = Z(p)\left(\slashed{p}-\Sigma(p)\right) \ , \eeq
and the Euclidianized gap equation in Landau gauge is given by
\beq \Sigma(p) =
3C_2(R)\int\frac{d^4k}{(2\pi)^4}\frac{\alpha\left((k-p)^2\right)}{(k-p)^2}\frac{\Sigma(k^2)}{Z(k^2)k^2
    + \Sigma^2(k^2)} \ , \eeq
where $Z(k^2)=1$ in this approximation and we linearize
the equation by neglecting $\Sigma^2(k^2)$ in the denominator. Upon converting it to a differential equation and assuming that the coupling
$\alpha(\mu) \approx \alpha_c$ is varying slowly ($\beta(\alpha) \simeq
0$) one gets the   
approximate (WKB) 
solutions \cite{Fukuda:1976zb}
\beq \Sigma(p) \propto p^{-\gamma(\mu)} \ , \qquad \Sigma(p) \propto
p^{\gamma(\mu) - 2} \ , \label{sol-to-gap-eq} \eeq
where the critical coupling is given in terms of the quadratic Casimir
of the fermions representation
\beq \alpha_c \equiv \frac{\pi}{3C_2(R)} \ . \label{critical-coupling}
\eeq
The anomalous dimension of the fermion mass operator is
\beq \gamma(\mu) = 1 - \sqrt{1-\frac{\alpha(\mu)}{\alpha_c}} 
\sim \frac{3C_2(R)\alpha(\mu)}{2\pi} \ . \label{admp}
\eeq
The first solution
corresponds 
to the running of an ordinary mass term ({\it  hard} mass) of nondynamical origin and the
second solution  to a {\it soft} mass dynamically generated. In fact in the second case one observes the $1/p^2$ behavior in the limit of large momentum. 

Within this approximation spontaneous symmetry breaking occurs when $\alpha$ reaches the critical coupling $\alpha_c$ given in Eq.~(\ref{critical-coupling}). {}From Eq.~(\ref{admp}) it is clear that $ \alpha_c $ is reached when 
$\gamma$  is of order unity \cite{Cohen:1988sq,Appelquist:1988yc}. Hence the symmetry breaking occurs when the soft and the hard mass terms scale as function of the energy scale in the same way. In Ref.~\cite{Appelquist:1988yc},  it was noted that in the lowest (ladder) order, the gap equation leads to the condition $\gamma(2-\gamma)=1$ for chiral symmetry breaking to occur. To all orders in perturbation theory this condition is gauge invariant and also equivalent nonperturbatively to the condition $\gamma=1$. However, to any finite order in perturbation theory these conditions are, of course, different. Interestingly the condition $\gamma(2-\gamma)=1$ leads again to the critical coupling $\alpha_c$  when using the perturbative leading order expression for the anomalous dimension which is $\gamma=\frac{3C_2(R)}{2\pi}\alpha$ .

%%%%%%%%%%%%%%%%%%%%%%%%%%%%%%%%%%%%%%%%%%%%%%%%%%%%%%%%%%%%%%%%%%

\section{Universal Electroweak Corrections for (Walking) Technicolor}
\label{electrocorrections}
Using our low energy effective theory the 7 parameters defined in \cite{Barbieri:2004qk} read:
\begin{eqnarray}
\hat{S} &=& \frac{(2 - \chi )\chi g^2}{2\tilde{g}^2+(2-2\chi+\chi^2)g^2}  
      \ , \\ 
\hat{T} &=& 0 \ , \\
W &=& M_W^2 \frac{g^2(M_A^2+(\chi-1)^2M_V^2)}{(2\tilde{g}^2+(2+(\chi-2
)\chi)g^2)M_A^2M_V^2} \ ,  \\
Y &=& M_W^2 \frac{g'^2((1+4y^2)M_A^2+(\chi -1)^2M_V^2)}{(2\tilde{g}^2+(2+4y^2+(\chi
-2)\chi)g'^2)M_A^2M_V^2}
 \ , \\
\hat{U} &=& 0 \ , \\
V &=& 0 \ , \\
X &=& g\,g' \, \frac{M_W^2}{M^2_A M^2_V}\frac{ M_A^2 - (\chi -1)^2M_V^2 }{\sqrt{(2\tilde{g}^2+(2-2\chi+\chi^2)g^2)(2\tilde{g}^2+(2+4y^2-2\chi+\chi^2)g'^2)}} 
 \ .
\end{eqnarray}
In these expressions the coupling constants $g$, $g'$ and $\tilde{g}$ are the ones in the  Lagrangian associated to the yet to be diagonalized spin one states. $W$, $Y$ and $X$ are sensitive to the ratio $M^2_{W}/M^2$ with $M^2$ the lightest of the massive spin one fields. 

\section{Spectrum of Strongly Coupled Theories: Higgsless versus Higgsfull theories}
\label{7}
Often, in the literature, a number of incorrect statements are made when discussing the spectrum of technicolor theories. Here we  will try to clarify first the situation in QCD and then show how to use new analytic means to gain control over the spectrum of strongly coupled theories with fermions in higher dimensional representations. 

 One approach 
 is based on studying the theory in the large number
of colors (N) limit \cite{'t Hooft:1973jz,Witten:1979kh}. At the same time one may
obtain more information by requiring the theory to model the 
(almost) spontaneous breakdown of chiral
 symmetry \cite{Nambu:1961tp,GellMann:1960np}. A standard test case, for ordinary QCD, is pion pion
scattering in the energy range up to about 1 GeV. Some time ago,
an attempt was made \cite{Sannino:1995ik,Harada:1995dc} to implement this
combined scenario.  We used pion pion scattering to provide some insight on the low lying hadronic spectrum of QCD. 
 
 Before turning to the spectrum of the lightest composite states in QCD we offer  a simple definition of Higgsless theory: {\it If the composite state with the same quantum numbers of the Higgs is not the lightest  particle in the spectrum after the Goldstones then the theory is Higgsless. } 
 In practice we  will use the massive spin one states to compare the mass of the composite Higgs with. 
 
 \subsection{The lightest composite scalars in QCD}
The scalar sector of QCD and any technicolor theory constitutes a complicated sector. {}For QCD, in \cite{Harada:2003em},  using the 't Hooft large N limit, chiral dynamics and unitarity constraints
the $f_0(600)$ resonance mass was found to be around 550 MeV. Other authors \cite{Pelaez:2003dy,Oller:1997ti,Uehara:2003ax} have found similar results.
 Such a low
value would make it different from a p-wave quark-antiquark state,
which is expected to be in the 1000-1400 MeV range.
We assume then that 
it is a four quark state (glueball states
 are expected to be in the 1.5
GeV range from lattice investigations).
 Four quark
states of diquark-quark type \cite{Jaffe:1976ig,Jaffe:1976ih} and meson-meson type
\cite{Weinstein:1982gc} have been
 discussed in the literature for many years.
Accepting this picture, however,  
poses a problem for the accuracy of the large N
inspired description of the scattering since
four quark states are  
predicted not to exist in the large N limit of QCD.
We shall take the point of view that a four quark 
type state is present since it allows a natural fit to
the low energy data.
 In practice, since the 
parameters
of the pion contact and rho exchange contributions are fixed, the 
sigma is the most important one for fitting and fits may 
even be achieved \cite{Harada:1996wr} if the vector meson
 piece is neglected. However
the well established, presumably four
 quark type, $f_0$(980) resonance
 must be included to achieve a fit in the region just
 around 1 GeV.

     There is by now a fairly large literature
 \cite{kyotoconf}
on the effect of light ``exotic" scalars in low energy meson meson
 scattering. There seems to be a consenesus, arrived at using rather
different approaches (keeping however, unitarity), that the sigma exists.

%%%%%%%
%%%%%%%
Here we  use two large N limits of QCD as well as our information on the low lying spectrum of QCD to extract information on the spectrum of the lightest states for strongly coupled theories with fermions in various representations of the underlying strongly coupled gauge group. Lifting the strongly coupled scale to the electroweak one for theories with underlying fermions in two index representations we  will show that the light scalar with the same quantum numbers of the Higgs is lighter than the lightest techni-vector meson. 

\subsection{Scalars in the 't Hooft Large N:  Higgsless theories}

We concentrate on the lightest scalar $f_0(600)$ and on the vector meson $\rho(770)$. The $q\bar{q}$ nature of the vector meson is clear. This means that its mass does not scale with the number of colors while its width decrease as $1/N$. We argued above that $f_0(600)$ is a multiquark state. In this case its mass scales with a positive power of N and its width remains constant or grows with N. In formulae: 
\begin{eqnarray}
m_{\rho} ^2&\sim &~~~~\Lambda_{QCD}^2 \ , \qquad \Gamma_{\rho} \sim \frac{1}{N} \\
m_{f_0}^2 &\sim &N^p \Lambda_{QCD}^2 \ , \qquad \Gamma_{f_0} \sim {N}^q \ ,
\end{eqnarray}
with $p>0$ and $q>-1$. 

Scaling up these results to the electroweak theory is straightforward. We first generalize the number of technidoublets gauged under the electroweak theory as well the number of technicolors $N_{TC}$, holding fixed the weak scale we have:
\begin{eqnarray}
M_{T\rho} &=& \frac{\sqrt{2}v_{weak}}{F_{\pi}}
\frac{\sqrt{3}}
{\sqrt{N_D\, N_{TC}}} m_{\rho}
 \\
M_{T f_{0}} &=& \frac{\sqrt{2}v_{weak}}{F_{\pi}\sqrt{N_D}} 
\left(\frac{N_{TC}}{\sqrt{3}}\right)^{\frac{p-1}{2}}m_{f_0}  \ ,
\end{eqnarray} 
where $N_D$ is the number of doublets, $v_{weak}$ is the electroweak scale and the extra $\sqrt{2}$ is due to our normalization of the pion decay constant. Note that for $p=0$ and $q=-1$  the $f_0(600)$ would scale like the $\rho$ and would then be regarded as a quark-antiquark meson at large $N$. However, as we mentioned, there are, by now, strong indications that this state  is not of $q\bar{q}$ nature and hence $p>0$ and $q>-1$. 

Let us choose for definitiveness $p=1$. Already for $N_{TC}\sim 6$, for any $N_D$ the scalar is heavier than the vector meson. Hence for fermions in the fundamental representation of the technicolor theory we expect {\it no scalars} lighter than the respective vector mesons for any $N_{TC}$ larger than or about $6$ technicolors. It is hence fair to call these theories Higgsless. Note that the previous statements may be altered if the theory features walking dynamics.

\subsection{Alternative Large N limits}
The previous results are in agreement with the common lore about the light spectrum of QCD-like theories. Interestingly even if for N=3 one has a scalar state lighter than the lightest vector meson it becomes heavier already for N$>$6.  Clearly the reason behind this is that, due to its multiquark nature, the lightest state possesses different scaling properties than the vector meson.  
The situation changes when we consider alternative extensions of QCD using higher dimensional representations.  At large $N$ different extensions capture different dynamical properties of QCD.  

\subsubsection{The Two Index Antisymmetric Fermions - Link to QCD}    
   Consider redefining the $N=3$ quark field with
 color index A (and flavor
index not written) as
\begin{equation}
        q_A= \frac{1}{2} \epsilon_{ABC}q^{\left[B,C\right]}\ ,\qquad  q^{\left[B,C\right]}=-q^{\left[C,B\right]},
\label{redefine}
\end{equation}
so that, for example, $q_1=q^{23}$ and similarly for the adjoint field,
${\bar q}^1={\bar q}_{23}$ etc. This is just a
 trivial change of variables. However for $N>3$ the resulting theory will be
 different since the two index
 antisymmetric quark representation has $N(N-1)/2$ rather
 than $N$ color components.
As was pointed out by Corrigan and Ramond
 \cite{Corrigan:1979xf}, who were
mainly interested in the problem of the
 baryons at large N, this shows
that the extrapolation of QCD to higher $N$ is not unique.
 Further investigation
of the properties of the alternative extrapolation model
introduced in \cite{Corrigan:1979xf} was 
carried out 
 by Kiritsis and Papavassiliou \cite{Kiritsis:1989ge}.
 
     It may be worthwhile to remark that gauge theories with
two index quarks have gotten a great deal of attention.
 Armoni, Shifman and Veneziano \cite{Armoni:2003gp} have
 proposed an interesting
 relation between certain sectors of the two index antisymmetric
 (and symmetric) theories at
 large number of colors and
sectors of super Yang-Mills (SYM).
Using a supersymmetric inspired
effective Lagrangian approach $1/N$ corrections were
investigated in \cite{Sannino:2003xe}. 

 Besides these two limits a third one for
 massless one-flavor QCD, which
 is in between the 't Hooft and Corrigan Ramond
ones, has been been
 proposed in \cite{Ryttov:2005na}.
 Here one first splits the
 QCD Dirac fermion into the two elementary Weyl fermions
 and afterwards assigns one of them
 to transform according to a rank-two antisymmetric
 tensor while the other
 remains in the fundamental representation of the
 gauge group. For three
 colors one reproduces
one-flavor QCD and for a generic number of colors the
theory is chiral. {}The generic $N$ is
 a particular case of the 
generalized
Georgi-Glashow (gGG) model \cite{Georgi:1985hf}.
The finite temperature phase transition and its relation with chiral symmetry has been investigated in \cite{Sannino:2005sk} while the effects of a nonzero
 baryon chemical potential were studied in \cite{Frandsen:2005mb}.
On the validity of the large N equivalence between different theories we refer the reader to \cite{Unsal:2006pj,Kovtun:2005kh}.

\begin{figure}[htbp]
\centering 
{
\includegraphics[width=
7.5cm,clip=true]{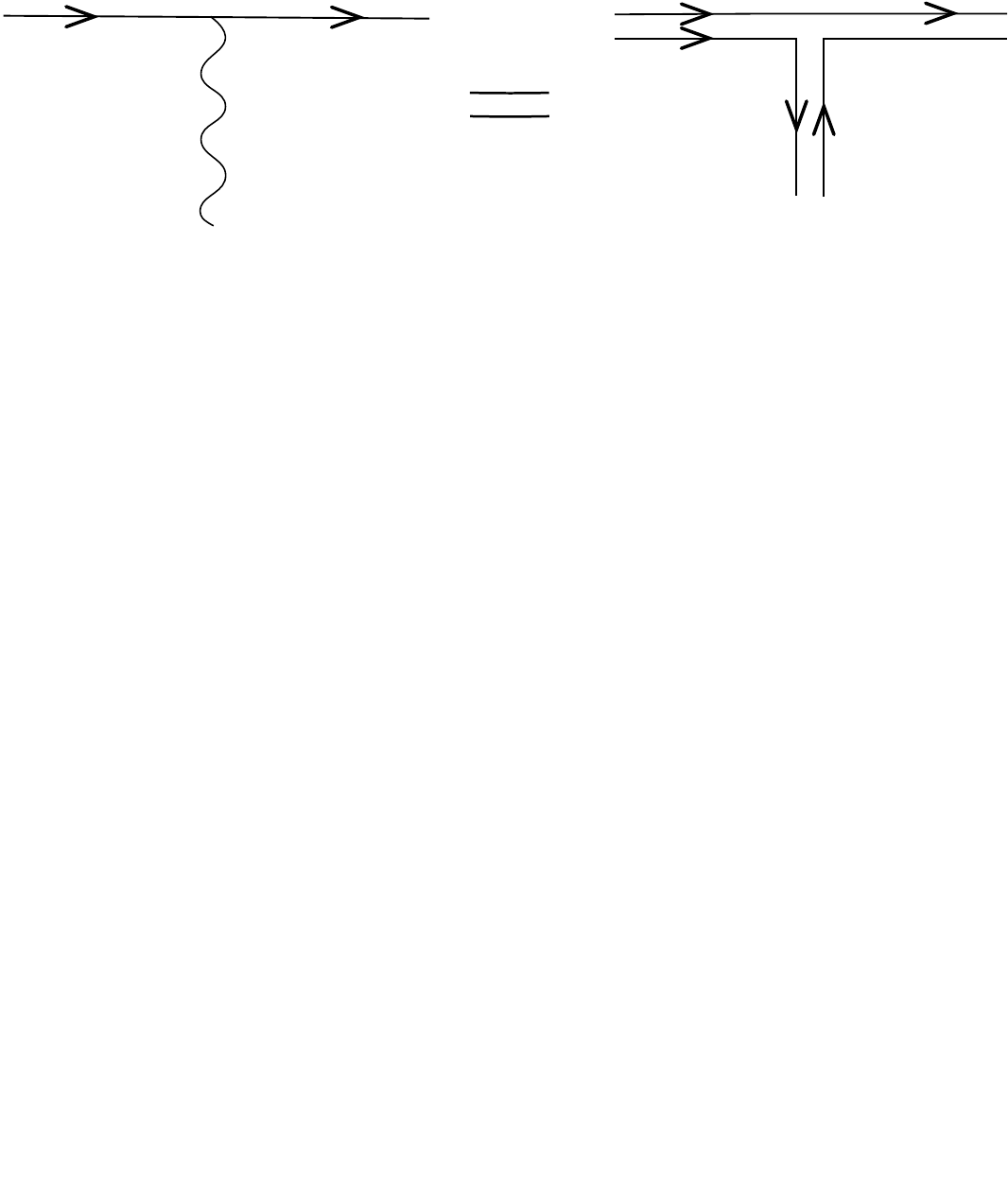}
%\leavevmode {
%\includegraphics[width=6cm,clip=true]{FigA}
}
\vskip -7cm
\caption[]
{Two index fermion - gluon vertex.} \label{FigA}
\end{figure}

   To illustrate the large N counting when quarks are designated
to transform according to the two index antisymmetric representation
of color SU(3) one may employ \cite{'t Hooft:1973jz} the mnemonic where
each tensor index of this group is represented by a directed line.
Then the quark-quark gluon interaction is pictured as in Fig. \ref{FigA}. 
The two index quark is pictured as two lines with arrows pointing in the 
same direction, as opposed to the gluon which has two lines with arrows 
pointing in opposite directions. The coupling constant representing this 
vertex is taken to be $g_t/\sqrt{N}$, where $g_t$ does not depend on $N$ and is kept fixed.

\begin{figure}[htbp]
\centering %\leavevmode
{
\includegraphics[width=
3.0cm,clip=true]{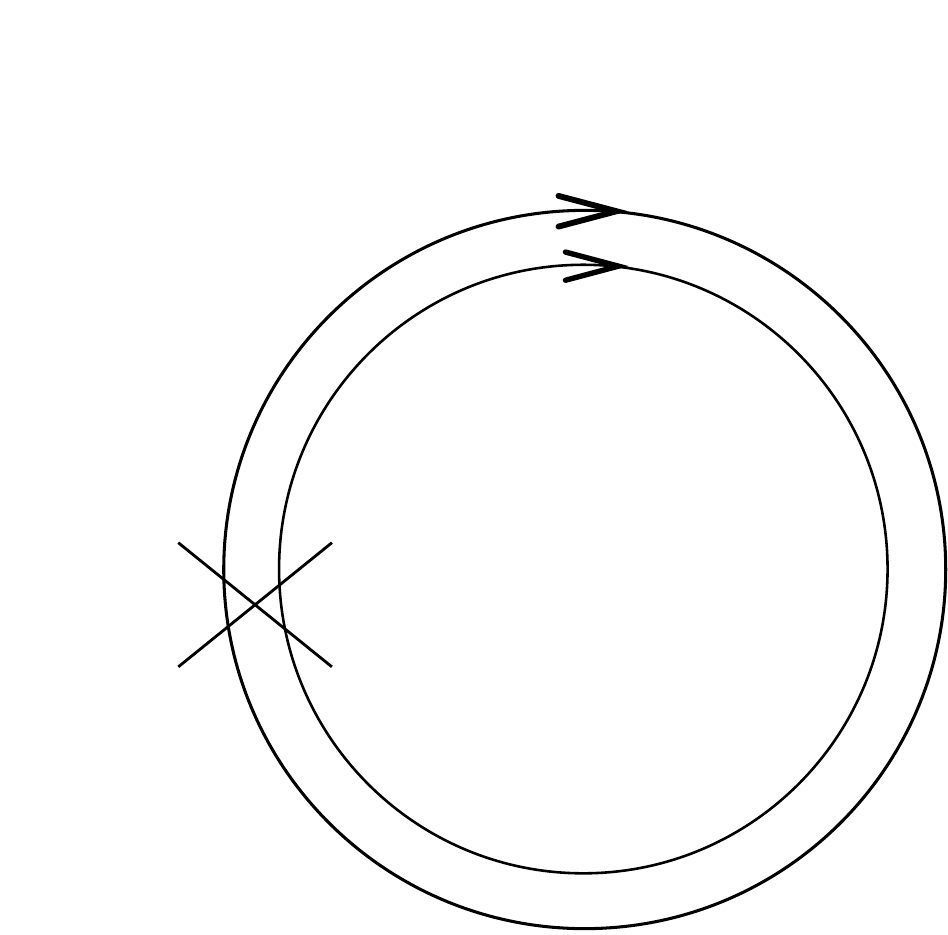}
%\leavevmode {
%\includegraphics[width=6cm,clip=true]{carhomp}
}
\caption[]
{Diagram for $F_\pi$ for the two index quark.} \label{FigB}
\end{figure}

    A ``one point function", like the pion decay constant, $F_\pi$
has as it's simplest diagram, Fig.~\ref{FigB}
          
The X represents a pion insertion and is associated with a
normalization factor for the color part of the pion's wavefunction,
\begin{eqnarray}
            \frac{\sqrt{2}}{\sqrt{N(N-1)}},
\label{wf}
\end{eqnarray}        
which scales for large N as $1/N$. The two circles each
carry a quark index so their factor scales as $N^2$ for
 large N; more
precisely, taking the antisymmetry into account, the factor is
\begin{equation}          
 \frac{N(N-1)}{2}.
\label{loop}
\end{equation}
The product of Eqs. (\ref{wf}) and (\ref{loop})
 yields the $N$ scaling for
$F_{\pi}$:
\begin{equation}
F_{\pi}^2(N)= \frac{N(N-1)}{6}F_{\pi}^2(3).
\label{fpiscaling}
\end{equation}
For large N, $F_\pi$ scales proportionately to $N$ rather
than to $\sqrt{N}$ as in the case of the 't Hooft extrapolation.

\begin{figure}[htbp]
\centering %\leavevmode
{
\includegraphics[width=
4.0cm,clip=true]{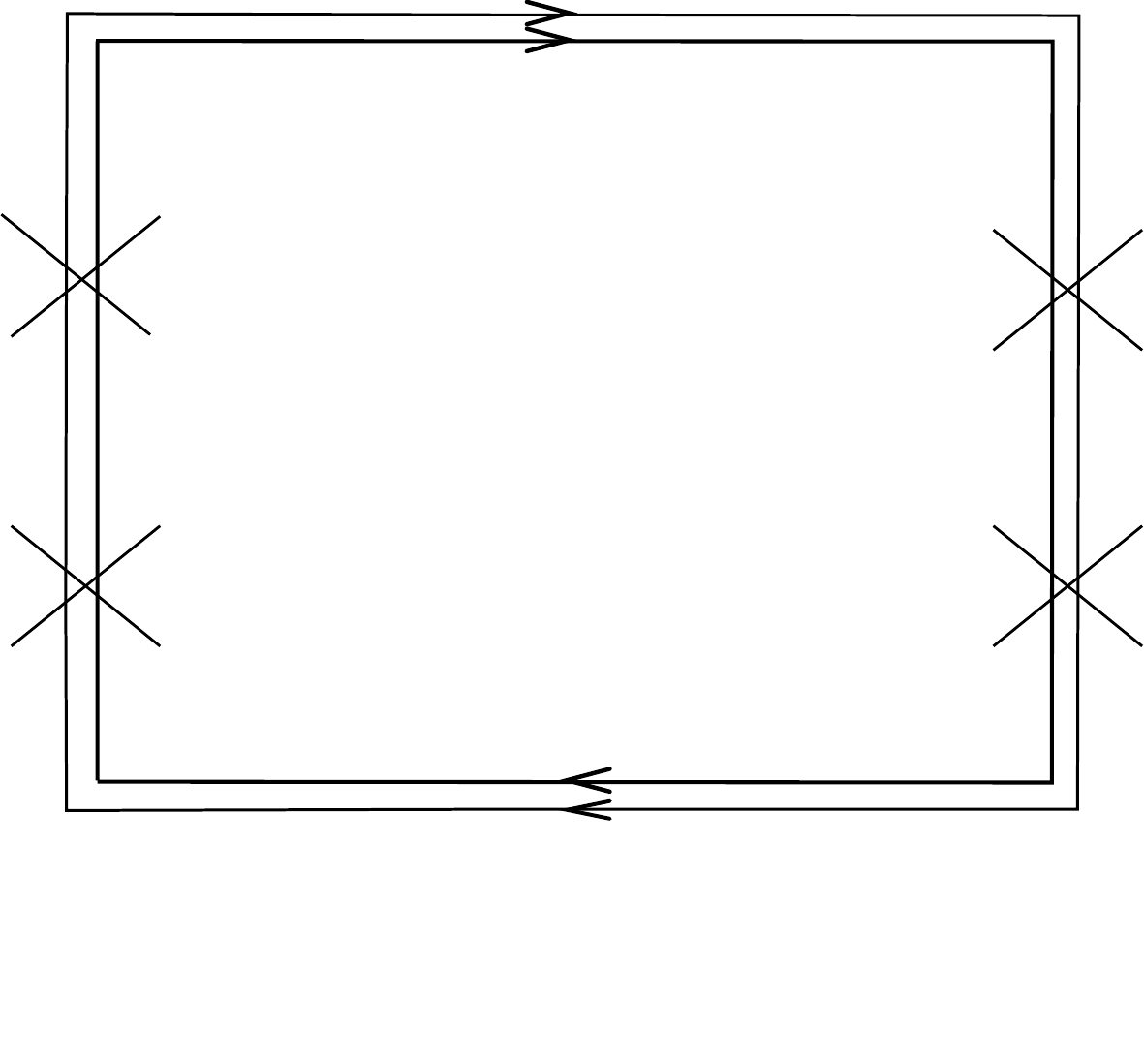}
%\leavevmode {
%\includegraphics[width=6cm,clip=true]{carhomp}
}
\vskip -1cm
\caption[]
{Diagram for the scattering amplitude, A with the 2 index quark. } \label{FigC}
\end{figure}

    Using this scaling the $\pi \pi$ scatttering amplitude, $A$ scales as,
\begin{equation}
A(N)=\frac{6}{N(N-1)}A(3),
\label{Ascaling}
\end{equation}
which, for large N scales as $1/N^2$ rather than
 as $1/N$  in
the 't Hooft extrapolation. This scaling law for
 large N may be verified
from the mnemonic in Fig.~\ref{FigC}, where there is
 an $N^2$ factor from the two loops
multiplied by four factors of $1/N$ from the X's.

\begin{figure}[htbp]
\centering %\leavevmode
{
\includegraphics[width=
8.5cm,clip=true]{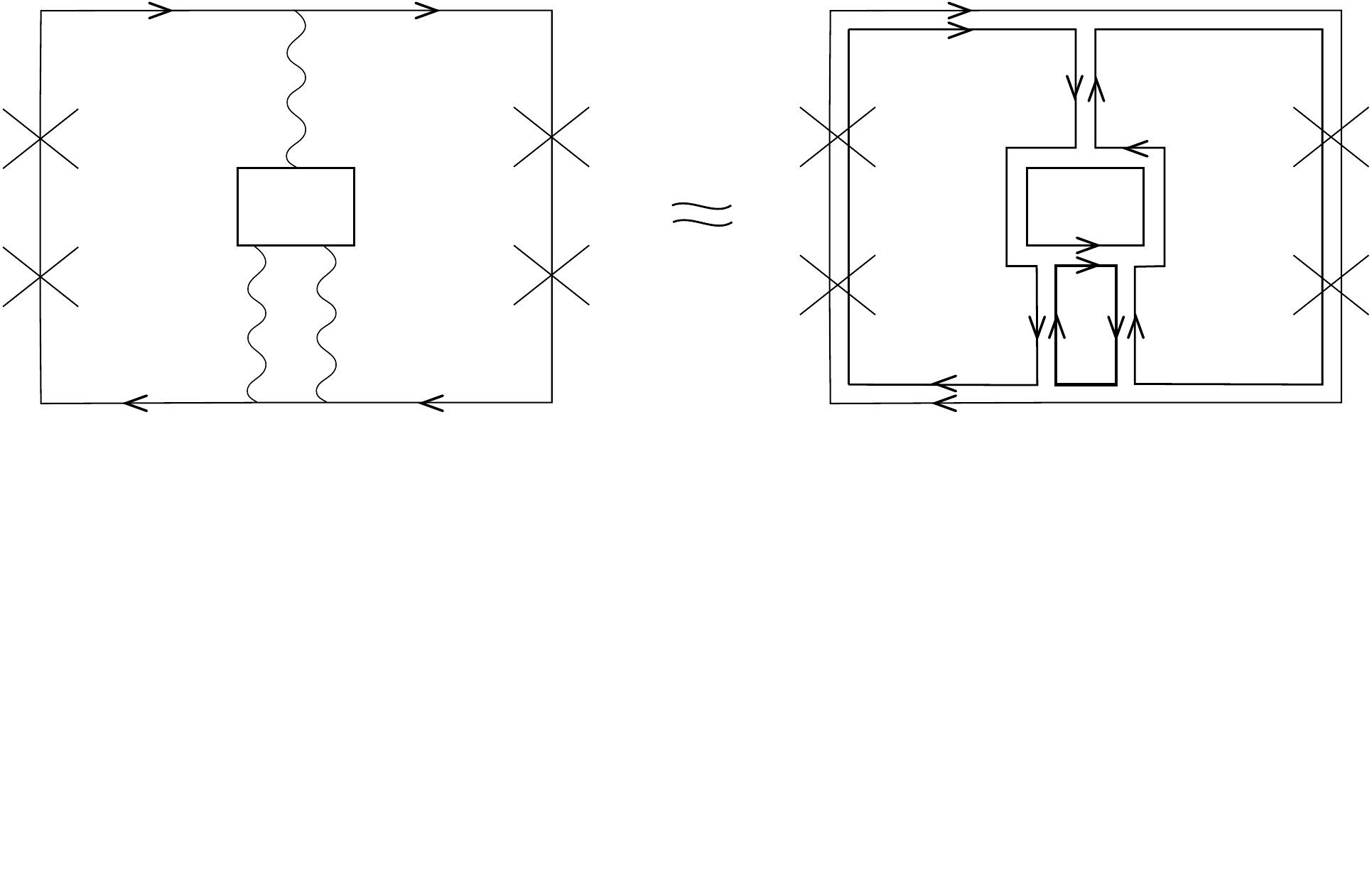}
%\leavevmode {
%\includegraphics[width=6cm,clip=true]{carhomp}
}
\vskip -3cm
\caption[]
{Diagram for the scattering amplitude, A including an internal 2 index quark loop. } \label{FigD}
\end{figure}

    There is still another different feature with respect to the 't Hooft expansion; consider the typical
$\pi \pi$ scattering diagram with an extra internal (two index)
quark loop, as shown in Fig.~\ref{FigD}.

 In this diagram there are 
four X's (factor from Eq.(\ref{wf})), five index loops (factor 
from Eq.(\ref{loop})) and six gauge coupling constants. These
combine to give a large N scaling behavior proportional
to $1/N^2$ for the $\pi \pi$ scattering amplitude. We see that
diagrams with an extra internal 2 index quark loop are not suppressed
compared to the leading diagrams. This is analogous,
as pointed out in \cite{Kiritsis:1989ge}, to the 
behavior of diagrams with an extra gluon loop in the 't Hooft
extrapolation scheme. Now, Fig.~\ref{FigD} is a diagram which can describe
a sigma particle exchange. Thus in the 2 index quark scheme,
``exotic" four quark resonances can appear at the leading order
in addition to the usual two quark resonances. The possibility of a sigma-type state appearing at leading order
means that one can construct a unitary $\pi \pi$ amplitude  
already at $N$ = 3 in the 2 antisymmetric index scheme. From the
 point of view of low energy $\pi \pi$ scattering, it seems to be unavoidable to say
that the 2 index scheme is  more realistic than the 't Hooft
scheme given the existence of a four quark type sigma.

  Of course, the usual 't Hooft extrapolation has a number of
 other things to recommend it. These include the fact that nearly all
meson resonances seem to be of the quark- antiquark type, the OZI rule
predicted holds to a good approximation and baryons emerge
 in an elegant way as solitons in the model. 

 A fair statement is that each extrapolation emphasizes
different aspects of $N$ = 3 QCD.
 In particular, the usual scheme
is not really a replacement for the true theory. That appears to be the
 meaning of the fact that the continuation to $N>3$ is not unique.  

\subsubsection{Quarks in two index symmetric color representation}

    Clearly the assignment of femions to the two index
 symmetric representation of
color SU(3) is very similar to the previous case.
 We denote the fields
as,
\begin{equation}
q_{\{AB \}}=q_{\{BA \}} \ .
\label{symquarks}
\end{equation}
There will be $N(N+1)/2$ different color states for the two index
 symmetric quarks. This means that
there is no value of $N$ for which
 the symmetric theory can be made to
 correspond to true QCD. On the other hand, for large N we can make the approximation 
\begin{equation}
A^{sym}(N) \approx A^{asym}(N),
\label{symasym}
\end{equation}
for the $\pi \pi$ scattering amplitude.

    As far as the large N counting goes,
 the mnemonics in Figs. \ref{FigA}-\ref{FigD}
are still applicable to the case of quarks
 in the two index symmetric color
representation. For not so large N, the scaling
 factor for the pion insertion is
\begin{eqnarray}
            \frac{\sqrt{2}}{\sqrt{N(N+1)}},
\label{symwf}
\end{eqnarray}        
and the pion decay constant scales as
\begin{equation}
F_{\pi}^{sym}(N)\propto \sqrt{\frac{N(N+1)}{2}}.
\label{symfpiscaling}
\end{equation}.

     With the identification $A^{QCD}=A^{asym}(3)$, the use of
Eq.(\ref{symasym})
enables us to estimate the large N
scattering amplitude as,
\begin{equation}
A^{sym}(N) \approx \frac{6}{N^2}A^{QCD}.
\label{bigN}
\end{equation}

In 
applications to minimal walking technicolor theories
this formula 
is useful for making estimates involving weak gauge bosons via
the Goldstone boson equivalence theorem \cite{Lee:1977eg}.

\begin{figure}[htbp]
\centering %\leavevmode
{
\includegraphics[width=
4.5cm,clip=true]{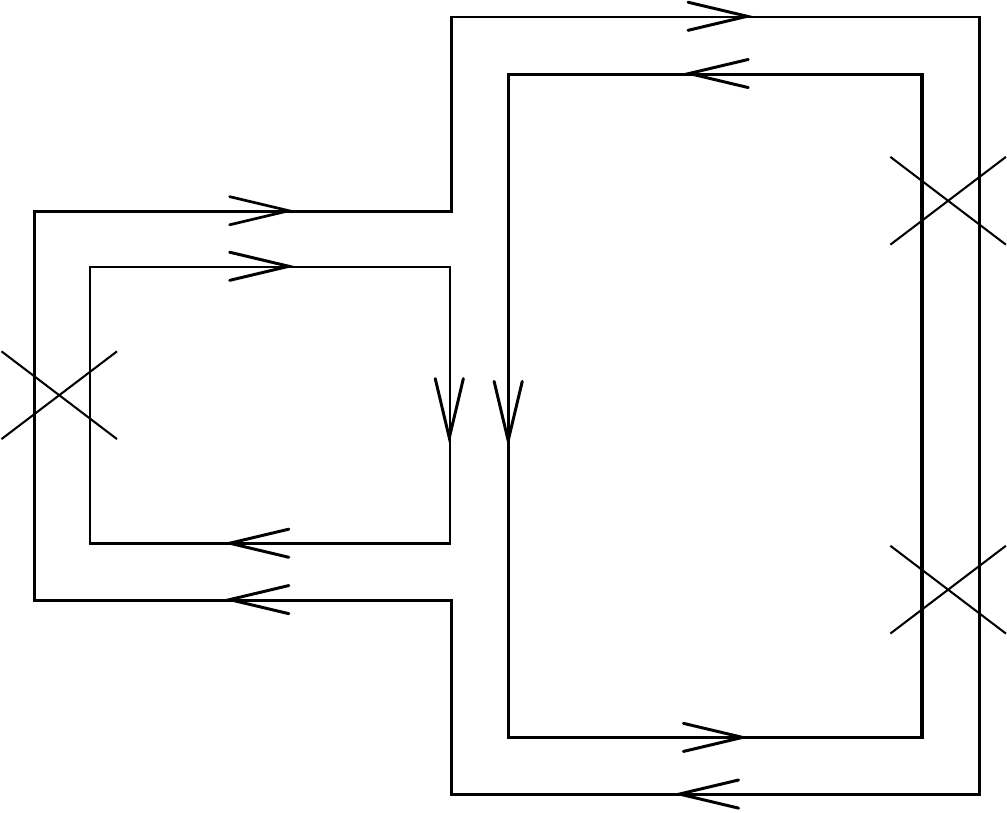}
%\leavevmode {
%\includegraphics[width=6cm,clip=true]{carhomp}
}
\caption[]
{Diagram for meson decay into two glueballs.} \label{FigE}
\end{figure}

     Finally we remark on the large N scaling rules for meson 
and glueball masses and decays in either the two index
antisymmetric or two index symmetric schemes. Both meson and
 glueball masses scale
as $(N)^0$. Furthermore, all six reactions of the type
\begin{equation}
a\rightarrow b + c,
 \label{abc}
\end{equation}
where a,b and c can stand for either a meson or a glueball,
scale as $1/N$. This is illustrated in Fig. \ref{FigE} for the case
of a meson decaying into two glueballs; note that the
glueball insertion scales as $1/N$ and that two interaction
 vertices are involved.

\vskip -.5 cm
 
 \subsubsection{Spectrum for Higher Dimensional Representations: Higgsfull theories}
Combining our knowledge of the QCD spectrum together with the rules above for the two index antisymmetric representation we deduce the following large N scaling:
\begin{eqnarray}
m_{\rho} ^2&\sim &\Lambda_{QCD}^2 \ , \qquad \Gamma_{\rho} \sim \frac{2}{N(N - 1)} \\
m_{f_0}^2 &\sim & \Lambda_{QCD}^2 \ , \qquad \Gamma_{f_0} \sim\frac{2}{N(N - 1)}  \ .
\end{eqnarray}
 The fact that in QCD  the state $f_0(600)$ is not narrow indicates that the unknown coefficient in the expression for the width, expected to be order one, is large. However, as we increase the number of colors we expect this state to become quickly narrow.  
Scaling up these results for a technicolor theory with $N_{TC}$  colors and fermions in the two index antisymmetric representation we have:
\begin{eqnarray}
M_{T\rho} &=& \frac{\sqrt{2}v_{weak}}{F_{\pi}}
\frac{\sqrt{3}\sqrt{2}}
{\sqrt{N_D\, N_{TC}(N_{TC}-1) }} m_{\rho}
 \\
M_{T f_{0}} &=&\frac{\sqrt{2}v_{weak}}{F_{\pi}}
\frac{\sqrt{3}\sqrt{2}}
{\sqrt{N_D\, N_{TC}(N_{TC}-1) }} m_{f_0}  \ .
\end{eqnarray} 
The input values here are the QCD masses for $f_0(600)$ and $\rho(770)$. Differently from the 't Hooft case the scalar will remain lighter than the associate technivector meson for any number of technicolors. Finally, increasing the number of technicolors and techniflavors we can achieve a very light scalar, lighter then its own technivector. Since in these theories one cannot differentiate a fermion-antifermion state from a multi fermion states we map the lightest scalar into the composite Higgs. 

So, even without invoking walking dynamics, higher dimensional representations provide a composite Higgs lighter than the technivector meson. These theories are Higgsfull for any number of colors. 

\begin{figure}[htbp]
\centering \leavevmode
{
\includegraphics[width=
7.5cm,clip=true]{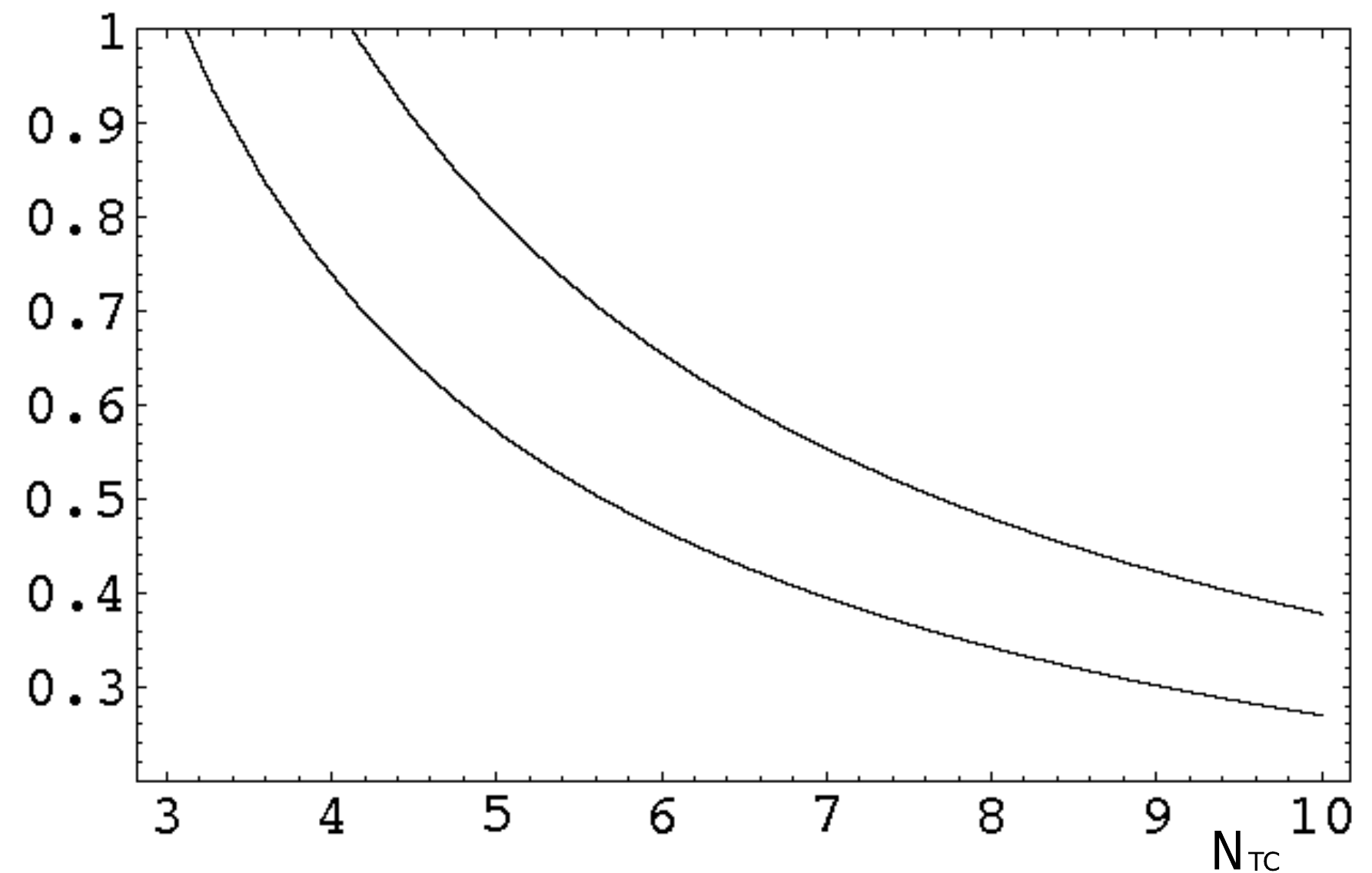}}\hskip .5cm 
\leavevmode{
\includegraphics[width=7.5cm,clip=true]{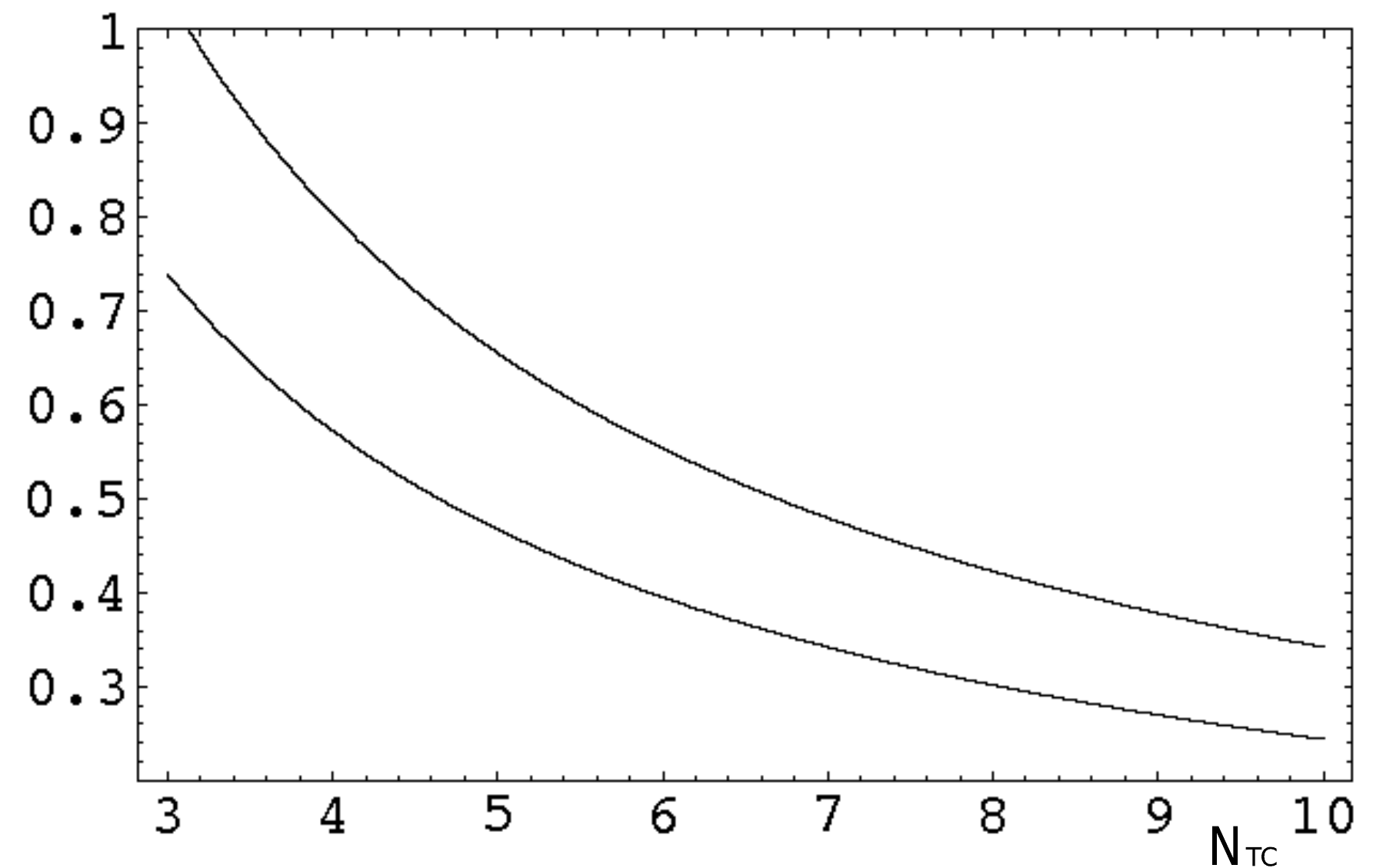}}
\caption[Corrigan-Ramond Scaling]
{Mass of the lightest vector meson (higher curve) and scalar meson (lower curve) as function of the number of colors in TeV units. At $N=3$ we match the spectrum to QCD for the two index antisymmetric representation. Here we use $N_D=2$. On the left panel we draw the spectrum for the two index antisymmetric extension of QCD while on the right we consider the two index symmetric representation. Note that now for any $N_D$ and $N_{TC}$ the scalar is always lighter than the associated vector meson. } \label{levelsplittingthooft}
\end{figure}

One can pass from the two index antisymmetric to the two index symmetric by replacing $N_{TC}-1$ with $N_{TC}+1$ in the expressions above and matching the result at infinite number of colors. In Fig.~\ref{levelsplittingthooft} the physical spectrum of spin one vector bosons and the lightest scalar is reported in TeV units in the case of two doublets ($N_D=2$ ) of technifermions for different number of colors. At $N=3$ we match the spectrum to QCD  for the two index antisymmetric representation. On the left panel we draw the spectrum for the two index antisymmetric extension of QCD while on the right we consider the two index symmetric representation normalized at large N with the two index antisymmetric one. {}For any $N_D$ and $N_{TC}$ the scalar is always lighter than the associated vector meson. In the case of the two index symmetric on approaches light masses a little faster when increasing the number of colors. 

Above we  demonstrated that i) It is possible to have composite theories which are Higgsfull ii) the resulting composite Higgs is light with respect to the TeV scale. The comparison with precision data must then be revised for these theories since the associated $S$ parameter constraint changes. Note that in the proof we  used only a straightforward geometrical scaling.

What happens to the mass of the composite Higgs in the case of walking?  By increasing the number of flavors all of the composite states from the chiral-symmetric broken side become massless when reaching the fixed point since the only invariant scale of the theory vanishes there \cite{Chivukula:1996kg}. This is supported by lattice simulations \cite{Catterall:2007yx}. We are, however, interested in the ratio between the masses of the various states to the pion decaying constant which is fixed to be the electroweak scale. Simple arguments suggest that if the transition is second order then there will be a light composite Higgs or else its mass to decay constant ratio will not vanish near the conformal point. In any event one can write a low energy effective action for the composite scalar with the quantum numbers of the Higgs -- treating it as a dilaton -- using trace and axial anomaly as well as chiral symmetry as done in \cite{Sannino:1999qe}. A similar analysis using trace anomaly has been also discussed in \cite{Goldberger:2007zk}. The resulting action contains, by construction, non-analitc powers of the composite Higgs field \cite{Sannino:1999qe} and must be treated as generating functional for the anomalous transformations of the underlying dynamics. 

The possibility of a light composite Higgs  in (walking) technicolor was first advocated in \cite{Hong:2004td,Dietrich:2005jn,Dietrich:2005wk,Dietrich:2006cm} and also proposed in \cite{Goldberger:2007zk} and \cite{Doff:2008xx}.  Since, as shown above using standard scaling arguments, it is possible to construct technicolor theories with a light composite Higgs it is relevant to study its phenomenological signatures \cite{Zerwekh:2005wh,Fan:2008jk}.

\newpage

%% THE BIBLIOGRAPHY

\end{document}